\documentclass[journal,10pt]{IEEEtran}
\usepackage{amsmath}
\usepackage{eqparbox}
\usepackage[usenames, dvipsnames]{color}
\usepackage{mathtools,amssymb,bm,mathabx,nccmath}
\usepackage{amstext}
\usepackage{amssymb}
\usepackage{graphicx}
\usepackage{booktabs}
\usepackage{longtable}
\usepackage{multicol}
\usepackage{algorithmicx}
\usepackage[ruled]{algorithm}
\usepackage{algpseudocode}
\usepackage{algpascal}
\usepackage{algc}
\usepackage{amsfonts}
\usepackage{dsfont}
\usepackage{array}
\usepackage[most]{tcolorbox}
\usepackage{cases}
\usepackage{pgfplots}
\usepackage{caption}
\usepackage{xr-hyper}
\usepackage{hyperref}
\usepackage{amsthm}
\DeclareCaptionLabelSeparator{periodspace}{.~}
\usepackage{float}
\usepackage{cite}

\usepackage{epstopdf}
\usepackage{adjustbox}
\theoremstyle{remark}

\newcommand\ASTART{\bigskip\noindent\begin{minipage}[b]{0.5\linewidth}}
	
	\newcommand\AENDSKIP{\end{minipage}\bigskip}
\newcommand\AEND{\end{minipage}}
\ifCLASSOPTIONcaptionsoff
\usepackage[nomarkers]{endfloat}
\let\MYoriglatexcaption\caption
\renewcommand{\caption}[2][\relax]{\MYoriglatexcaption[#2]{#2}}
\fi
\theoremstyle{plain}
\newtheorem{thm}{\textbf{Theorem}}
\newtheorem{lem}{\textbf{Lemma}}
\newtheorem{prop}{\textbf{Proposition}}

\theoremstyle{definition}

\newtheorem{rem}{\textbf{Remark}}

\newcommand*{\rom}[1]{\expandafter\@slowromancap\romannumeral #1@}
\def\change{black}

\newcommand{\RN}[1]{%
\textup{\uppercase\expandafter{\romannumeral#1}}%
}

\usepackage{standalone}
\graphicspath{{./Figures/}}
\usepackage{times}
\usepackage[latin1]{inputenc}
\usepackage[T1]{fontenc}
\usepackage[english]{babel}
\usepackage{graphicx}
\usepackage{amsmath, amsfonts, dsfont,amssymb, graphicx, array, tabularx, booktabs,multicol}
\usepackage{amsthm}
\usepackage{epsf,epsfig}
\usepackage{setspace}

\usepackage{multirow}
\usepackage{url}
\usepackage{color}
\usepackage[nolist,printonlyused]{acronym}      
\usepackage{comment}
\usepackage{cite}
\usepackage{bm}
\usepackage{tikz}
\usetikzlibrary{decorations.pathreplacing}

\usepackage{hyperref}
\usepackage{mathtools}

\usepackage{blindtext}
\usepackage{xcolor}
\newcommand{\gf}[1]{\textcolor{black}{{#1}}}

\newtheorem{cor}{Corollary}

\newcommand{\mx}[1]{\mathbf{#1}}
\newcommand{\bs}[1]{\boldsymbol{#1}}
\providecommand{\keywords}[1]{\textbf{\textit{Index terms---}} #1}
\usepackage{subcaption}
\definecolor{amber}{rgb}{1.0, 0.49, 0.0}
\definecolor{ao}{rgb}{0.0, 0.5, 0.0}

\def\R2#1{\textcolor{black}{#1}}
\def\R3#1{\textcolor{black}{#1}}

\renewcommand{\triangleq}{\mathbin{\setstackgap{S}{0pt}\stackMath\Shortstack{\smalltriangleup\\ =}}}
\usepackage[usestackEOL]{stackengine} 

\begin{document}

\title{Sampling}
\singlespacing
\title{Exploiting Spatial and Temporal Correlations in Massive MIMO Systems Operating Over Non-Stationary Aging Channels}
\singlespacing
\author{
 Sajad Daei $^{\dag}$, G\'{a}bor Fodor$^{\star\dag}$, Mikael Skoglund $^{\dag}$\\
 \small $^\dag$KTH Royal Institute of Technology, Stockholm, Sweden. \\
\small $^\star$Ericsson Research, Stockholm, Sweden.\\

}
\maketitle
\pagestyle{plain}


\begin{acronym}
  \acro{2G}{Second Generation}
  \acro{3G}{3$^\text{rd}$~Generation}
  \acro{3GPP}{3$^\text{rd}$~Generation Partnership Project}
  \acro{4G}{4$^\text{th}$~Generation}
  \acro{5G}{5$^\text{th}$~Generation}
  \acro{AA}{Antenna Array}
  \acro{AC}{Admission Control}
  \acro{AD}{Attack-Decay}
  \acro{ADSL}{Asymmetric Digital Subscriber Line}
  \acro{DASE}{Detetministic Averaged Spectral Efficiency}
	\acro{AHW}{Alternate Hop-and-Wait}
  \acro{AMC}{Adaptive Modulation and Coding}
	\acro{AP}{Access Point}
  \acro{APA}{Adaptive Power Allocation}
  \acro{AR}{autoregressive}
  \acro{ARMA}{Autoregressive Moving Average}
  \acro{ATES}{Adaptive Throughput-based Efficiency-Satisfaction Trade-Off}
  \acro{AWGN}{additive white Gaussian noise}
  \acro{BB}{Branch and Bound}
  \acro{BD}{Block Diagonalization}
  \acro{BER}{bit error rate}
  \acro{BF}{Best Fit}
  \acro{BLER}{BLock Error Rate}
  \acro{BPC}{Binary power control}
  \acro{BPSK}{binary phase-shift keying}
  \acro{BPA}{Best \ac{PDPR} Algorithm}
  \acro{BRA}{Balanced Random Allocation}
  \acro{BS}{base station}
  \acro{LoS}{line of sight}
  \acro{NLoS}{non-line of sight}
  \acro{AoA}{angle of arrival}
  \acro{AoD}{angle of departure}
  \acro{CAP}{Combinatorial Allocation Problem}
  \acro{CAPEX}{Capital Expenditure}
  \acro{CBF}{Coordinated Beamforming}
  \acro{CBR}{Constant Bit Rate}
  \acro{CBS}{Class Based Scheduling}
  \acro{CC}{Congestion Control}
  \acro{CDF}{Cumulative Distribution Function}
  \acro{CDMA}{Code-Division Multiple Access}
  \acro{CL}{Closed Loop}
  \acro{CLPC}{Closed Loop Power Control}
  \acro{CNR}{Channel-to-Noise Ratio}
  \acro{CPA}{Cellular Protection Algorithm}
  \acro{CPICH}{Common Pilot Channel}
  \acro{CoMP}{Coordinated Multi-Point}
  \acro{CQI}{Channel Quality Indicator}
  \acro{CRM}{Constrained Rate Maximization}
	\acro{CRN}{Cognitive Radio Network}
  \acro{CS}{Coordinated Scheduling}
  \acro{CSI}{channel state information}
  \acro{CSIR}{channel state information at the receiver}
  \acro{CSIT}{channel state information at the transmitter}
  \acro{CUE}{cellular user equipment}
  \acro{D2D}{device-to-device}
  \acro{DCA}{Dynamic Channel Allocation}
  \acro{DE}{Differential Evolution}
  \acro{DFT}{Discrete Fourier Transform}
  \acro{DIST}{Distance}
  \acro{DL}{downlink}
  \acro{DMA}{Double Moving Average}
	\acro{DMRS}{Demodulation Reference Signal}
  \acro{D2DM}{D2D Mode}
  \acro{DMS}{D2D Mode Selection}
  \acro{DPC}{Dirty Paper Coding}
  \acro{DRA}{Dynamic Resource Assignment}
  \acro{DSA}{Dynamic Spectrum Access}
  \acro{DSM}{Delay-based Satisfaction Maximization}
  \acro{ECC}{Electronic Communications Committee}
  \acro{EFLC}{Error Feedback Based Load Control}
  \acro{EI}{Efficiency Indicator}
  \acro{eNB}{Evolved Node B}
  \acro{EPA}{Equal Power Allocation}
  \acro{EPC}{Evolved Packet Core}
  \acro{EPS}{Evolved Packet System}
  \acro{E-UTRAN}{Evolved Universal Terrestrial Radio Access Network}
  \acro{ES}{Exhaustive Search}
  \acro{FDD}{frequency division duplexing}
  \acro{FDM}{Frequency Division Multiplexing}
  \acro{FER}{Frame Erasure Rate}
  \acro{FF}{Fast Fading}
  \acro{FSB}{Fixed Switched Beamforming}
  \acro{FST}{Fixed SNR Target}
  \acro{FTP}{File Transfer Protocol}
  \acro{GA}{Genetic Algorithm}
  \acro{GBR}{Guaranteed Bit Rate}
  \acro{GNSS}{Global Navigation Satellite System}
  \acro{GLR}{Gain to Leakage Ratio}
  \acro{GOS}{Generated Orthogonal Sequence}
  \acro{GPL}{GNU General Public License}
  \acro{GRP}{Grouping}
  \acro{HARQ}{Hybrid Automatic Repeat Request}
  \acro{HMS}{Harmonic Mode Selection}
  \acro{HOL}{Head Of Line}
  \acro{HSDPA}{High-Speed Downlink Packet Access}
  \acro{HSPA}{High Speed Packet Access}
  \acro{HTTP}{HyperText Transfer Protocol}
  \acro{ICMP}{Internet Control Message Protocol}
  \acro{ICI}{Intercell Interference}
  \acro{ID}{Identification}
  \acro{IETF}{Internet Engineering Task Force}
  \acro{ILP}{Integer Linear Program}
  \acro{JRAPAP}{Joint RB Assignment and Power Allocation Problem}
  \acro{UID}{Unique Identification}
  \acro{IID}{Independent and Identically Distributed}
  \acro{IIR}{Infinite Impulse Response}
  \acro{ILP}{Integer Linear Problem}
  \acro{IMT}{International Mobile Telecommunications}
  \acro{INV}{Inverted Norm-based Grouping}
	\acro{IoT}{Internet of Things}
  \acro{IP}{Internet Protocol}
  \acro{IPv6}{Internet Protocol Version 6}
  \acro{ISD}{Inter-Site Distance}
  \acro{ISI}{Inter Symbol Interference}
  \acro{ITU}{International Telecommunication Union}
  \acro{JOAS}{Joint Opportunistic Assignment and Scheduling}
  \acro{JOS}{Joint Opportunistic Scheduling}
  \acro{JP}{Joint Processing}
	\acro{JS}{Jump-Stay}
    \acro{KF}{Kalman filter}
  \acro{KKT}{Karush-Kuhn-Tucker}
  \acro{L3}{Layer-3}
  \acro{LAC}{Link Admission Control}
  \acro{LA}{Link Adaptation}
  \acro{LC}{Load Control}
  \acro{LOS}{Line of Sight}
  \acro{LP}{Linear Programming}
  \acro{LS}{least squares}
  \acro{LTE}{Long Term Evolution}
  \acro{LTE-A}{LTE-Advanced}
  \acro{LTE-Advanced}{Long Term Evolution Advanced}
  \acro{M2M}{Machine-to-Machine}
  \acro{MAC}{Medium Access Control}
  \acro{MANET}{Mobile Ad hoc Network}
  \acro{MC}{Modular Clock}
  \acro{MCS}{Modulation and Coding Scheme}
  \acro{MDB}{Measured Delay Based}
  \acro{MDI}{Minimum D2D Interference}
  \acro{MF}{Matched Filter}
  \acro{MG}{Maximum Gain}
  \acro{MH}{Multi-Hop}
  \acro{MIMO}{multiple input multiple output}
  \acro{MINLP}{Mixed Integer Nonlinear Programming}
  \acro{MIP}{Mixed Integer Programming}
  \acro{MISO}{multiple input single output}
  \acro{ML}{maximum likelihood}
  \acro{MLWDF}{Modified Largest Weighted Delay First}
  \acro{MME}{Mobility Management Entity}
  \acro{MMSE}{minimum mean squared error}
  \acro{LMMSE}{linear MMSE}
  \acro{MOS}{Mean Opinion Score}
  \acro{ASE}{averaged spectral efficiency}
  \acro{MPF}{Multicarrier Proportional Fair}
  \acro{MRA}{Maximum Rate Allocation}
  \acro{MR}{Maximum Rate}
  \acro{MRC}{maximum ratio combining}
  \acro{MRT}{Maximum Ratio Transmission}
  \acro{MRUS}{Maximum Rate with User Satisfaction}
  \acro{MS}{mobile station}
  \acro{MSE}{mean squared error}
  \acro{MSI}{Multi-Stream Interference}
  \acro{MTC}{Machine-Type Communication}
  \acro{MTSI}{Multimedia Telephony Services over IMS}
  \acro{MTSM}{Modified Throughput-based Satisfaction Maximization}
  \acro{MU-MIMO}{multi-user multiple input multiple output}
  \acro{MU}{multi-user}
  \acro{NAS}{Non-Access Stratum}
  \acro{NB}{Node B}
  \acro{NE}{Nash equilibrium}
  \acro{NCL}{Neighbor Cell List}
  \acro{NLP}{Nonlinear Programming}
  \acro{NLOS}{Non-Line of Sight}
  \acro{NMSE}{Normalized Mean Square Error}
  \acro{NORM}{Normalized Projection-based Grouping}
  \acro{NP}{non-polynomial time}
  \acro{NRT}{Non-Real Time}
  \acro{NSPS}{National Security and Public Safety Services}
  \acro{O2I}{Outdoor to Indoor}
  \acro{OFDMA}{orthogonal frequency division multiple access}
  \acro{OFDM}{orthogonal frequency division multiplexing}
  \acro{OFPC}{Open Loop with Fractional Path Loss Compensation}
	\acro{O2I}{Outdoor-to-Indoor}
  \acro{OL}{Open Loop}
  \acro{OLPC}{Open-Loop Power Control}
  \acro{OL-PC}{Open-Loop Power Control}
  \acro{OPEX}{Operational Expenditure}
  \acro{ORB}{Orthogonal Random Beamforming}
  \acro{JO-PF}{Joint Opportunistic Proportional Fair}
  \acro{OSI}{Open Systems Interconnection}
  \acro{PAIR}{D2D Pair Gain-based Grouping}
  \acro{PAPR}{Peak-to-Average Power Ratio}
  \acro{P2P}{Peer-to-Peer}
  \acro{PC}{Power Control}
  \acro{PCI}{Physical Cell ID}
  \acro{PDF}{probability density function}
  \acro{PDPR}{pilot-to-data power ratio}
  \acro{PER}{Packet Error Rate}
  \acro{PF}{Proportional Fair}
  \acro{P-GW}{Packet Data Network Gateway}
  \acro{PL}{Pathloss}
  \acro{PPR}{pilot power ratio}
  \acro{PRB}{physical resource block}
  \acro{PROJ}{Projection-based Grouping}
  \acro{ProSe}{Proximity Services}
  \acro{PS}{Packet Scheduling}
  \acro{PSAM}{pilot symbol assisted modulation}
  \acro{PSO}{Particle Swarm Optimization}
  \acro{PZF}{Projected Zero-Forcing}
  \acro{QAM}{Quadrature Amplitude Modulation}
  \acro{QoS}{Quality of Service}
  \acro{QPSK}{Quadri-Phase Shift Keying}
  \acro{RAISES}{Reallocation-based Assignment for Improved Spectral Efficiency and Satisfaction}
  \acro{RAN}{Radio Access Network}
  \acro{RA}{Resource Allocation}
  \acro{RAT}{Radio Access Technology}
  \acro{RATE}{Rate-based}
  \acro{RB}{resource block}
  \acro{RBG}{Resource Block Group}
  \acro{REF}{Reference Grouping}
  \acro{RLC}{Radio Link Control}
  \acro{RM}{Rate Maximization}
  \acro{RNC}{Radio Network Controller}
  \acro{RND}{Random Grouping}
  \acro{RRA}{Radio Resource Allocation}
  \acro{RRM}{Radio Resource Management}
  \acro{RSCP}{Received Signal Code Power}
  \acro{RSRP}{Reference Signal Receive Power}
  \acro{RSRQ}{Reference Signal Receive Quality}
  \acro{RR}{Round Robin}
  \acro{RRC}{Radio Resource Control}
  \acro{RSSI}{Received Signal Strength Indicator}
  \acro{RT}{Real Time}
  \acro{RU}{Resource Unit}
  \acro{RUNE}{RUdimentary Network Emulator}
  \acro{RV}{Random Variable}
  \acro{SAC}{Session Admission Control}
  \acro{SCM}{Spatial Channel Model}
  \acro{SC-FDMA}{Single Carrier - Frequency Division Multiple Access}
  \acro{SD}{Soft Dropping}
  \acro{S-D}{Source-Destination}
  \acro{SDPC}{Soft Dropping Power Control}
  \acro{SDMA}{Space-Division Multiple Access}
  \acro{SE}{spectral efficiency}
  \acro{SER}{Symbol Error Rate}
  \acro{SES}{Simple Exponential Smoothing}
  \acro{S-GW}{Serving Gateway}
  \acro{SINR}{signal-to-interference-plus-noise ratio}
  \acro{SI}{Satisfaction Indicator}
  \acro{SIP}{Session Initiation Protocol}
  \acro{SISO}{single input single output}
  \acro{SIMO}{single input multiple output}
  \acro{SIR}{signal-to-interference ratio}
  \acro{SLNR}{Signal-to-Leakage-plus-Noise Ratio}
  \acro{SMA}{Simple Moving Average}
  \acro{SNR}{signal-to-noise ratio}
  \acro{SORA}{Satisfaction Oriented Resource Allocation}
  \acro{SORA-NRT}{Satisfaction-Oriented Resource Allocation for Non-Real Time Services}
  \acro{SORA-RT}{Satisfaction-Oriented Resource Allocation for Real Time Services}
  \acro{SPF}{Single-Carrier Proportional Fair}
  \acro{SRA}{Sequential Removal Algorithm}
  \acro{SRS}{Sounding Reference Signal}
  \acro{SU-MIMO}{single-user multiple input multiple output}
  \acro{SU}{Single-User}
  \acro{SVD}{Singular Value Decomposition}
  \acro{TCP}{Transmission Control Protocol}
  \acro{TDD}{time division duplexing}
  \acro{TDMA}{Time Division Multiple Access}
  \acro{TETRA}{Terrestrial Trunked Radio}
  \acro{TP}{Transmit Power}
  \acro{TPC}{Transmit Power Control}
  \acro{TTI}{Transmission Time Interval}
  \acro{TTR}{Time-To-Rendezvous}
  \acro{TSM}{Throughput-based Satisfaction Maximization}
  \acro{TU}{Typical Urban}
  \acro{UE}{User Equipment}
  \acro{ULA}{Uniform Linear Array}
  \acro{UEPS}{Urgency and Efficiency-based Packet Scheduling}
  \acro{UL}{uplink}
  \acro{UMTS}{Universal Mobile Telecommunications System}
  \acro{URI}{Uniform Resource Identifier}
  \acro{URM}{Unconstrained Rate Maximization}
  \acro{UT}{user terminal}
  \acro{VR}{Virtual Resource}
  \acro{VoIP}{Voice over IP}
  \acro{WAN}{Wireless Access Network}
  \acro{WCDMA}{Wideband Code Division Multiple Access}
  \acro{WF}{Water-filling}
  \acro{WiMAX}{Worldwide Interoperability for Microwave Access}
  \acro{WINNER}{Wireless World Initiative New Radio}
  \acro{WLAN}{Wireless Local Area Network}
  \acro{WMPF}{Weighted Multicarrier Proportional Fair}
  \acro{WPF}{Weighted Proportional Fair}
  \acro{WSN}{Wireless Sensor Network}
  \acro{WWW}{World Wide Web}
  \acro{XIXO}{(Single or Multiple) Input (Single or Multiple) Output}
  \acro{ZF}{zero-forcing}
  \acro{ZMCSCG}{Zero Mean Circularly Symmetric Complex Gaussian}
\end{acronym}

\begin{abstract}
This work investigates a multi-user, multi-antenna uplink wireless system, \gf{in which} multiple users transmit signals to a base station. Prior research has explored the potential for linear growth in spectral efficiency by employing multiple transmit and receive antennas. This gain depends heavily on the quality of channel state information and \gf{the number of} uncorrelated antennas. However, spatial correlations, arising from closely-spaced antennas and channel aging effects -- stemming from the difference between the channel \gf{state} at pilot and data time instances -- can substantially counteract these benefits, and degrade the transmission rate, especially in non-stationary environments. To address these challenges, this work introduces a real-time beamforming framework to compensate for the spatial correlation and channel aging effects. \gf{First}, a channel estimation scheme leveraging temporal channel correlations and considering mobile device velocity and antenna spacing \gf{is developed}. Subsequently, an expression approximating the average spectral efficiency -- \gf{which depends} on pilot spacing, pilot and data powers, and beamforming vectors -- is obtained. By maximizing this expression, optimal parameters are identified. Numerical results demonstrate the effectiveness of the proposed approach compared to prior works. Interestingly, the optimal pilot spacing remains unaffected by large-scale channel parameters and the velocities of \gf{interfering} users. The impact of interference components also diminishes with an increasing number of transmit antennas.
\end{abstract}  


\section{Introduction}
In the past two decades, the integration of multiple antennas at both the transmitter and receiver nodes has emerged as a crucial advancement, offering significant advantages in achieving high \ac{SE} \cite{rhee2004optimality,rusek2012scaling,jafar2001channel,jafar2004transmitter,lu2014overview,soysal2009optimality,Hassibi:03}. Notably, massive \ac{MIMO} technology, characterized by a substantial number of antennas at the \ac{BS}, enhances the coverage and \ac{SE} in the uplink of wireless communications. The studies by \cite{rusek2012scaling,rhee2004optimality} highlight a linear growth in ergodic capacity with the number of transmit antennas compared to the single transmit antenna case. 
\gf{Notably,} this linear gain is achieved in high \ac{SINR} regimes with uncorrelated  transmit and receive antennas, \gf{especially} when perfect \ac{CSIT} is available \cite{Hassibi:03,Hoydis:13, Fodor:16}. 

Furthermore, the correlation  between transmit antennas in uplink communications due to insufficient antenna spacing in small mobile devices can \gf{degrade} the \ac{SE}. This necessitates optimal signal design strategies at the transmitter. Beamforming, a scalar coding strategy where the transmit covariance matrix is unit-rank, has been identified as \gf{a possible solution} \cite{rhee2004optimality,jafar2004transmitter}. In beamforming, the symbol stream undergoes coding and multiplication by different coefficients at each antenna before transmission. In the context of \ac{MU-MIMO}, studies such as \cite{rhee2004optimality} demonstrate that user-specific beamforming achieves the sum-capacity with partial channel side information, as discussed in \cite{jafar2004transmitter,jorswieck2004channel}.

In \gf{multi-cell} single-antenna uplink scenarios, 
works such as \cite{marzetta2010noncooperative,papazafeiropoulos2015deterministic,Hoydis:13} suggest that linear decoders and precoders behave nearly optimally. Notably, the impact of intra-cell interference diminishes as the number of \ac{BS} antennas becomes sufficiently large. Additionally, some studies have derived deterministic approximations of the \ac{SINR} for uplink scenarios with \ac{MRC} and \ac{MMSE} receivers under block-fading channel assumptions, assuming that the number of \ac{BS} antennas and users tend to infinity at the same rate.

While the advantages highlighted in the mentioned studies concerning multiple antennas are evident, these benefits depend critically on the perfect knowledge of the channel state at the transmitter. 
\gf{Indeed, the 
crucial role of \ac{CSIT} has been analyzed and emphasized in several seminal works;
see e.g., \cite{Hassibi:03, Kobyashi:11, Khalilsarai:23,daei2024improved}.}

However, existing models often assume block-fading, where the channel remains constant during training and data \gf{transmission} times -- a condition that may not hold in practice, particularly in non-stationary wireless channels as modeled in \cite{non_stationary_model, Banerjee:20, Iimori:21, Loschenbrand:22, Song:22, cheng2022channel, zou2023joint, bian2021general, daei2024towards}. 
Furthermore, due to the relative movement of mobile users in wireless communications, delayed \ac{CSIT}, known as channel aging, may occur, resulting from changes in the channel at pilot and data time slots. Neglecting this channel aging effect can negatively impact the capacity of massive \ac{MU-MIMO} systems, motivating several papers to leverage temporal correlation structures to improve channel estimates
\cite{Fodor:2021,Fodor:22,Abeida:10,Hijazi:10,Truong:13, Kong:2015, Yuan:20, Kim:20, Fodor:23, daei2024towards,daei2024optimaltransmitter}.

Studies exploring the impact of channel aging, such as \cite{Truong:13, Fodor:23,daei2024towards,daei2024optimaltransmitter}, have introduced closed-form expressions for the deterministic equivalent \ac{SINR} under channel aging, considering \ac{SIMO} uplink and \ac{MISO} downlink scenarios. The inherent challenges of pilot spacing and frame size dimensioning, discussed in \cite{Savazzi:09, Savazzi:09B, Fodor:23}, highlight the need to balance the resources for \ac{CSIR} acquisition and communication in the presence of channel aging, considering power budgets and evolving channel characteristics.

Motivated by these insights and utilizing tools from random matrix theory\cite{bai2008clt}, this paper introduces an asymptotically tight deterministic expression for the average \ac{SE} in the \ac{MU-MIMO} scenario, featuring multiple antennas at both uplink users and the \ac{BS}. The proposed framework capitalizes on the temporal dynamics of the channel within a multi-user context, emphasizing the importance of tailored \ac{CSIT} acquisition for \ac{MIMO} systems operating over non-stationary wireless channels. Specifically, the emphasis lies in the uplink of a \ac{MU-MIMO} system, where each user has multiple antennas \gf{operating over} non-stationary aging channels. Delving into the intricate trade-offs related to imperfect channel knowledge, temporal correlation, and channel aging contributes to a holistic understanding of \ac{MIMO} system performance in dynamic (and specifically non-stationary) wireless environments. Furthermore, the paper establishes a real-time beamforming framework to harness the evolving nature of both spatial and temporal correlations. Numerical experiments show the efficacy of the proposed approach in exploiting both spatial and temporal correlations.

\subsection{Contributions and Key Differences \gf{Compared With} Prior Works}
We summarize the contributions of this paper in the following list:
\begin{enumerate}
\item \textbf{Deterministic spectral efficiency in \ac{MU-MIMO} systems}: {We propose a novel, explicit deterministic expression that accurately characterizes the expected spectral efficiency in multi-user MIMO (MU-MIMO) systems with multiple antennas. Unlike previous works that rely on first-order deterministic equivalent approximations, such as those in \cite{Hoydis:13, wagner2012large,tulino2004random,adhikary2013joint, couillet2011deterministic, asgharimoghaddam2018decentralizing, daei2024towards, Fodor:23}, or the use-and-then-forget (UTF) bounds in \cite{marzetta2016fundamentals, ngo2013energy, marzetta2006much}, our proposed bound takes into account second-order determinist expressions that provides significantly improved accuracy.
This bound remains precise across a broad range of scenarios, including both sub-6 GHz and millimeter-wave (mm-wave) channels, and is particularly effective as the number of \ac{BS} antennas increases. Moreover, the number of users can be even much larger that the number of BS antennas. The theoretical validity of our approach is supported by mathematical analysis and extensive numerical simulations, demonstrating its superiority over existing methods in predicting the expected spectral efficiency.}

 \item \textbf{Optimal beamforming design for enhanced spectral efficiency}: {To leverage the advantages of employing multiple antennas at the transmitter, we investigate a beamforming vector designed to reshape the data symbols. The determination of optimal beamforming vectors involves maximizing the per-slot deterministic spectral efficiency. We find a closed-form relation for the optimal real-time deterministic beamformer. The utilization of these optimal beamformers significantly enhances the resulting deterministic equivalent spectral efficiency. 
This differs from \cite{couillet2012random}, which proposes a random beamforming approach.}

 \item \textbf{Optimal frame design and power allocation}: We propose an optimization framework aimed at maximizing the proposed deterministic-equivalent averaged spectral efficiency while satisfying some power constraints. This endeavor involves identifying optimal values for parameters such as the number of frames, frame sizes, pilot and data powers.
 \item \textbf{Proposing a new algorithm for resource optimization} By utilizing an alternative optimization technique and employing steepest ascent, we introduce an effective algorithm for the optimization of transmitter parameters, including pilot spacing, pilot and data powers, and beamformers.
\item    \textbf{Transmitter-centric optimization approach}: 
Our findings indicate that interference components, such as path loss and Doppler frequencies, do not exert an influence on the deterministic equivalent spectral efficiency when a large number of transmit antennas are employed. Additionally, interference components do not impact the optimal frame design, suggesting that all optimization tasks related to frame design can be performed at the transmitter. This eliminates the necessity of transmitting optimal parameters through feedback control loops from the receiver to the transmitter, as discussed in \cite{Truong:13,Fodor:23}.
In scenarios without incorporating transmit beamforming design and with single-antenna users, as discussed in \cite{daei2024towards}, interference may adversely affect the deterministic spectral efficiency. While previous works, including \cite{Fodor:23, Savazzi:09, Savazzi:09B}, advocate for achieving optimal pilot spacing at the receiver side and transmitting optimal values to the transmitter through feedback or control loops, it is essential to acknowledge that this approach may waste resources and make delays leading to the potential reduction in spectral efficiency.
    
    \item \textbf{Closed-form expressions for the correlation matrix in {MIMO} channels}:
    We formulate explicit closed-form expressions for the correlation matrix needed for \ac{MIMO} channels. Our findings demonstrate that these expressions are accessible in practical non-stationary MIMO time-varying environments. Specifically, in stationary environments, the derived formulas explicitly rely on the spacing between transmit and receive antenna elements, as well as the velocity of users.
\end{enumerate}
    
\subsection{Outline}
The paper is organized as follows. In Section \ref{sec:model}, we state the considered model and assumptions for the channel provided in Section \ref{sec:channel_model} and for the uplink pilot and data model provided in Section \ref{sec:pilot_data_model}. Section \ref{Sec:Proposed_scheme} outlines the roadmap for obtaining the instantaneous \ac{SINR} and introduces an alternative deterministic expression, leveraging random matrix theory to approximate the random \ac{SINR}. In Section \ref{Sec:Proposed_scheme}, we propose a novel optimization problem that takes into account this deterministic SE and finds the optimal values of beamforming vectors, frame sizes, number of frames and pilot and data powers. Moreover, we propose a heuristic algorithm to find the optimal values of frame size, number of frames and pilot and data powers. Finally, the paper is concluded in Section \ref{sec:conclusion}.

\subsection{Notation}
$\mx{I}_N$ denotes the identity matrix of size $N$. $\textbf{vec}$ stands for the column stacking vector operator that transforms a matrix $\mx{X}\in\mathbb{C}^{M\times N}$ into its vectorized version $\mx{x}\triangleq \textbf{vec}(\mx{X})\in\mathbb{C}^{M N\times 1}$. 
The cross covariance matrix between vectors $\mx{x}$ and $\mx{y}$ is shown by $\mx{C}_{\mx{x},\mx{y}}\triangleq \mathds{E}\big[ (\mx{x}-\mathds{E}[\mx{x}])(\mx{y}-\mathds{E}[\mx{y}])^{\rm H}\big]$. The auto-covariance of a random vector $\mx{x}$ is shown by $\mx{C}_{\mx{x}}\triangleq \mathds{E}\Big\{ (\mx{x}-\mathds{E}[\mx{x}])(\mx{x}-\mathds{E}[\mx{x}])^{\rm H}\Big\}$. $\mx{e}_k\in\mathbb{R}^N$ refers to a vector that has all components equal to zero except for the $k$-th component. $\mx{1}_N\in\mathbb{R}^{N\times 1}$ is an all-one vector of size $N$. $j\triangleq\sqrt{-1}$ is the imaginary unit. $\lambda_{\max}(\mx{A})$ stands for the maximum eigenvalue of the matrix $\mx{A}$. $[{A}]_{k,l}$ or $A[k,l]$ is used to denote the $(k,l)$-th element of the matrix $\mx{A}$. {\color{\change} The complex conjugate of a matrix $\mx{B}\in\mathbb{C}^{M\times N}$ is shown by $\mx{B}^{*}$. $I_0(\cdot)$ is the zeroth-order modified Bessel function of the first kind and is defined as $I_0(x)\triangleq \tfrac{1}{\pi}\int_{0}^{\pi} {\rm e}^{x \cos(t)}{\rm d}t$. The zeroth-order Bessel function of the first kind is related to the modified Bessel function as $J_0(x)=I_0(j x)$. The Frobenius inner product between two matrices $\mx{A},\mx{B}\in\mathbb{C}^{M\times N}$ is defined as $\langle \mx{A},\mx{B} \rangle\triangleq \sum_{i=1}^{M}\sum_{j=1}^{N} A[i,j] B^{*}[i,j]$  }. $\gamma\xrightarrow{\mathds{P}} \gamma^\circ $ means that $\gamma$ converges in probability to $\gamma^\circ$, i.e. $\mathds{P}[|\gamma-\gamma^\circ|>\epsilon]=0~\forall \epsilon>0$. $\mu_n\xrightarrow{a.s.}\mu $ implies that the spectral distribution of $\mu_n$ converges to the spectral distribution of $\mu$ in the large limit as $n\rightarrow \infty$. To simplify notation, throughout the paper, we tag User-1 as the intended user, and will sometimes
drop index $k=1$ when referring to the tagged user. Finally, the spectral norm and Frobenius norm of a matrix $\mx{A}$ is denoted by $\|\mx{A}\|$ and $\|\mx{A}\|_F$, respectively.

\section{System Model}\label{sec:model}
We consider a single-cell uplink communication system, where multiple users, each equipped with multiple antennas, transmit data symbols to a multi-antenna \ac{BS} with a \ac{ULA} (see Figure \ref{fig:rician_model}). The number of antennas employed at each user is denoted by $N_t$ and the number of antennas at the \ac{BS} is denoted by $N_r$. Each user transmits $M$ data frames with sizes $q_1,..., q_M$, as illustrated in Figure \ref{fig:framework}. Each frame consists of one pilot (training) time slot for \ac{CSIT}, with the remaining slots dedicated to data transmission. {We adopt a narrowband model, where multiple subcarriers fall within the coherent bandwidth and experience similar channel conditions, eliminating the need for a frequency-domain index. These subcarriers support Zadoff-Chu sequence-based pilots, enabling user separation in the code domain, ensuring pilot orthogonality while avoiding interference. This approach is consistent with prior works \cite{Fodor:16}, \cite{Fodor:22}, and \cite{Hoydis:13}.} In what follows, we describe the channel model and the received signal model at pilot and data time slots. { The time slot $t$ refers to the symbol time $t T_s$, where $T_s$ is the symbol period. Without loss of generality, we assume the symbol period $T_s=1$.}

\subsection{Channel Model}\label{sec:channel_model}
\begin{figure}
    \centering
    \includegraphics[scale=.23]{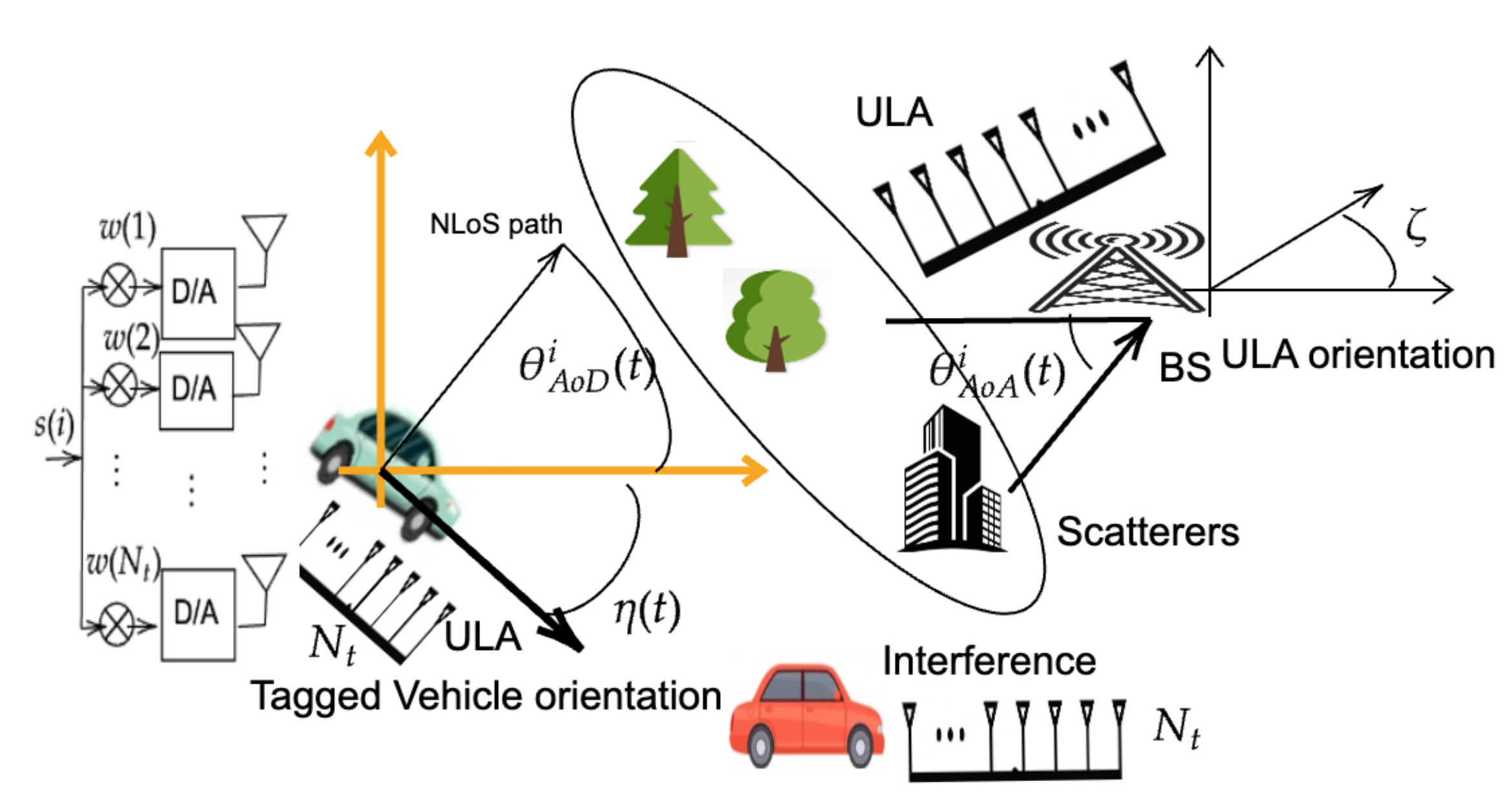}
    \caption{A typical example of multi-antenna multi-user wireless system. Left image: This image shows the beamforming strategy where a single symbol $s(t)$ at time slot $t$ is transmitted simultaneously from $N_t$ number of transmit antennas in the tagged user. {The vector $\mx{w}\triangleq[w(1),...,w(N_t)]^T$ is the beamforming vector}. Right image: This image shows a typical time-varying channel in vehicular networks where there are $L(t)$ number of scatterers between the tagged user and the \ac{BS} at time $t$.}
    \label{fig:rician_model}
\end{figure}

{\color{\change}
Consider a wireless communication scenario depicted in Figure~\ref{fig:rician_model}, where a tagged \ac{UE} is moving at time $t$ with velocity $\nu(t)$ in the direction given by the scalar angle $\eta(t)$. 
The \ac{BS}, equipped with \gf{a} \ac{ULA}, is oriented along the direction $\zeta$. The propagation environment between the tagged \ac{UE} (User 1) and the \ac{BS} consists of $L(t)$ scatterers at time $t$. Each scatterer induces a propagation path characterized by \ac{AoD} $\theta_{\mathrm{AoD}}^i(t)$, \ac{AoA} $\theta_{\mathrm{AoA}}^i(t)$, and Doppler frequency $f_d^i(t)$ given by
\begin{align}
f_d^i(t) = \frac{\nu(t) \cos\bigl(\eta(t) - \theta_{\mathrm{AoD}}^{i}(t)\bigr)}{\lambda}, \quad i=1, \dots, L(t),
\end{align}
where $\lambda \triangleq \frac{c}{f_c}$ denotes the wavelength corresponding to the carrier frequency $f_c$, and $c=3\times 10^8~\mathrm{m/s}$ is the speed of light. The inter-element spacing at the transmit and receive arrays are denoted by $d_T$ and $d_R$, respectively. Mathematically, the \ac{MIMO} channel vector ${\mathbf{h}}(t)$ at time $t$ between the tagged user (User 1) and the \ac{BS} represents the small-scale fading and can be expressed as
\begin{align}\label{eq:mimo_channel}
{\mathbf h}(t)=
\sum_{i=1}^{L(t)}
{\rm e}^{\mathrm{j}\left(2\pi f_d^{i}(t)t+\beta_i\right)}
\mathbf a_R\bigl(\theta_{\mathrm{AoA}}^{i}(t)\bigr)
\otimes
\mathbf a_T\bigl(\theta_{\mathrm{AoD}}^{i}(t)\bigr),
\end{align}
where $\beta_i$ are independent and identically distributed (i.i.d.) random variables, each uniformly distributed on the interval $[-\pi,\pi]$. Due to the symmetry and uniformity of the distribution of these phases, it follows that $\mathds{E}[{\rm e}^{\mathrm{j}\beta_i}]=0$ and thus the resulting channel has zero mean, i.e. $\mathds{E}[\mx{h}(t)]=\mx{0}$. The receive and transmit steering vectors are defined as
\begin{align}
    & \mathbf a_R(\theta)
   =\!\Bigl[\,1,\;
             {\rm e}^{\mathrm{j}\frac{2\pi d_R}{\lambda}\cos(\zeta-\theta)},\dots,
             {\rm e}^{\mathrm{j}\frac{2\pi d_R}{\lambda}(N_r-1)\cos(\zeta-\theta)}
      \Bigr]^{\!\top}\!,\\
      &\mathbf a_T(\theta)
   =\!\Bigl[\,1,\;
             {\rm e}^{\mathrm{j}\frac{2\pi d_T}{\lambda}\cos(\eta-\theta)},\dots,
             {\rm e}^{\mathrm{j}\frac{2\pi d_T}{\lambda}(N_t-1)\cos(\eta-\theta)}
      \Bigr]^{\!\top}.
\end{align}

At a given time $t$, the MIMO channel matrix between user $k$ and the BS is shown by $\mathbf{H}_k(t)\in\mathbb{C}^{N_r\times N_t}$. The corresponding vectorized channel is obtained by stacking the columns of $\mathbf{H}_k(t)$ into the vector $\mathbf{h}_k(t)=\mathrm{vec}\big(\mathbf{H}_k(t)\big)\in\mathbb{C}^{N\times 1}$, with $N\triangleq N_t N_r$.

The covariance matrix of the channel for user $k$ at time $t$ is defined as $\mathbf{C}_{\mathbf{h}_k}(t)\triangleq\mathds{E}\left[\mathbf{h}_k(t)\mathbf{h}_k^{\mathsf{H}}(t)\right],
$
while the cross-covariance matrix and the correlation matrix between times $t_1$ and  $t_2$ are given by $\mx{C}_{\mx{h}_k}(t_1,t_2)\triangleq \mathds{E}[{\mx{h}_k}(t_1){\mx{h}^H_k}(t_2)]$ and $\mx{P}_{\mx{h}_k}(t_1,t_2)\triangleq \mx{C}_{\mx{h}_k}^{-\tfrac{1}{2}}(t_1)\mx{C}_{\mx{h}_k}(t_1,t_2)\mx{C}_{\mx{h}_k}^{-\tfrac{1}{2}}(t_2)$, respectively. 

These definitions clearly establish the basis for the subsequent analytical and practical evaluations of the time-varying channel characteristics. More explicitly, the correlation matrix and covariance matrix directly appear in our channel estimator in Section \ref{sec:channel_estimnate}, and in the \ac{SINR} and \ac{SE} calculations in Section \ref{Sec:SINR}.}

\subsection{Uplink Pilot and Data Model}\label{sec:pilot_data_model}
\begin{figure}
    \centering
\includegraphics[scale=.2]{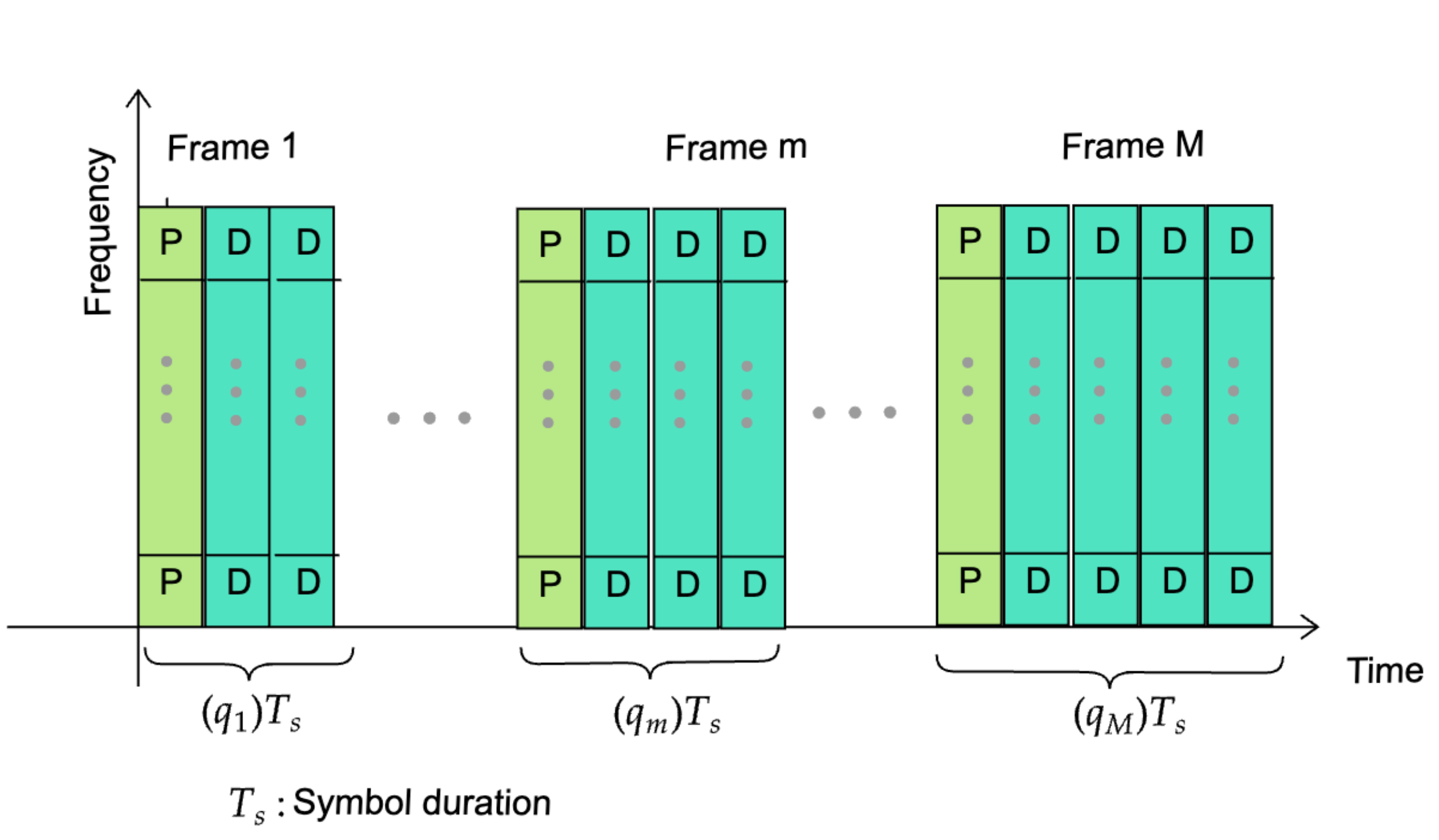}
    \caption{A schematic figure for multi-frame data transmission. $q_1,..., q_M$ specifies the data length of frames $1,..., M$. The first time slot of each frame is considered for pilot transmission and the rest is used for data transmission. }
    \label{fig:framework}
\end{figure}

We consider $M$ frames $m=1,..., M$ in which frame $m$ consists of $q_m$ time slots. The first time slot of each frame is devoted to pilot transmission and the rest is employed to transmit data.
{
Within frame $m$ out of a total of $M$ frames, each communication user transmits $\tau_p$ pilot symbols during the first time slot. This is followed by $q_m - 1$ data time slots, as depicted in Figure~\ref{fig:framework}. We assume that there are at least $K N_t$ pilot symbols in the frequency domain, where $K$ is the total number of users and each user has $N_t$ transmit antennas. These pilots are {mutually orthogonal in the code domain}, meaning each user $k$ employs a pilot matrix $\mathbf{S}_i$ that satisfies $ \mathbf{S}_i^\mathsf{T} \,\mathbf{S}_j 
   \;=\;
   \begin{cases}
       \tau_p\mathbf{I}_{N_t}, & \text{if } i = j,\\[6pt]
       \mathbf{0}, & \text{if } i \neq j.
   \end{cases}$.
This ensures that when one user's pilot is processed (e.g., by correlating the received signal with its own pilot matrix), all other users' signals vanish, {preventing pilot contamination}. Consequently, for the \textit{tagged} user (User 1), each of its $N_t$ antennas occupies its own set of $\tau_p$ pilot symbols in the first time slot. Since the pilots of other users remain orthogonal to this set, User 1's channel estimation can be performed independently, free from interference caused by other users' pilots.
}

Define $\delta_m\triangleq \left(\sum_{l=1}^m q_l\right)+1$. As we assumed $T_s=1$, the total transmission duration for all frames is $\delta_M-1$ time slots. The time slots of frame $m$ are from $\delta_{m-1}$ to $\delta_{m}-1$.
User-$k$ transmits $\tau_p$ pilot symbols at power $P_{p,k}$, and data symbols in slot $t$ at power $P_{d,k}$,for $k=1 \ldots K$. 
Assuming that each user uses a pilot matrix $\mx{S}_i\in\mathbb{C}^{\tau_p\times N_t}$, the received pilot signal from the tagged user (User 1) at the \ac{BS} is given by:
\begin{align}\label{eqn:received_training_seq}
\mx{Y}_{1,{\rm p}}(t)
&=
\alpha_1\sqrt{P_{{\rm p},1}} \mx{H}_1(t) \mx{S}_1^{\top } +\mathbf{N}_{\rm p}(t) ~~ \in \mathbb{C}^{N_r \times \tau_{\rm p}},
\end{align}
where { $\mx{S}_1$ is the pilot matrix of the tagged user}, $\alpha_1$ represents the
large scale fading factor for the tagged user, $P_{{\rm p},1}$ denotes the pilot power of the tagged user at each pilot time slot,
and $\mathbf{N}(t)\in \mathbb{C}^{N_r \times \tau_{\rm p}}$
is the 
\ac{AWGN} with element-wise variance $\sigma_{\rm p}^2$. 
It is beneficial to write the equation \eqref{eqn:received_training_seq} in a matrix-vector form as follows:
\begin{align}
    \mx{y}_{1,{\rm p}}(t)\triangleq\textbf{vec}\left(\mathbf{Y}_p(t)\right)=\alpha_1\sqrt{P_{{\rm p},1}} \overline{\mx{S}}\mathbf{h}_1(t) +\mx{n}_{\rm p}(t) ~\in \mathbb{C}^{\tau_{\rm p} N_r \times 1},
\end{align}
where
$\mathbf{ y}_{1,{\rm p}}(t)$, $\mx{ n}_p(t) \in \mathbb{C}^{\tau_{\rm p} N_r \times 1}$, 
$\overline{\mx{S}} \triangleq  \mx{S}_1\otimes \mathbf{I}_{N_r}\in \mathbb{C}^{\tau_{\rm p} N_r \times N}$ are such that $\overline{\mx{S}}^{\rm H}\overline{\mx{S}}=\tau_{\rm p}\mx{I}_{N}$, $\mx{h}(t)\triangleq{\rm vec}(\mx{H}(t))$ {
and $N\triangleq N_t N_r$}\footnote{Sometimes, we drop the index referring to the time slot $t$ for notational simplicity.}.
We also assume that the pilot power corresponding to each pilot time slot is obtained as $P_{{\rm p},1}=\tfrac{{P_{{\rm p},1}}_{\max}}{M}$, where ${P_{{\rm p},1}}_{\max}$ denotes the maximum pilot power of the tagged user.

At data time slot $t$ of frame $m$, the received signal at the BS can be stated as follows:
\begin{align}
    &\scalebox{.8}{$\mathbf{y}_{\rm d}(t)=   \underbrace{{\alpha_1} \mx{H}_1(t) \mx{x}_1(t)}_{\text{tagged user}}+ \underbrace{\sum_{k=2}^K \alpha_{k} \mx{H}_k(t) \mx{x}_{k}(t)}_{\text{co-scheduled MU-MIMO users}}$}
+\mathbf{n}_{\rm d}(t)\in\mathbb{C}^{N_r \times 1},
\label{eq:data_measurements}
\end{align}
where $\alpha_1$ is path loss component, $\mx{x}_{k}(t)\triangleq \mx{w}_k(t) s_k(t)\in\mathbb{C}^{N_t\times 1}$ is the transmitted signal by the $k$-the user, $s_k(t)\in\mathbb{C}^{1\times 1}$ is the transmitted data symbol of user $k$ with transmit data power $P_{{\rm d},k}=\mathds{E}[|s_k(t)|^2]$. {$s_k(t) \in \mathbb{C}$ is distributed as a zero-mean Gaussian random variable}. $\mx{w}_k(t)\in\mathbb{C}^{N_t\times 1}$ is the real-time beamforming vector of $k$-th user corresponding to time slot $t$. These beamforming vectors will be later optimized to enhance spectral efficiency. Furthermore, $\mathbf{n}_{\rm d}(t)~\sim \mathcal{CN}\left(\mx{0},\sigma_d^2 \mathbf{I}_{N_r}\right)$
is the \ac{AWGN} at the receiver. {\color{\change} It is important to note that in practical scenarios involving RF impairments, the assumption of white noise with a diagonal covariance matrix should be replaced by a colored noise model characterized by a non-diagonal and potentially rank-deficient covariance matrix. However, for analytical simplicity, we adopt the \ac{AWGN} assumption in this paper.} Here, the data power of each data time slot is obtained as $P_{{\rm d},k}=\tfrac{{P_{{\rm d},k}}_{\max}}{\delta_M -1-M}$, where ${P_{{\rm d}, k}}_{\max}$ denotes the maximum data power of user $k$.
The total power allocated for pilot and data transmission must not exceed the overall power budget, denoted by $P_{\rm tot}$. For notational simplicity, we sometimes omit the dependence on the time slot $t$ when referring to channel $\mx{h}$, and beamforming vectors, that is, $\mx{h}_k\triangleq \mx{h}_k(t)$ and $\mx{w}_k\triangleq \mx{w}_k(t)$. 

\section{Proposed Scheme}
\label{Sec:Proposed_scheme}

In this section, we provide a road-map of our approach for optimizing beamforming directions, pilot spacing, and power control. First in 
Section \ref{sec:correlation_matrix_in_MIMO}, we establish an explicit link between the velocity of mobile users and the correlation matrix at two different time instants. Then, we provide a \ac{LMMSE} estimate the time-varying channel as a function of frame size in presence of temporal prior information in Section \ref{sec:channel_estimnate}. {\color{\change}The channel estimates directly depend on the channel covariance and correlation matrices obtained in Section \ref{sec:correlation_matrix_in_MIMO}.} Next, in Section \ref{sec:data_estimate}, given the channel estimates, we provide an optimal \ac{MMSE} receiver in order to estimate the data message of the tagged user. Next, in Section \ref{Sec:SINR}, based on channel estimate and data estimate, we calculate the instantaneous slot-by-slot \ac{SINR} of the tagged user over fast fading non-stationary channels. Finally, we present a theorem that provides a deterministic expression for the instantaneous \ac{SINR}, referred to as the deterministic \ac{SINR}.

\subsection{Correlation Matrix \gf{of} MIMO Channels}\label{sec:correlation_matrix_in_MIMO}
In this section, we obtain an explicit relation between the correlation matrix that reflects the temporal correlation information of the channel and the velocity of users. {\color{\change}The correlation matrices play a pivotal role in effectively capturing the dynamic spatial-temporal structure of the wireless channel. Specifically, they serve as fundamental components in the channel estimation algorithm presented in Section~\ref{sec:channel_estimnate}, providing critical prior knowledge that significantly enhances estimation accuracy. Moreover, accurate knowledge of these matrices is instrumental in deriving precise expressions for \ac{SE}, as outlined in Section~\ref{Sec:SINR}. Furthermore, as demonstrated in Section~\ref{Sec:SINR}, the optimal frame size, the optimal allocation of pilot resources, and the optimal power distribution are intricately linked to these correlation structures. Therefore, a rigorous characterization of the correlation matrices, as provided in this section, forms the basis for optimizing system performance under realistic and non-stationary propagation scenarios.

The correlation matrix of the channel is typically influenced by several factors, including the propagation geometry, user velocity, and antenna characteristics. 
To obtain the correlation matrix, we first make several assumptions for the non-stationary channel model:
\begin{enumerate}\label{assumptions}
  \renewcommand\theenumi{\Roman{enumi}}
  \renewcommand\labelenumi{\theenumi)}
    \item Random phases: The phases $\beta_i$s are i.i.d. and independent of all angles and Doppler shifts\label{A1}
    \item  Ray \gf{statistical symmetry}:  For any fixed $t$, the triple
      $\bigl(\theta^{i}_{\!\rm AoA}(t),\theta^{i}_{\!\rm AoD}(t),
             f^{i}_{d}(t)\bigr)$
      is i.i.d.\ over path index~$i$ with known time-varying \acp{PDF}
      $f_{\rm AoA}(\theta,t)$ and $f_{\rm AoD}(\theta,t)$.\label{A2}
\end{enumerate}
Based on these assumptions, we find the correlation matrix in the following proposition. 
\begin{prop}[Correlation matrix of MIMO channels]
\label{prop.rician}
Consider the multi-path \ac{MIMO} channel provided in \eqref{eq:mimo_channel}. Suppose $\theta_{\rm AoA}(t)$ and $\theta_{\rm AoD}(t)$
      are independent random processes with known \acp{PDF}
      $f_{\rm AoA}(\theta(t))$ and $f_{\rm AoD}(\theta(t))$. 


Define the receive / transmit matrices
\begin{align*}
&\mathbf R_{\mx{a}_R}(t_{1},t_{2})\triangleq\mathds{E}
     \bigl[\mathbf a_R(\theta_{\rm AoA}(t_{1}))
           \mathbf a_R^{\mathsf{H}}(\theta_{\rm AoA}(t_{2}))\bigr],\nonumber\\
&\mathbf R_{\mx{a}_T}(t_{1},t_{2})\triangleq\mathds{E}
     \bigl[\mathbf a_T(\theta_{\rm AoD}(t_{1}))
           \mathbf a_T^{\mathsf{H}}(\theta_{\rm AoD}(t_{2}))\bigr],\nonumber\\
 &T_T(t_{1},t_{2})[q_1,q_2]\triangleq\mathds{E}_{\theta_{\rm AoD}}[{\rm e}^{\mathrm{j} \Delta\phi}], q_1,q_2=0,..., N_t-1,\nonumber\\
&\Delta \phi\triangleq \tfrac{2\pi}{\lambda}\big[(t_1\nu(t_1)+d_T q_1)\cos(\eta(t_1)-\theta_{\rm AoD}(t_1))-\nonumber\\
&
~~~~~~~~(t_2\nu(t_2)+d_T q_2)\cos(\eta(t_2)-\theta_{\rm AoD}(t_2))\big].
\end{align*}
Then, given the assumptions \eqref{A1} and \eqref{A2}, the channel correlation matrix is obtained as
\begin{align}\label{eq:cor1}
    &\mx{P}_{\mx{h}}(t_1,t_2) =\rho_{L} \bigl[\mx{R}_{\mx{a}_R}^{-1/2}(t_1)\,\mx{R}_{\mx{a}_R}(t_1,t_2)\,\mx{R}_{\mx{a}_R}^{-1/2}(t_2)\bigr] \;\otimes\nonumber\\
    &\bigl[\,\mx{R}_{\mx{a}_T}^{-1/2}(t_1)\,\mx{T}_T(t_1,t_2)\,\mx{R}_{\mx{a}_T}^{-1/2}(t_2)\bigr],
\end{align}
where $\rho_L\triangleq \frac{\min\{L(t_1), L(t_2)\}}{\sqrt{L(t_1)\,L(t_2)}}$, $\mx{R}_{\mx{a}_R}(t)\triangleq \mathbf R_{\mx{a}_R}(t,t)$ and $\mx{R}_{\mx{a}_T}(t)\triangleq \mathbf R_{\mx{a}_T}(t,t)$.


\end{prop}

Proof: See Appendix \ref{proof.prop_rician}.

The formulas provided in \eqref{eq:cor1} can be fully calculated by numerical integration. The instantaneous velocity profile $\nu(t)$ is readily available at the \ac{UE} through on-board \ac{GNSS} sensors, whereas the time-varying angular \acp{PDF} (e.g. $f_{\rm AoD}(\theta,t)$) can be estimated from the beam-power histogram produced by standard 5G beam-management procedures. 
This model aligns with non-isotropic scattering (more detailed can be found in \cite{non_stationary_model, baddour2004accurate, jakes1974mobile}).
\begin{cor}
\label{corr.von_miss}(Special cases: Stationary environments and von Mises distribution for \ac{AoA} and \ac{AoD})
    In the stationary cases where the statistics of the user and environment do not change with time, then the parameters $\nu(t)=\nu, \eta(t)=\eta,L(t)=L, \theta_{\rm AoA}(t)=\theta_{\rm AoA}, \theta_{\rm AoD}(t)=\theta_{\rm AoD}$ are fixed. Define $\tau\triangleq t_2-t_1$ and $n_R\triangleq p_2-p_1, n_T\triangleq q_2-q_1, k_T\triangleq\tfrac{2\pi d_T}{\lambda}, 
    k_R\triangleq\tfrac{2\pi d_R}{\lambda},\psi_{\tau}\triangleq \tfrac{2\pi \nu \tau}{\lambda}$. Assume that the \ac{AoD} and \ac{AoA} follow the von Mises distribution with the \acp{PDF} given by
    \begin{align}\label{eq:pdfs_von}
    &f_{\rm AoD}(\theta_{\rm AoD})=\frac{{\rm e}^{\kappa_{\rm AoD}\cos(\theta_{\rm AoD}-\theta^c_{\rm AoD})}}{2\pi J_0(j\kappa_{\rm AoD})},\nonumber\\
    &f_{\rm AoA}(\theta_{\rm AoA})=\frac{{\rm e}^{\kappa_{\rm AoA}\cos(\theta_{\rm AoA}-\theta^c_{\rm AoA})}}{2\pi J_0(j\kappa_{\rm AoA})},
\end{align}
    where $\theta^c_{\rm AoD}$ is the central \ac{AoD} and $\kappa_{\rm AoD}$ is a measure of dispersion from the central \ac{AoD}. Similarly, $\theta_{\rm AoA}^c$ and $\kappa_{\rm AoA}$ are defined for the \ac{AoA}. Given these assumptions and by using \eqref{eq:cor1}, the channel correlation matrix $\mx{P}_{\mx{h}}(t_1,t_2)$ is obtained as
\begin{align}
 \mx{P}_{\mx{h}}(t_1,t_2) =\mx{I}_{N_r}\otimes \mx{R}_{a_T}^{-\frac12}\mx{T}(\tau)\mx{R}_{a_T}^{-\frac12},  
\end{align}
where 
\begin{equation}
\begin{aligned}
R_{a_T}[q_1,q_2]
&= \frac{I_0\Bigl(
     \sqrt{%
       \begin{aligned}
         &\kappa_{\mathrm{AoD}}^2
          - k_T^2 n_T^2 \\[-0.5ex]
         &\quad
          +2\,\mathrm{j}\,\kappa_{\mathrm{AoD}}\;k_T n_T\,
           \cos\!\bigl(\eta - \theta_{\mathrm{AoD}}^c\bigr)
       \end{aligned}
     }
   \Bigr)}
   {I_0\!\bigl(\kappa_{\mathrm{AoD}}\bigr)},\\[1ex]
T_T(\tau)[q_1,q_2]
&=\frac{I_0\Bigl(
     \sqrt{%
       \begin{aligned}
         &\kappa_{\mathrm{AoD}}^2
          - \bigl(k_T n_T + \psi_\tau\bigr)^2+ \\
         &
          2\,\mathrm{j}\,\kappa_{\mathrm{AoD}}\,
           \bigl(k_T n_T + \psi_\tau\bigr)\,
           \cos\!\bigl(\eta - \theta_{\mathrm{AoD}}^c\bigr)
       \end{aligned}
     }
   \Bigr)}
   {I_0\!\bigl(\kappa_{\mathrm{AoD}}\bigr)}.
\end{aligned}
\end{equation}

    

\end{cor}
\begin{cor}\label{corr.uniform}(Special cases: Stationary environment and uniform angles).
    In the case where the \ac{AoD} and \ac{AoA} are distributed uniformly between $[-\pi,\pi]$, we have $\kappa_{\rm AoD}=\kappa_{\rm AoA}=0$. The correlation matrix in this case becomes in the form of:
\begin{align}
 \mx{P}_{\mx{h}}(\tau) =\mx{I}_{N_r}\otimes \mx{R}_{a_T}^{-\frac12}\mx{T}(\tau)\mx{R}_{a_T}^{-\frac12},  
\end{align}
where 
\begin{equation}
\begin{aligned}
R_{a_T}[q_1,q_2]
  &= J_0\bigl(k_T\,(q_2 - q_1)\bigr),\\
T_T(\tau)[q_1,q_2]
  &= J_0\bigl(k_T\,(q_2 - q_1) + \psi_{\tau}\bigr).
\end{aligned}
\end{equation}

 In the case of having a single transmit antenna, it holds that $d_T=0$ and it holds that

 \begin{align}
    R_{a_T}=1, T(\tau)= J_0(\psi_{\tau}),
 \end{align}
and 
\begin{align}
  \mx{P}_{\mx{h}}(\tau)=J_0(\psi_{\tau}) \mx{I}_{N_r},  
\end{align}
    which aligns with the well-established Jakes model provided in \cite[Eq. 4]{baddour2004accurate} and \cite{jakes1974mobile}.
\end{cor}
Corollaries \ref{corr.von_miss} and \ref{corr.uniform} are proved in Appendix \ref{proof.prop_rician}.}
\subsection{LMMSE Channel Estimation with Prior Information}\label{sec:channel_estimnate}
We assume that in each data slot $t$ of frame $m$, the \ac{BS} utilizes the pilot signals of the current and previous frames to estimate the channel of the tagged user in data time slots of the current frame. According to \eqref{eqn:received_training_seq}, the observed measurements at the pilot time slots corresponding to frames $m-1$ and $m$ are given by:
{
\begin{align}\label{eqn:measurements_pilot}
&\mx{Y}_{1,{\rm p}}\left(\delta_{m-2}\right)=
\alpha_1 \sqrt{P_{{\rm p},1}}\mx{H}_1\left(\delta_{m-2}\right) \mathbf{S}_1^{\top } +\mathbf{N}_p\left(\delta_{m-2}\right),\nonumber\\
&\mx{Y}_{1,{\rm p}}\left(\delta_{m-1}\right)=
\alpha_1 \sqrt{P_{{\rm p},1}}\mx{H}_1\left(\delta_{m-1}\right) \mathbf{S}_1^{\top } +\mathbf{N}_p\left(\delta_{m-1}\right).\nonumber\\
\end{align}
By stacking the above relations in matrix-vector form, we have that
\begin{align}\label{eqn:measurements_pilot_vectorized}
   & {\mx{y}}_{1,{\rm p}}\triangleq \begin{bmatrix}
       \textbf{vec}(\mx{Y}_{1,{\rm p}}(\delta_{m-2}))\\
       \textbf{vec}(\mx{Y}_{1,{\rm p}}(\delta_{m-1}))
   \end{bmatrix}=
   &\widetilde{\mx{S}}{\mx{h}_{1,{\rm p}}}+{\bs{\epsilon}_{\rm p}}\in\mathbb{C}^{2N_r\tau_{\rm p} \times 1},
\end{align}
where ${\mx{h}_{1,{\rm p}}}=\begin{bmatrix}
       \mx{h}_1(\delta_{m-2})\\
       \mx{h}_1(\delta_{m-1})
   \end{bmatrix}\in\mathbb{C}^{2N\times 1}$, $\widetilde{\mx{S}}=\mx{S}_1\otimes \mx{I}_{2N_r}$ and ${\bs{\epsilon}_{\rm p}}\sim \mathcal{CN}(\mx{0},\sigma^2_{{\rm p},1} \mx{I}_{2N_r\tau_{\rm p}})$. Note that as a convention, we assumed that $q_l=0~\forall l\le 0$.}
Having the measurements in pilot time slots provided in \eqref{eqn:measurements_pilot_vectorized}, one can estimate the channel at data time slot $t$ in frame $m$ as is stated in the following lemma.
\begin{lem}(\cite[Lemma 1 and Corollary 1]{daei2024towards})
\label{lem:mmsechannel}	
The \ac{LMMSE} estimate of the channel vector at time slot $t$ of frame $m$ given the measurements $\mx{y}_{\rm p}$ in \eqref{eqn:measurements_pilot_vectorized} is obtained as
\begin{align}
    &\widehat{\mx{h}}_1(t)=\tfrac{1}{\alpha_1\sqrt{P_{\rm p}}\tau_{\rm p}} \mx{E}(\mx{q},t) \left(\mx{M}_m(\mx{q})+\tfrac{\sigma^2_{{\rm p},1}}{\alpha_1^2 P_{{\rm p},1} \tau_{\rm p}}\mx{I}_{2N}\right)^{-1}\widetilde{\mx{S}}^{\mathsf{H}}{\mx{y}}_{1,{\rm p}},
\end{align}
Moreover, the covariance matrix of the \textup{LMMSE} channel estimate is provided by:

\begin{align}
\label{eq:rmmse}
\mx{C}_{\widehat{\mx{h}}_1}(t)
=& \mx{E}_m(\mx{q}, t) \left(\mx{M}_m(\mx{q})+\tfrac{\sigma^2_{{\rm p},1}}{\tau_{\rm p} \alpha_1^2 P_{{\rm p},1}}\mx{I}_{2N}\right)^{-1} \mx{E}_m^{\mathsf{H}}(\mx{q}, t).
\end{align}
Here,
\begin{align}
&    \mx{E}_m(\mx{q},t)\triangleq
{\begin{bmatrix}
    \mx{C}_{\mx{h}_1}(\delta_{m-2},t)&\mx{C}_{\mx{h}_1}(\delta_{m-1},t) 
\end{bmatrix}};\label{eqn:E_m}\\
& \mx{M}_m(\mx{q})\triangleq  {\begin{bmatrix}
    \mx{C}_{\mx{h}_1}(\delta_{m-2})&\mx{C}_{\mx{h}_1}(\delta_{m-2},\delta_{m-1})\\
    \mx{C}_{\mx{h}_1}(\delta_{m-1},\delta_{m-2})&\mx{C}_{\mx{h}_1}(\delta_{m-1})
    \end{bmatrix}}\label{eqn:M_m}.
\end{align}
\end{lem}
\subsection{MMSE Data Estimate}\label{sec:data_estimate}
In this section, we aim to calculate the optimum receiver $\mx{g}_1\in\mathbb{C}^{N_r\times 1}$ that estimates the transmitted data symbol of the tagged user which is denoted by $s_1(t)\in\mathbb{C}$. We assume that the transmitted data vector of the tagged user has zero mean with variance $P_{{\rm d},1}$.
Specifically, the BS estimates the data transmitted vector of the tagged user at time slot $t$ based on channel estimates provided at $t$. 
Then, the optimum \ac{MMSE} receiver combiner is provided via the following proposition whose proof is provided in Appendix \ref{proof.prop_data_estimate}.
\begin{prop}\label{prop.mmse_receiver}
 {\color{\change}Let $\mathbf P_{\mathrm c}\!\in\!\{0,1\}^{N\times N}$ denote the \emph{commutation matrix}, defined by
\begin{align*}
   &\scalebox{.85}{$ P_{\mathrm c}[k',k]\triangleq
\begin{cases}
  1, &\displaystyle
      k'=(j-1)N_t+i,\;
      k =(i-1)N_r+j,\\
  0, &\text{o.w.},
\end{cases}
\quad
\begin{aligned}
  &1\le i\le N_t,\\[-2pt]
  &1\le j\le N_r ,
\end{aligned}$}
\end{align*}
\gf{which} has the following properties:
\begin{itemize}
    \item $\mathbf P_{\mathrm c}\,\operatorname{vec}(\mathbf X)
          =\operatorname{vec}\!\bigl(\mx{X}^{\mathsf T}\bigr),
          \qquad\forall\,\mx{X}\in\mathbb C^{N_r\times N_t},$
          \item $\mathbf P_{\mathrm c}^{-1}= \mathbf P_{\mathrm c}^{\mathsf T}$,
          \item $\mathbf P_{\mathrm c}(\mx{A}\otimes\mx{B})\mathbf P_{\mathrm c}=\mx{B}\otimes\mx{A}$,
for all conformable matrices~$\mx{A},\mx{B}$.
\end{itemize}
 }
Then, the \ac{MMSE} optimum receiver given the prior information $\widehat{\mx{h}}_1(t)$ is obtained as follows:
\begin{align}\label{eq:G_star}
   &\mx{g}_1^{\star}(t)\triangleq \mathop{\arg\min}_{\mx{g}_1\in\mathbb{C}^{ N_r\times 1}} \mathds{E}_{s_1|\widehat{\mx{h}}_1(t)}[|\mx{g}_1^H\mx{y}-s_1|^2] =\nonumber\\
   &\alpha_1\sqrt{P_{{\rm d}, 1}} \mx{F}(t)^{-1}\widehat{\mx{H}}_1(t)\mx{w}_1(t),
\end{align}
where
\begin{align}
    &\scalebox{.8}{$\mx{F}(t)\triangleq\sum_{k=1}^K \alpha_k^2\mathcal{A}_k(\mx{D}_k)+\sigma^2_d\mx{I}_{N_r}, \mx{D}_k\triangleq\mx{Q}_k+ \widehat{\mx{h}}_k(t)\widehat{\mx{h}}_k(t)^H,$}\\
    & \scalebox{.8}{$\mx{D}_k\triangleq\begin{bmatrix}
         \mx{D}_k^{{(1,1)}}&\hdots&\mx{D}_k^{(1,N_r)}\\
         \vdots&\ddots&\hdots\\
         \mx{D}_k^{{(N_r,1)}}&\hdots&\mx{D}_k^{(N_r,N_r)}
     \end{bmatrix}, \widehat{\mx{H}}_k\triangleq {\rm vec}^{-1}(\widehat{\mx{h}}_k),\mx{Q}_k\triangleq\mx{C}_{\mx{h}_k(t)}-\mx{C}_{\widehat{\mx{h}}_k(t)}$},
\label{eq:D_k}\\
    &\scalebox{.9}{$\mathcal{A}_k(\mx{D}_k)\triangleq\begin{bmatrix}
        \langle {\mx{D}^{\prime}}_k^{(1,1)},\mx{C}_{\mx{x}_k} \rangle& \hdots& \langle {\mx{D}^{\prime}}_k^{(1,N_r)},\mx{C}_{\mx{x}_k} \rangle\\  
        \vdots &\ddots&\vdots\\
         \langle {\mx{D}^{\prime}}_k^{(N_r,1)},\mx{C}_{\mx{x}_k} \rangle& \hdots& \langle {\mx{D}^{\prime}}_k^{(N_r,N_r)},\mx{C}_{\mx{x}_k} \rangle
    \end{bmatrix},$}.\label{eq:A_operator}\\
    &{\mx{D}^{\prime}}_k\triangleq \mx{P}_{\rm c}^H \mx{D}_k \mx{P}_{\rm c}.
\end{align}
\end{prop}
These
results serve as a starting point for deriving the achievable \ac{SINR} and spectral
efficiency in the next subsection.

\subsection{\ac{SINR} and \ac{SE} Calculations}
\label{Sec:SINR}
In this subsection, we will calculate the instantaneous \ac{SINR} based on channel and data estimates provided in Sections \ref{sec:channel_estimnate}
and \ref{sec:data_estimate}.
The following theorem calculates the instantaneous SINR for the tagged user.
\begin{thm}\label{thm.Instantanous_SINR}
Given the channel estimate $\widehat{\mx{h}}_1(t)$ of the tagged user, the receiver utilizes the {MMSE} combiner to estimate the data symbols of the tagged user during time slot $t$. Consequently, the instantaneous {SINR} of the data symbol for the tagged user at time slot $t$ is derived as follows:
\begin{align}\label{eq:Instantanous_SINR_simplified}
     &\gamma(\mx{q},t,\widehat{\mx{h}}_1(t))= \alpha_1^2  \widehat{\mx{h}}_1^{\mathsf{H}}\Big(\mx{C}_{\mx{x}_1}\otimes \left(\mx{F}(t)- \alpha_1^2 \mathcal{A}_1(\widehat{\mx{h}}_1 \widehat{\mx{h}}_1^{\mathsf{H}})\right)^{-1} \Big)\widehat{\mx{h}}_1\nonumber\\
     &~~~~~~~~~~~~~\triangleq\alpha_1^2  \widehat{\mx{h}}_1^{\mathsf{H}} \left( \mx{C}_{\mx{x}_1} \otimes \mx{F}_1^{-1}(t)\right) \widehat{\mx{h}}_1,
     \end{align}
     where
     \begin{align}\label{eq:F_1_def}
         \mx{F}_1\triangleq \mx{F}_1(t)\triangleq \mx{F}(t)-\alpha_1^2 \mathcal{A}_1 \Big(\widehat{\mx{h}}_1\widehat{\mx{h}}_1^{\mathsf{H}}\Big).
     \end{align}
     

\end{thm}
Proof: See Appendix \ref{proof.thm.instantnaous_sinr}.

The latter result immediately leads to finding the random {SE} as follows:
\begin{align}\label{eq:random_SE}
\textup{SE}\Big(\mx{q},t,
\widehat{\mx{h}}_1(t)\Big)\triangleq\log\Big(1+\gamma(\mx{q},t,\widehat{\mx{h}}_1(t))\Big),
\end{align}
which is a random variable. In the following theorem, utilizing concentration inequality findings from the tools of random matrix theory, as provided in \cite{bai2010spectral}, we provide a deterministic equivalent formulation for the SE. This expression offers a suitable approximation for the average SE in scenarios where the number of BS antennas is sufficiently large.
{
{\color{\change}
\begin{thm}\label{thm.stiel}
Let $\bs{\Theta}\;\triangleq\;\sum_{k=1}^K \alpha_k^2\,\mathcal{A}_k(\mx{Q}_k),  \rho_{\rm d}\;\triangleq\;\sigma_{\rm d}^2$.
Define
\[
  \bs{\omega} \;\triangleq\; [\omega_2,\dots,\omega_K]^T,
\]
as the unique solution to the fixed-point system
\begin{align}\label{eq:fixedpoint_iter}
  \omega_k &=\;\bigl\langle \alpha_k^2\,\mx{C}_{\widehat{\mx{h}}_k},\;
                    \mx{C}_{\mx{x}_k}\otimes \mx{T}(\rho_d)\bigr\rangle,
  \quad k=2,\ldots,K,
\end{align}
where
\begin{align}
    \mx{T}(\rho_d)\triangleq \left(\sum_{k=2}^K \tfrac{\alpha_k^2 \mathcal{A}_k(\mx{C}_{\widehat{\mx{h}}_k})}{1+\omega_k}+\bs{\Theta}+\rho_{\rm d} \mx{I}_{N_r}\right)^{-1}. \label{eq:T_rho}
\end{align}
Also, define the vector
\begin{align}
     &\bs{\omega}'\triangleq[\omega'_2,..., \omega'_{K}]^T=(\mx{I}_{K-1}-\mx{J})^{-1} \mx{v}',\nonumber\\
\end{align}
where 
\begin{align}
     &\scalebox{.98}{$J_{k,l}\triangleq \frac{{\rm tr}\Big(\alpha_k^2\alpha_l^2\mathcal{A}_k(\mx{C}_{\widehat{\mx{h}}_k}) \mx{T}(\rho_d) \mathcal{A}_l(\mx{C}_{\widehat{\mx{h}}_l})\mx{T}(\rho_d)\Big)}{(1+\omega_l)^2}, k,l=2,..., K$},\nonumber\\
    &v'_k={\rm tr}\Big(\alpha_k^2\alpha_1^2\mathcal{A}_k(\mx{C}_{\widehat{\mx{h}}_k}) \mx{T}(\rho_d) \mathcal{A}_1(\mx{C}_{\widehat{\mx{h}}_1}) \mx{T}(\rho_d)\Big), k=2,..., K.
\end{align}
Assume that $\mx{C}_{\widehat{\mx{h}}_k}, k=1,..., K$ have uniformly bounded spectral norms with respect to $N$. 
 Then, when $N_r$ is sufficiently large ($N_r\rightarrow \infty$), the instantaneous {SE} provided in \eqref{eq:random_SE} is concentrated around a deterministic expression given by:
\begin{align}\label{eq:capacity_bound} 
    {\textup{SE}}^{\circ}(\mx{q}, t,\mx{w}(t))\triangleq \log_2\Big(1+\mathds{E}[\gamma]\Big)-\frac{{\rm var}(\gamma)}{2{\rm ln}(2)(1+\mathds{E}[\gamma])^2},
\end{align}
where 
\begin{align}\label{eq:var_term}
 &\scalebox{.9}{$\mathds{E}[\gamma]\xrightarrow{a.s.} P_{{\rm d}, 1}\alpha_1^2
  \left\langle \mx{C}_{\widehat{\mx{h}}_1(t)}, \mx{w}(t)){\mx{w}(t)}^H\otimes \mx{T}(\rho_d) \right\rangle $}  ,\nonumber\\
&\scalebox{.9}{$ 
{\rm var}(\gamma)\xrightarrow{a.s.} 2 P_{{\rm d},1}\alpha_1^2 \langle \mx{C}_{\mx{h}_1(t)}, \mx{w}(t){\mx{w}(t)}^H\otimes\mx{T}'(\rho_d)\rangle
$},  
\end{align}
and
\begin{align}
  \mx{T}(\rho_d)
  &=\Bigl(\sum_{k=2}^K \tfrac{\alpha_k^2\,\mathcal{A}_k(\mx{C}_{\widehat{\mx{h}}_k})}
                                     {1+\omega_k}
         +\bs{\Theta}
         +\rho_{\rm d}\,\mx{I}_{N_r}\Bigr)^{-1},\\
  \mx{T}'(\rho_d)
  &=\mx{T}(\rho_d)\,
    \Bigl(\alpha_1^2\,\mathcal{A}_1(\mx{C}_{\widehat{\mx{h}}_1})
          +\sum_{k=2}^K\!
             \frac{\omega'_k\,\alpha_k^2\,\mathcal{A}_k(\mx{C}_{\widehat{\mx{h}}_k})}
                  {(1+\omega_k)^2}\Bigr)\,
    \mx{T}(\rho_d).
\end{align}
\end{thm}
Proof: See Appendix \ref{proof.thm.stiel}.
}
{
\begin{rem}(Relevance of Theorem \ref{thm.stiel} to prior arts)
 To compare with prior works, we assume for simplicity that $N_t=1, \alpha_i=1,i=1,..., K$ and thus there is no beamforming vectors. The output of the \ac{MMSE} linear detector for the tagged user (User 1) is given by:
 \begin{align}\label{eq:linear_detector_known}
 r_1 =\mx{g}_1^H \mx{h}_1 s_1+\sum_{l\neq 1}\mx{g}_1^H \mx{h}_l s_l+\mx{g}_1^H \mx{n}_d.    
 \end{align}
Let $\mx{h}_i=\widehat{\mx{h}}_i+\widetilde{\mx{h}}_i$.  We rewrite \eqref{eq:linear_detector_known} as follows:
 \begin{align}\label{eq:linear_detect_unknown}
   r_1 =\mx{g}_1^H \widehat{\mx{h}}_1 s_1+\sum_{l\neq 1}\mx{g}_1^H \widehat{\mx{h}}_l s_l+\sum_{l=1}^K \mx{g}_1^H \widetilde{\mx{h}}_l s_l+\mx{g}_1^H \mx{n}_d.   
 \end{align}
Given the channel estimates are known and the channel estimate errors are unknown to the receiver, an achievable rate is obtained in \cite{Hassibi:03} which is a lower-bound for the ergodic capacity (see e.g. \cite[Eq: 2.42]{marzetta2016fundamentals}) and is given as follows:
\begin{align}
&\scalebox{.95}{$ C\ge \mathds{E}\Big[\log_2\Big(1+\frac{P_{d_1}|\mx{g}_1^H \widehat{\mx{h}}_1|^2}{\sum_{l\neq 1}|\mx{g}_1^H \widehat{\mx{h}}_l|^2 P_{d_l}+\sum_{l= 1}^K \mathds{E}[|\mx{g}_1^H \widetilde{\mx{h}}_l|^2] P_{d_l}+\sigma_d^2\|\mx{g}_1\|_2^2}\Big)\Big]$} \nonumber\\
&\triangleq \mathds{E}[\log_2(1+ {\rm SINR})]\triangleq\mathds{E}[{\rm SE}],
\end{align}
where ${\rm SINR}$ and ${\rm SE}$ are respectively called the effective instantaneous \ac{SINR} and \ac{SE}.

The authors in \cite{marzetta2016fundamentals} proposed an approach to find a lower-bound for the above expected spectral efficiency i.e.  $\mathds{E}[{\rm SE}]$ known as Use and Then Forget (UTF) by rewriting \eqref{eq:linear_detector_known} as follows (see e.g. \cite[Eq. 9]{jose2011pilot}, \cite[Eq. 40]{bjornson2017massive}, \cite[Sec. 3.2.2]{marzetta2016fundamentals}):
\begin{align}\label{eq:linear_detector_estimate_utf}
  &\scalebox{.8}{$ r_1 =\underbrace{{\mathds{E}[\mx{g}_1^H \mx{h}_1] s_1}}_{\text{Desired signal}}+\underbrace{\sum_{l=1}^K \mx{g}_1^H {\mx{h}}_l s_l-\mathds{E}[\mx{g}_1^H {\mx{h}}_1]s_1}_{\text{Interference}}+\underbrace{\mx{g}_1^H \mx{n}_d }_{\text{noise}}  $}.
 \end{align} 
This leads to the following lower-bound on the ergodic capacity:
\begin{align}\label{eq:utf_bound}
   &\scalebox{.8}{${\rm UTF~lower~ bound}=\log_2\Big(1+\frac{P_{d_1}(\mathds{E}[\mx{g}_1^H {\mx{h}}_1])^2}{\sum_{i=1}^K P_{d_i}\mathds{E}[|\mx{g}_1^H {\mx{h}}_i|^2]-P_{d_1}(\mathds{E}[\mx{g}_1^H {\mx{h}}_1])^2+\sigma_d^2\mathds{E}[\|\mx{g}_1\|_2^2]}$}\Big).
\end{align}
The UTF lower-bound in initially provided in \cite{marzetta2006much} based on the approach of \cite{Hassibi:03}  has been also exploited in several recent works such as \cite{marzetta2016fundamentals,bjornson2017massive,bjornson2016massive} and serves as a simple lower-bound to approximate the expected spectral efficiency. This strategy is explained in details in \cite[Lemma 2]{bjornson2016massive} or \cite[Eq. 9]{jose2011pilot} for \ac{MRC} combiner. The numerator and denominator of the UTF lower-bound are calculated based on Monte-Carlo simulations and there is no closed-form expression that predicts the behavior of the lower-bound in the asymptotic case for \ac{MMSE} receiver.
Hoydis et.al in \cite{Hoydis:13} found an approximation for the expected SINR $\mathds{E}[{\rm SINR}]$ by deriving deterministic equivalent expressions for the numerator and denominator of the instantaneous SINR in the limit. This means implicitly passing the expectation into both numerator and denominator. Then, they use Jensen's upper-bound to approximate the expected spectral efficiency as $\log_2(1+\mathds{E}[{\rm SINR}])$. 
It is worth mentioning that the latter operation does not necessarily lead to a lower-bound or upper-bound for the expected SINR. Sometimes, it overestimates $\mathds{E}[{\rm SE}]$ and sometimes it underestimates $\mathds{E}[{\rm SE}]$. This bound also serves as an upper-bound for UTF lower-bound in \eqref{eq:utf_bound} and sometimes predicts the UTF bound very well. Additionally, their approach which is based on \cite{wagner2012large,couillet2011deterministic} requires the number of users be much less than the number of antennas i.e, $K\ll N$.
Moreover, Ngo et.al in \cite[Eq. 25]{ngo2013energy} has found a lower-bound for $\mathds{E}[\text{SE}]$ in case of \ac{MMSE} receiver as follows:
\begin{align}\label{eq:ngo_lower_bound}
 \mathds{E}[\text{SE}]\ge \log_2\Big(1+\frac{1}{\mathds{E}[\frac{1}{\gamma_{\rm lower}}]}\Big)   
\end{align}
where $\gamma_{\rm lower}\triangleq \frac{1}{[(\mx{I}_{K}+\frac{1}{\sigma_d^2}\mx{H}^H_{\rm tot}\mx{H}_{\rm tot} )^{-1}]_{1,1}}-1$ and $\mx{H}_{\rm tot}\triangleq [\mx{h}_1, \mx{h}_2,..., \mx{h}_K]\in\mathbb{C}^{N\times K}$. Couillet et.al in  
 \cite{couillet2012random} proposed another bound based on Jensen's upper-bound given by
 \begin{align}\label{eq:jensen_upper}
   \mathds{E}[{\rm SE}]\le \log_2(1+\mathds{E}[{\rm SINR}]),  
 \end{align}
 and proved that when both transmit and receive number of antennas goes to infinity, this bound serves a good approximation for the expected spectral efficiency but overestimates the spectral efficiency in medium and small number of antennas.
\end{rem}
}
The deterministic equivalent \ac{SE} provided in \eqref{eq:capacity_bound} depends on the beamforming vectors $\mx{w}_1(t),..., \mx{w}_K(t)$ of size $N_t\times 1$. We further find the optimal value of the beamforming vector of the tagged user at time slot $t$ in the following proposition.
\begin{prop}\label{prop.optimal_beamforming}
Partition the matrix $\mx{C}_{\widehat{\mx{h}}_1}$ as below:

\begin{align}\label{eq:R_z_partition}
\scalebox{.8}{$\mx{C}_{\widehat{\mx{h}}_1}=\begin{bmatrix}
      \mx{C}_{\widehat{\mx{h}}_1}^{1,1}&\hdots&\mx{C}_{\widehat{\mx{h}}_1}^{1,N_r}\\
      \vdots&\ddots&\vdots\\
      \mx{C}_{\widehat{\mx{h}}_1}^{N_r,1}&\hdots&\mx{C}_{\widehat{\mx{h}}_1}^{N_r,N_r}
  \end{bmatrix}.$}
\end{align}
Define the linear operator $\mathcal{G}(\cdot):\mathbb{C}^{N\times N}\rightarrow \mathbb{C}^{N_t\times N_t}$ as
\begin{align}\label{eq:g_operator}
    \mathcal{G}(\mx{C}_{\widehat{\mx{h}}_1})=\sum_{i,l=1}^{N_r}T(i,l)\mx{C}_{\widehat{\mx{h}}_1}^{i,l},
\end{align}
where $ \mx{T}\triangleq\mx{T}(\rho_d)$ is given in \eqref{eq:T_rho}.
Then, the optimal real-time beamformer ${{\mx{w}_1}(t)}^\star$ that maximizes the deterministic equivalent \ac{SE} provided in \eqref{eq:capacity_bound} is the eigenvector of $\mx{T}(\rho_d)$ that corresponds to the maximum eigenvalue of $\mx{T}(\rho_d)$.
\end{prop}
Proof: See Appendix \ref{proof.prop.beamform}.

In what follows, we provide an optimization problem that finds the optimal values of $M$, $q_m$, ${P_{{\rm p},1}}_{\max}$ and ${P_{{\rm d},1}}_{\max}$. The objective function that we use is obtained by replacing the optimal ${{\mx{w}_1^\star}(t)}$ into the formula \eqref{eq:capacity_bound} and taking average over all time slots, which is named \ac{DASE}, and is given by
\begin{align}\label{eq:DASE}
   {\rm DASE} =\tfrac{\sum_{l=1}^{\delta_M -1}\text{SE}^{\circ}(\mx{q}, l,{\mx{w}_1}^{\star}(l))}{\delta_M -1}.
\end{align}
By having this objective function, the proposed optimization problem is as follows:

\begin{align}\label{eq:opt_problem}
&{\max}_{\mx{q},M, {P_{\rm p}}_{\max}, {P_{{\rm d}}}_{\max}} \tfrac{\sum_{l=1}^{\delta_M -1}\text{SE}^{\circ}(\mx{q}, l,{\mx{w}_1^\star}(l))}{\delta_M -1},
~~\nonumber\\
&{\rm s.t.}~~{P_{\rm p}}_{\max}+{P_{{\rm d}}}_{\max}\le P_{\rm tot},
\end{align}
where ${P_{{\rm d}}}_{\max}={P_{{\rm d},1}}_{\max}$ and ${P_{\rm p}}_{\max}={P_{{\rm p},1}}_{\max}$ denote the maximum data 
 and pilot power of the tagged user, respectively.

\begin{rem}(Critical factors in the proposed optimization problem \eqref{eq:opt_problem}.)
{The numerical experiments presented in Section \ref{Sec:simulations} indicate that neither the Doppler frequency of interfering users nor the interference path significantly impact the determination of the optimal frame design ($\mx{q}^\star$ and $M^\star$). Furthermore, as the number of transmit antennas increases, the variations in spectral efficiency (SE) of the tagged user due to different interference path losses and Doppler frequencies become negligible. This observation highlights the transmitter-centric nature of our approach, allowing all optimization tasks to be efficiently executed at the transmitter side rather than the receiver (BS) side.
More specifically, users are assumed to predict their own velocities for future time instances, enabling them to determine the optimal frame size and number of frames using our proposed strategy without requiring knowledge of the time-varying covariance matrices. However, obtaining the optimal pilot and data power allocation necessitates predicting these covariance matrices over time. This can be achieved {\color{\change}either through analytical expressions provided in Section \ref{sec:correlation_matrix_in_MIMO} (see \eqref{eq:cov_elem})} or through advanced Kalman filtering techniques \cite{onlineCCS} or machine learning models \cite{koopa_theory_ML}.} 
\end{rem}
The optimization problem \eqref{eq:opt_problem} is a \ac{NP}-hard mixed-integer nonlinear problem. In this section, we present a heuristic algorithm named OptResource for determining the optimal values of frame size, number of frames, and pilot and data powers. The pseudocode for this algorithm is outlined in Algorithm \ref{alg.optweights}. For each number of frames and frame size, an estimation of the beamforming vectors is obtained based on an estimate of the pilot and data powers, using Proposition \ref{prop.optimal_beamforming} in Line 10. Subsequently, a projected gradient ascent is employed to identify the optimal values of pilot and data powers (Lines 12 to 19).
The projected gradient ascent involves two steps: the update step and projection. In the update step (Line 15), a regular gradient ascent is performed. However, the resulting power variables after this step may fall outside the feasible region (where the sum of pilot and power must be less than the total power budget). To address this, the updated power variables are projected in Line 17 to the nearest point within the feasible region. After determining the optimal pilot and data powers, Line 10 is again used to update the beamforming, and an alternative optimization (AO) approach is applied to jointly find the optimum beamforming and powers.
The algorithm iterates through all possible values of $M$ and $q_m$ to identify the combination leading to the maximum spectral efficiency. The final outputs of OptResource are $\mx{q}^\star, M^\star, {P_{{\rm p},1}}_{\max}^{\star}, {P_{{\rm d},1}}_{\max}^{\star}, \mx{W}_{\rm B}^\star=[{\mx{w}_1^\star}(1)),..., {\mx{w}_1^\star({\delta_{M^\star}-1})}]$. 
{
The computational complexity of the projected gradient ascent depends on the dimensionality of the optimization vector, which, in our case, is two. Additionally, computing the DASE (Equation \ref{eq:DASE}) at each time slot requires implementing a fixed-point algorithm with a computational complexity of $\mathcal{O}(K-1)$. Furthermore, obtaining the covariance matrices $\mx{C}_{\widehat{\mx{h}}_l}$ for each $l=1,..., K$ involves the solution of Equation \ref{eq:rmmse}, which requires $\mathcal{O}(N^3)$ computations. In summary, the overall computational complexity is $\mathcal{O}({q}_{\max} (K-1) N^3)$.}

{

\begin{rem}(Convergence analysis)
 The convergence of Algorithm \ref{alg.optweights} is directly linked to the convergence properties of computing the deterministic spectral efficiency in Equation \eqref{eq:capacity_bound}. A detailed convergence analysis is presented in the following proposition, which is adapted from \cite[Appendix I.B]{Wagner:2012} and Vitali's convergence theorem \cite{pap2002convergence}:
 \begin{prop}
   Let $\omega_l^{(p)}, l=2,..., K$ be the $p$-th iteration of the following fixed-point equation
   \begin{align}
       &\scalebox{.93}{$\omega_k^{(p+1)} =\left\langle \mx{C}_{\widehat{\mx{h}}_k}(t),\left(\sum_{l=2}^K\tfrac{\mx{C}_{\widehat{\mx{h}}_l}(t)}{1+\omega_l^{(p)}}+\bs{\Theta}+\rho_{\rm d} \mx{I}_{N_r}\right)^{-1}\right\rangle$}
   \end{align}
   with the initial point $\omega_l^{(0)}=\frac{1}{\rho_{\rm d}}$. Then, it holds that
\begin{align}
   \left| \omega_l^{(p+1)}\!-\!\omega_l^{(p)}\right| \le  \frac{1}{\rho_d^2} \sup_{2\le l\le K}\left|\omega_l^{(p)}\!-\!\omega_l^{(p-1)}\right|
\end{align}
and for $l=2,..., K$ the iteration converges to $\lim_{p\rightarrow \infty}\omega_l^{(p)}=\omega_l,$ defined in \eqref{eq:fixedpoint_iter}. 
 \end{prop}
\end{rem}}
\alglanguage{pseudocode}
{
 \resizebox{.5\textwidth}{!}{
\begin{minipage}{.8\textwidth}
\begin{algorithm}[H]
	\caption{Proposed algorithm for resource optimization in MIMO systems}\label{alg.optweights}
	\begin{algorithmic}[1]
		\Procedure {OptResource}{$\mx{P}_{\mx{h}_i}(t_1,t_2)$, $\rm Tol$, $\mx{q}_{\max}$, $M_{\max}$, ${\rm maxiter}$, ${\rm maxiter}_{\rm AO}$, $P_{\rm tot}$}
  \State Define  $\bs{\varrho}\triangleq[{P_{{\rm p},1}}_{\max},{P_{{\rm d},1}}_{\max}]^{\top },$
  \State  Pick an initial point $\bs{\varrho}^0$ for maximum pilot and data power that satisfies $\mx{1}^{\top }\bs{\varrho}^0 \le P_{\rm tot}$ 
 
  \State $M=1, {\rm SE}^\star=0, \bs{\varrho}^\star\leftarrow\bs{\varrho}^0$
  \While{$M\le M_{\max}$}
\For{$i_1=1~{\rm to} \lceil\tfrac{q_{\max}}{M}\rceil $}

\State \vdots
\For{$i_M=1~{\rm to} \lceil \tfrac{q_{\max}}{M}\rceil $}
\For{${\rm it}=1~{\rm to}~{\rm maxiter}_{\rm AO}$}
 \State Find an estimate of the beamforming vectors $\mx{W}_{\rm B}\triangleq [\mx{w}_1(1),..., \mx{w}_1({\delta_M-1})]$ by Proposition \ref{prop.optimal_beamforming} as follows:
 
  $\mx{W}_{\rm B}^{\star}\leftarrow\mathop{\arg\max}_{\mx{W}_{\rm B}} f_1(\bs{\varrho}^\star, \mx{W}_{\rm B})={\rm SE}^{\circ}(\mx{q},\bs{\varrho}^\star,\mx{W}_{\rm B})$
\State Objective function for power:
\State $f(\bs{\varrho})\triangleq {\rm DASE}(i_1,...,i_M, M, \bs{\varrho},\mx{W}_{\rm B}^\star)$ in \eqref{eq:DASE}
	
  \State $k\leftarrow 1$
        \While{$\|{\bs{\varrho}}^{k}-{\bs{\varrho}}^{k-1}\|_2 >\rm Tol$ and $|f({\bs{\varrho}}^{k})-f({\bs{\varrho}}^{k-1})|>\rm Tol$ and $k<{\rm maxiter}$} 

Compute ascent direction by calculating $\tfrac{\partial {\rm SE}}{\partial {\bs{\varrho}}}$ at the point ${\bs{\varrho}}^k$	\\
 Update according to
         $\widetilde{{\bs{\varrho}}}^{k+1}\leftarrow {\bs{\varrho}}^{k}+\mu \tfrac{\partial {\rm SE}}{\partial{\bs{\varrho}}} $\\
 Project onto the feasible set
 \\
${\bs{\varrho}}^{k+1}=\mathop{\arg\min}_{\mx{x}}\|\mx{x}-\widetilde{{\bs{\varrho}}}^{k+1}\|_2~~{\rm s.t.} ~\mx{1}^{\top } \mx{x}\le P_{\rm tot}$
  \State $k\leftarrow k+1$
	\EndWhile
\State ${\bs{\varrho}}^{\star} \leftarrow {\bs{\varrho}}^k$
\State $SE=f(\bs{\varrho}^\star)$

\If{$SE> {SE}^\star$}

\State ${SE}^{\star}\leftarrow SE$
\State $\mx{q}^\star\leftarrow[i_1,,..., i_M]^{\top }$
\State $M^{\star}\leftarrow M$
  \EndIf	
  \EndFor
  \EndFor
  
\State   \vdots

  \EndFor
  \State $M\leftarrow M+1$
  \EndWhile
\Statex $[{P_{{\rm p},1}}_{\max}^\star , {P_{{\rm d},1}}_{\max}^\star]^{\top }\leftarrow{\bs{\varrho}}^\star$
		\EndProcedure
		\Statex Outputs: $\mx{q}^\star, M^\star,{P_{{\rm p},1}}_{\max}^\star , {P_{{\rm d},1}}_{\max}^\star,\mx{W}_{\rm B}^\star=[{\mx{w}_1(1)}^\star,..., \mx{w}_1({\delta_{M^\star}-1})]$.
	\end{algorithmic}
\end{algorithm}
\end{minipage}}}

\begin{figure}[h]
    \centering
    
    \begin{subfigure}[]{0.24\textwidth}
        \centering
 \includegraphics[scale=.23,trim={2cm 0cm 0cm 0cm}]{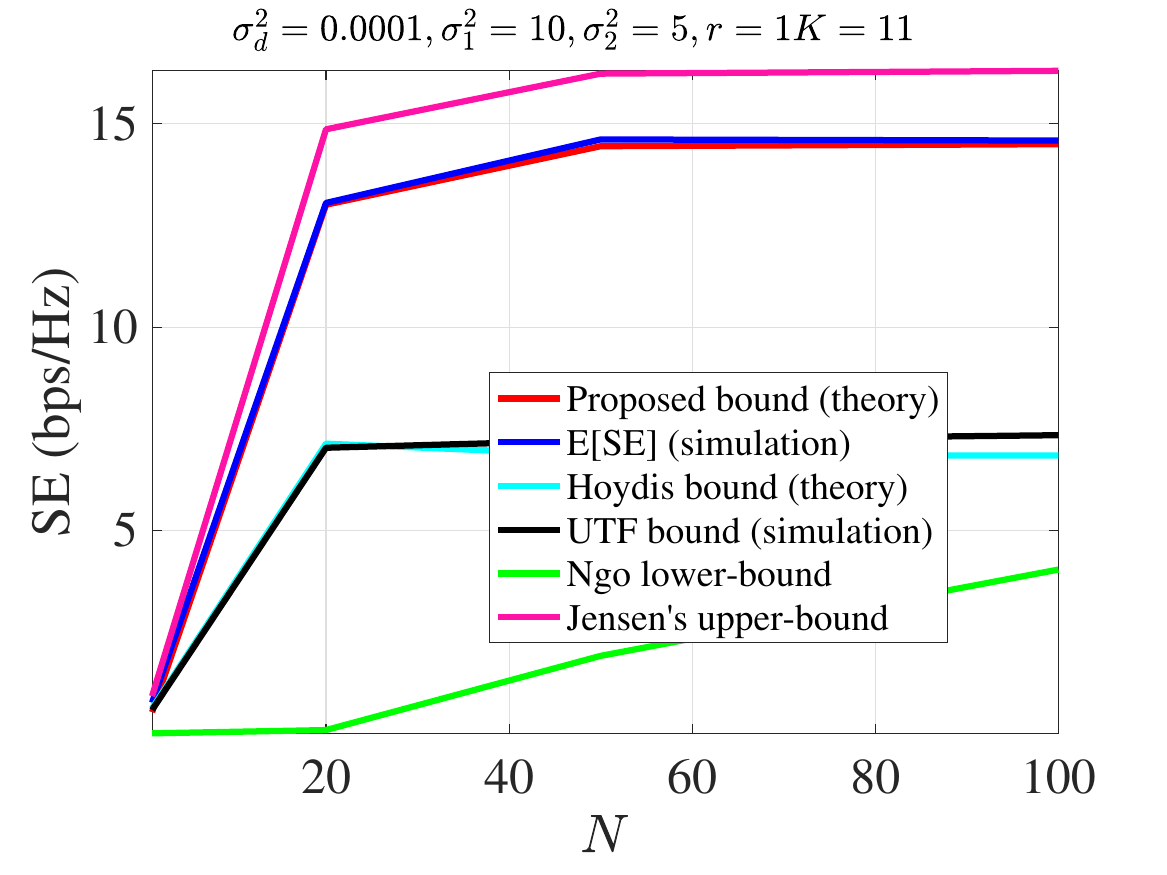}
        \caption{}
        \label{fig:SEbound(a)}
    \end{subfigure}
       \hfill
    \begin{subfigure}[]{0.24\textwidth}
        \centering
 \includegraphics[scale=.23,trim={2cm 0cm 0cm 0cm}]{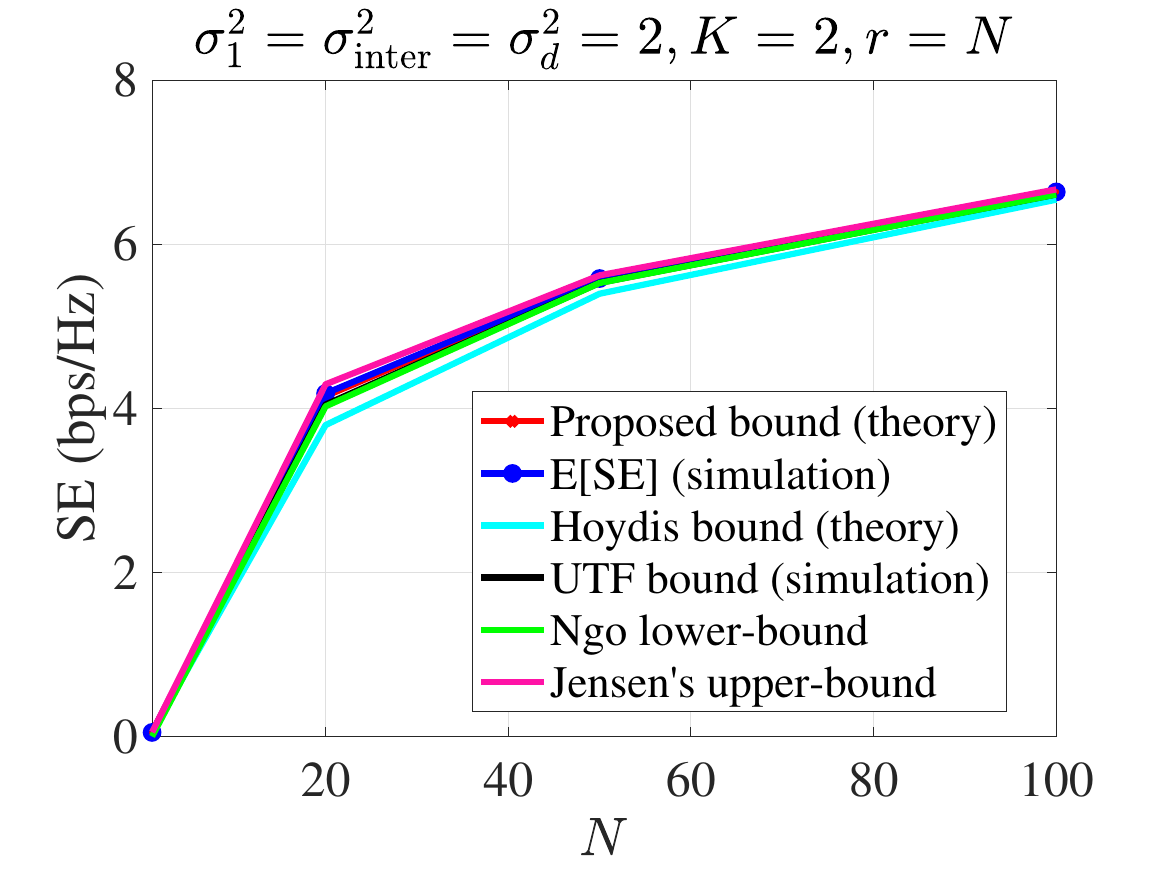}
        \caption{}
      \label{fig:SEbound(b)}
    \end{subfigure}
     \hfill
     \begin{subfigure}[]{0.24\textwidth}
        \centering
\includegraphics[scale=.23,trim={2cm 0cm 0cm 0cm}]{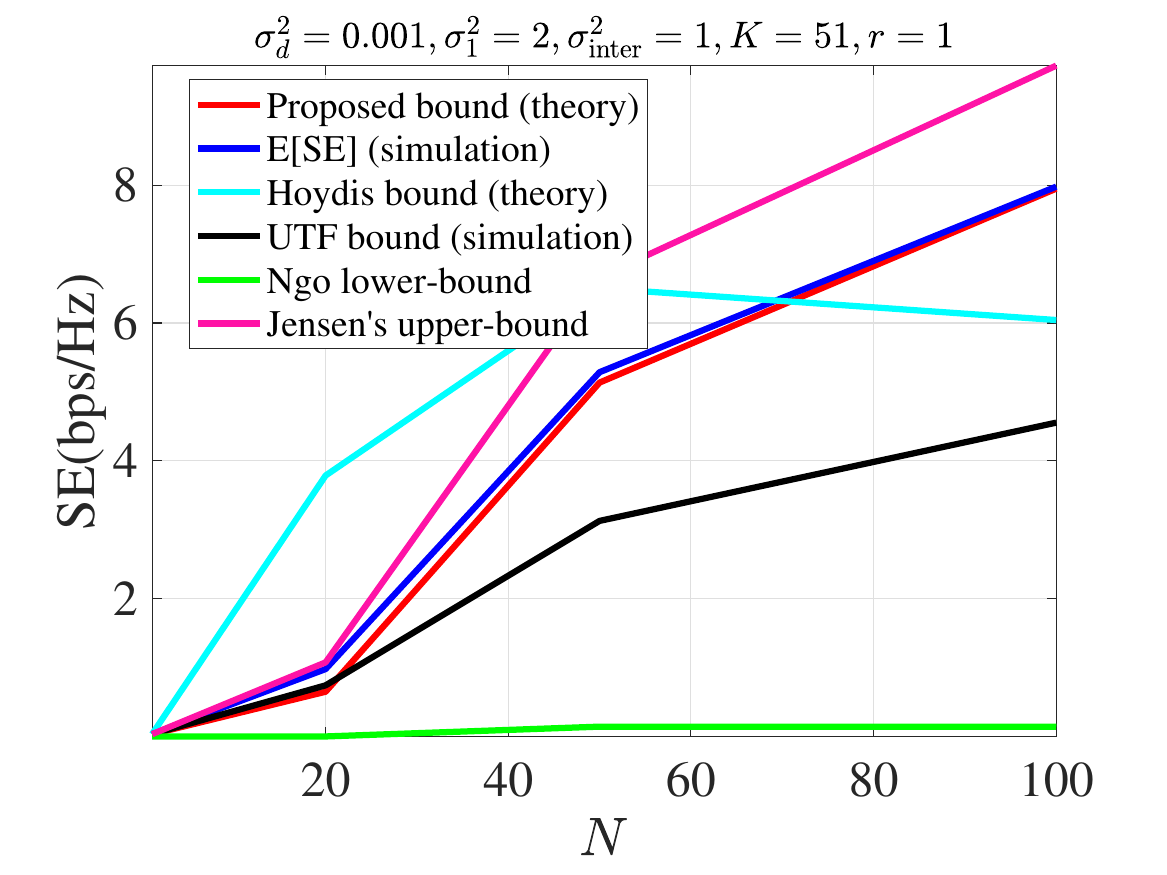}
        \caption{}
        \label{fig:SEbound(c)}
    \end{subfigure}
    \hfill
    \begin{subfigure}[]{0.24\textwidth}
        \centering
 \includegraphics[scale=.23,trim={2cm 0cm 0cm 0cm}]{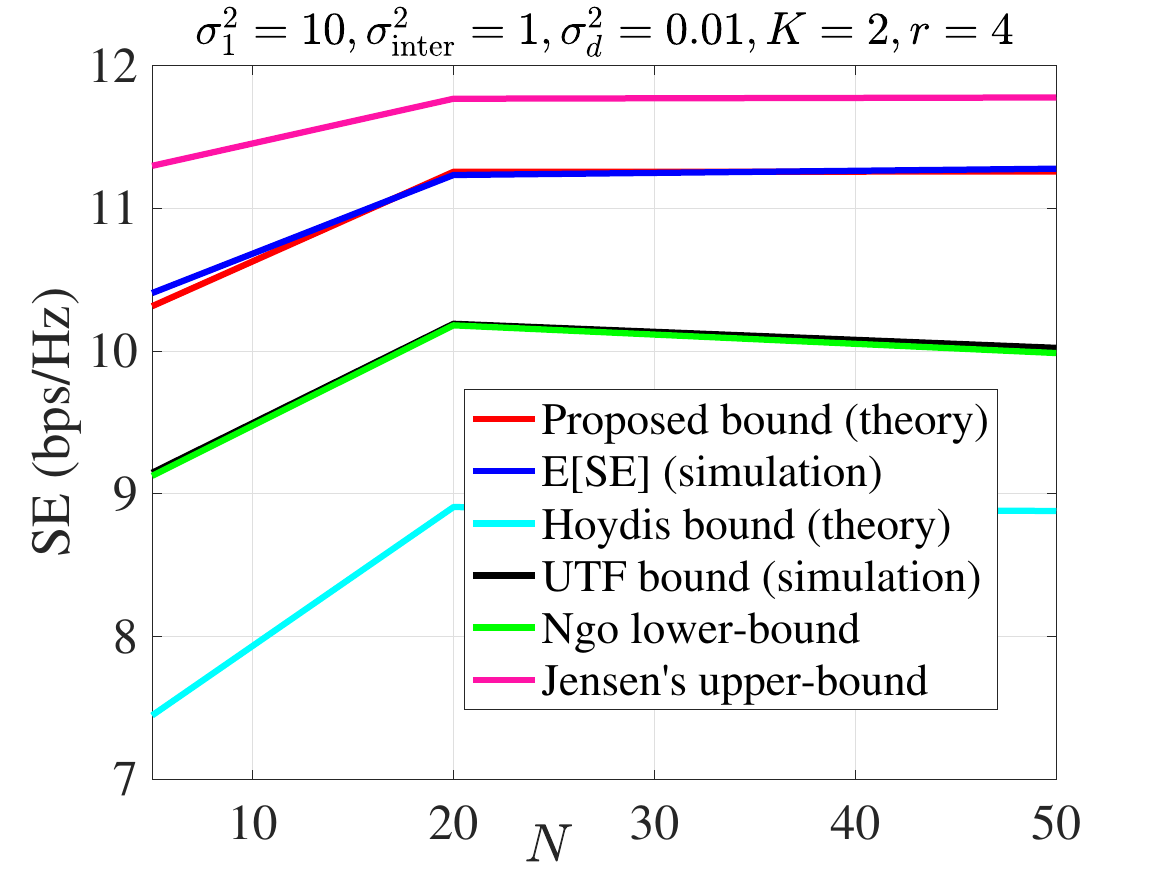}
        \caption{}
        \label{fig:SEbound(d)}
    \end{subfigure}
  \hfill
    \begin{subfigure}[]{0.24\textwidth}
        \centering
 \includegraphics[scale=.23,trim={2cm 0cm 0cm 0cm}]{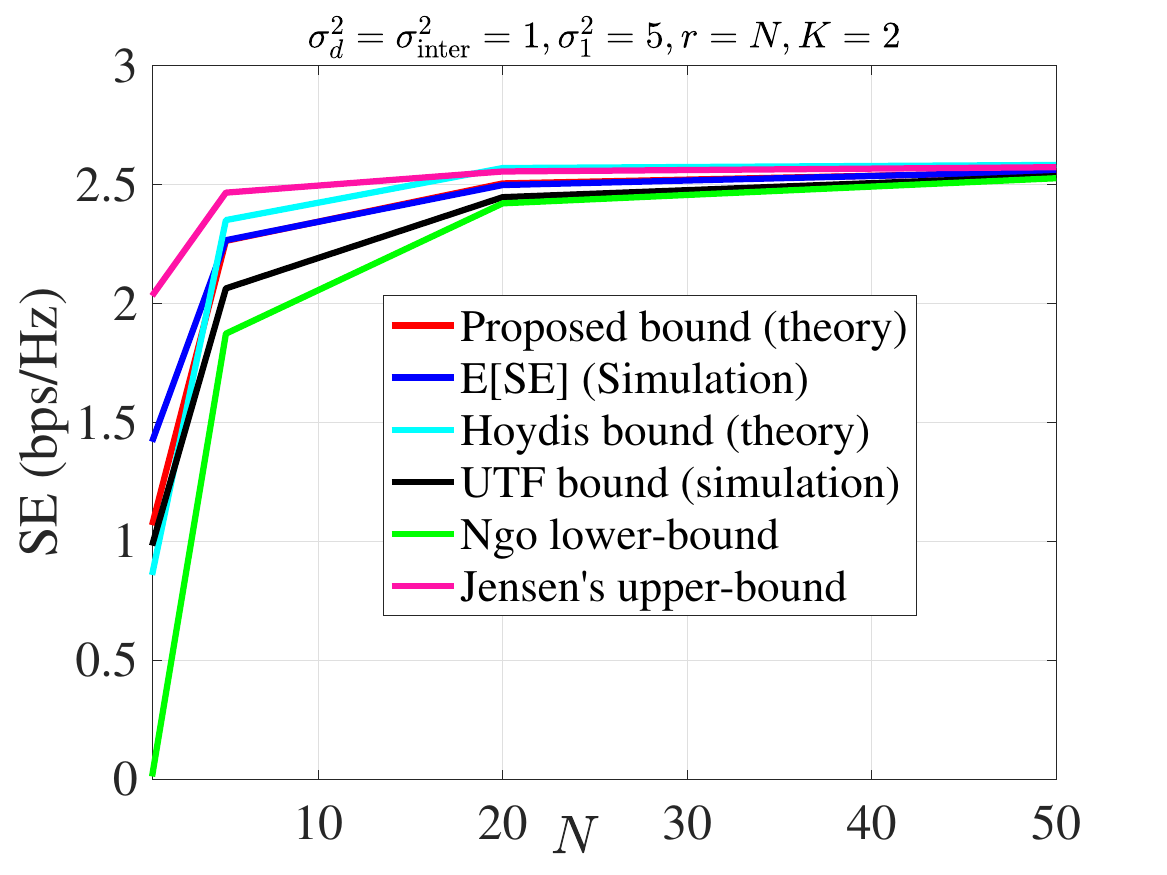}
        \caption{}
        \label{fig:SEbound(e)}
    \end{subfigure}
    \hfill
     \begin{subfigure}[]{0.24\textwidth}
        \centering
 \includegraphics[scale=.23,trim={2cm 0cm 0cm 0cm}]{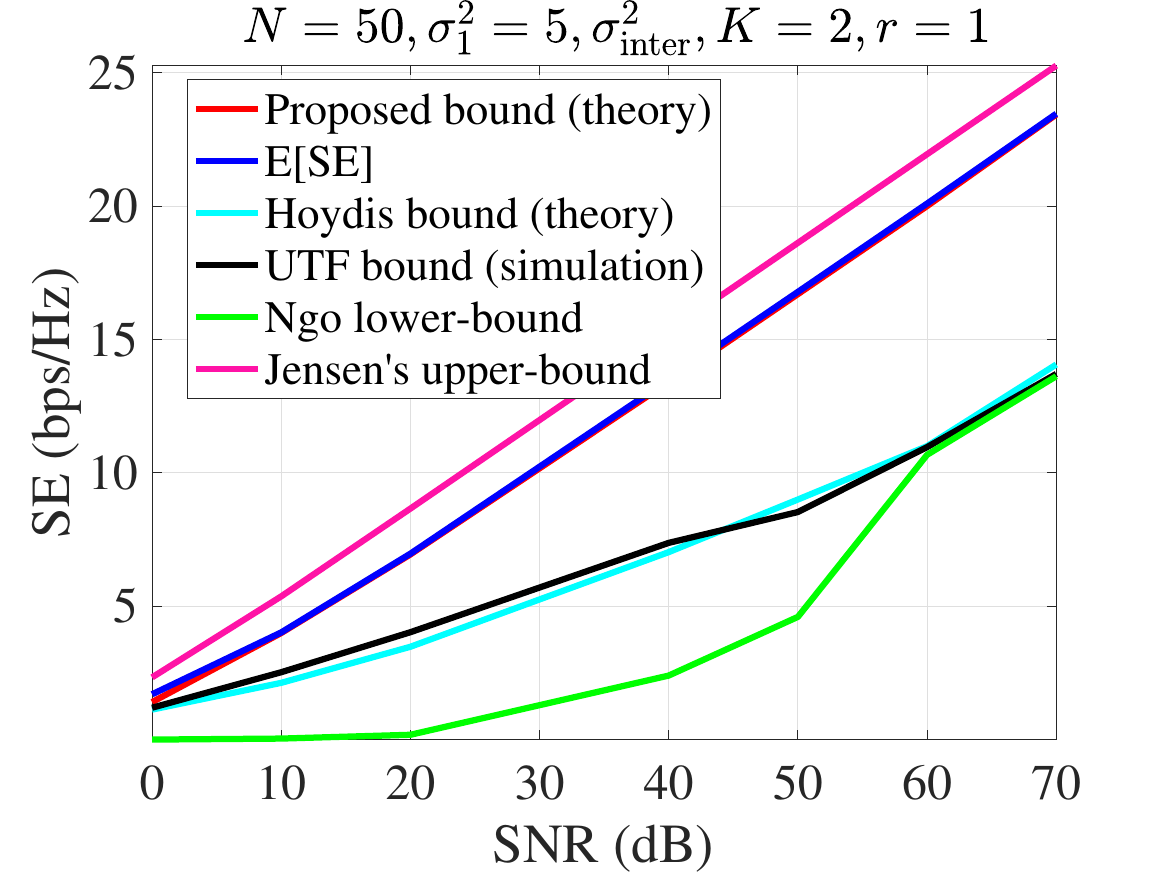}
        \caption{}
        \label{fig:SEbound(f)}
    \end{subfigure}
     \hfill
     \begin{subfigure}[]{0.24\textwidth}
        \centering
 \includegraphics[scale=.23,trim={2cm 0cm 0cm 0cm}]{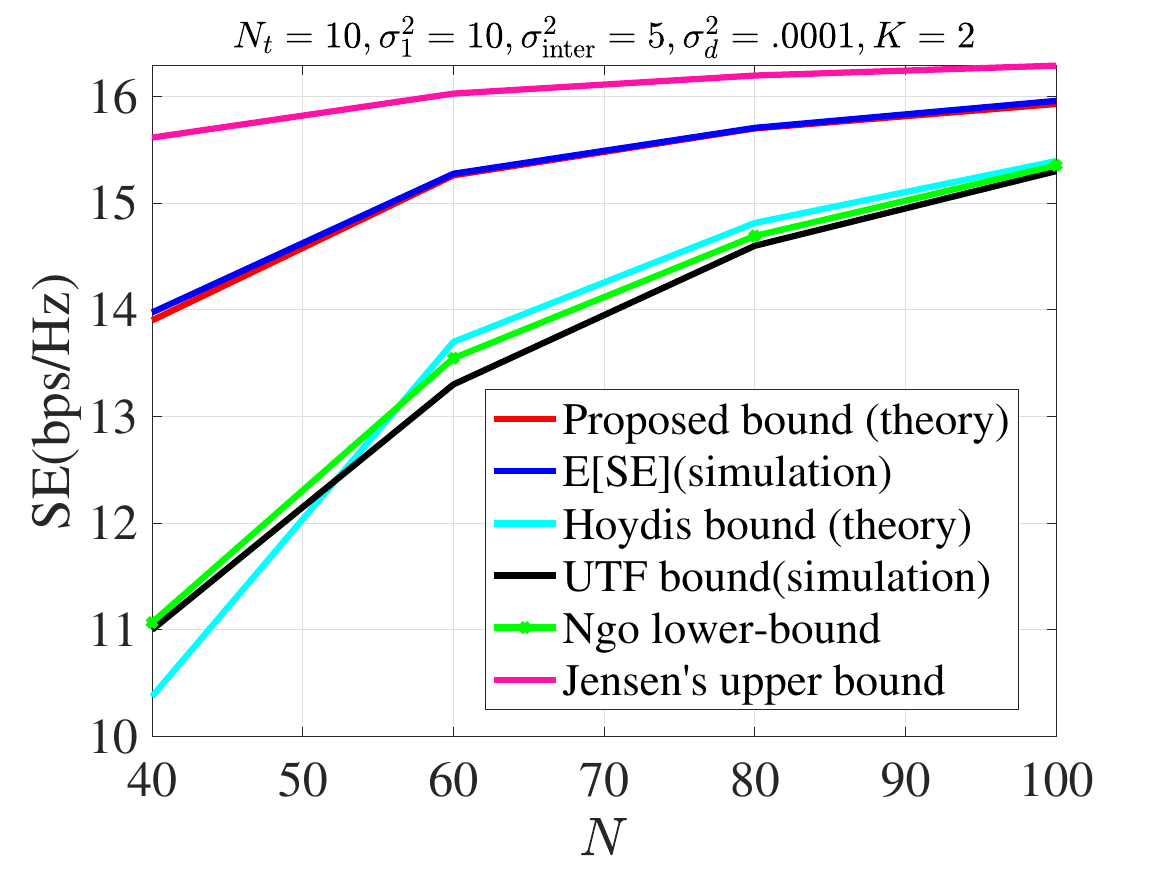}
        \caption{}
        \label{fig:SEbound(g)}
    \end{subfigure}
     \hfill
     \begin{subfigure}[]{0.24\textwidth}
        \centering
 \includegraphics[scale=.23,trim={2cm 0cm 0cm 0cm}]{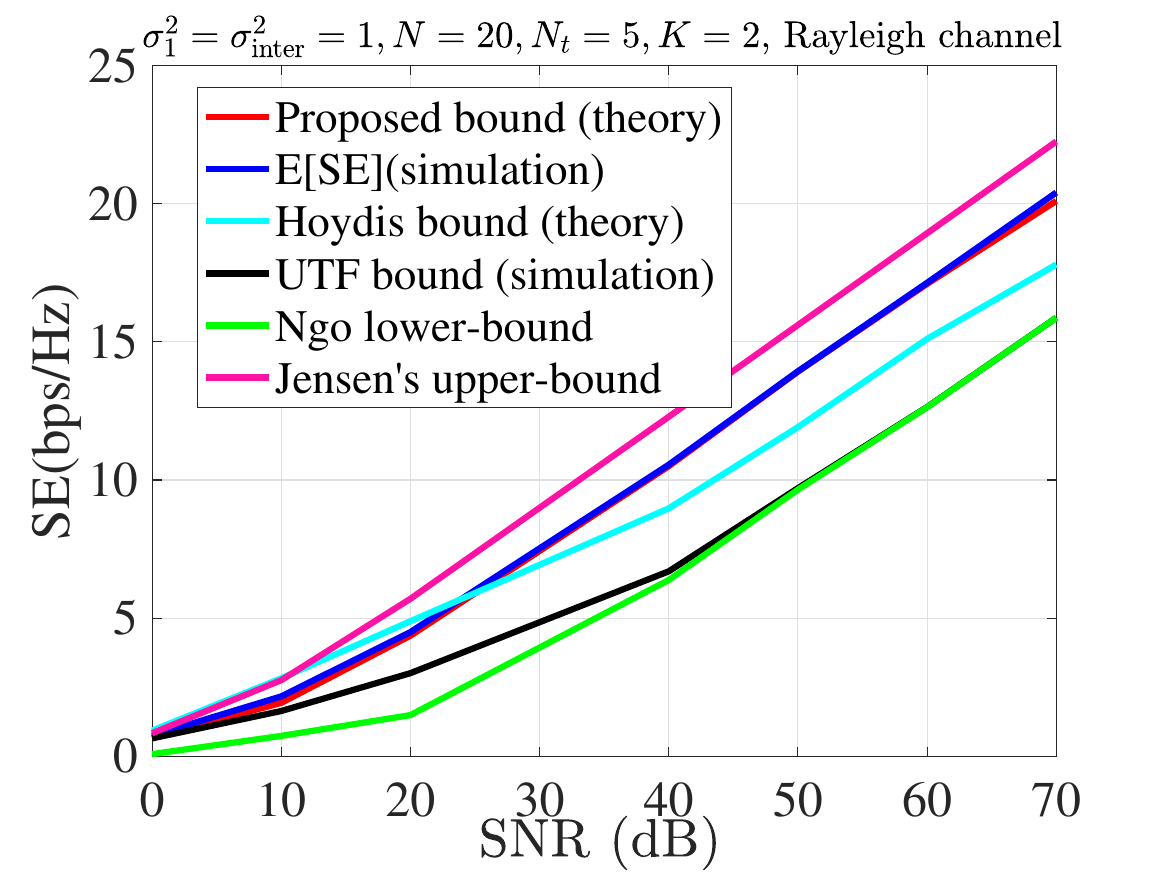}
        \caption{}
        \label{fig:SEbound(h)}
    \end{subfigure}
    \caption{{This figure evaluates the performance of our proposed capacity bound in Theorem \ref{thm.stiel} with state-of-the-art methods: UTF bound in \eqref{eq:utf_bound} calculated by Monte-Carlo simulations, the bound provided in \cite{Hoydis:13} using random matrix theory tools, the Ngo lower-bound provided in \cite[Eq.25]{ngo2013energy} and the Jensen's upper-bound in \eqref{eq:jensen_upper} computed using Monte-Carlo simulations and approximated using random matrix theory tools in \cite{couillet2011deterministic,couillet2012random,wagner2012large}.   An i.i.d. Rayleigh channel with independent largely-spaced antennas is considered in Figures \ref{fig:SEbound(b)} and \ref{fig:SEbound(e)} while a mm-wave channel with correlated antennas is considered in Figures \ref{fig:SEbound(a)}, \ref{fig:SEbound(c)}, \ref{fig:SEbound(d)}, \ref{fig:SEbound(f)}}. $r$ shows the rank of the channel. In Figures \ref{fig:SEbound(g)}, \ref{fig:SEbound(h)}, Rayleigh fading is assumed with high spatial correlations in the transmit antennas and independent receive antennas.}
     \label{fig:capacity_bounds}
\end{figure}
\begin{figure}[h]
    \centering
    
    \begin{subfigure}[]{0.24\textwidth}
        \centering
 \includegraphics[scale=.23,trim={2cm 0cm 0cm 0cm}]{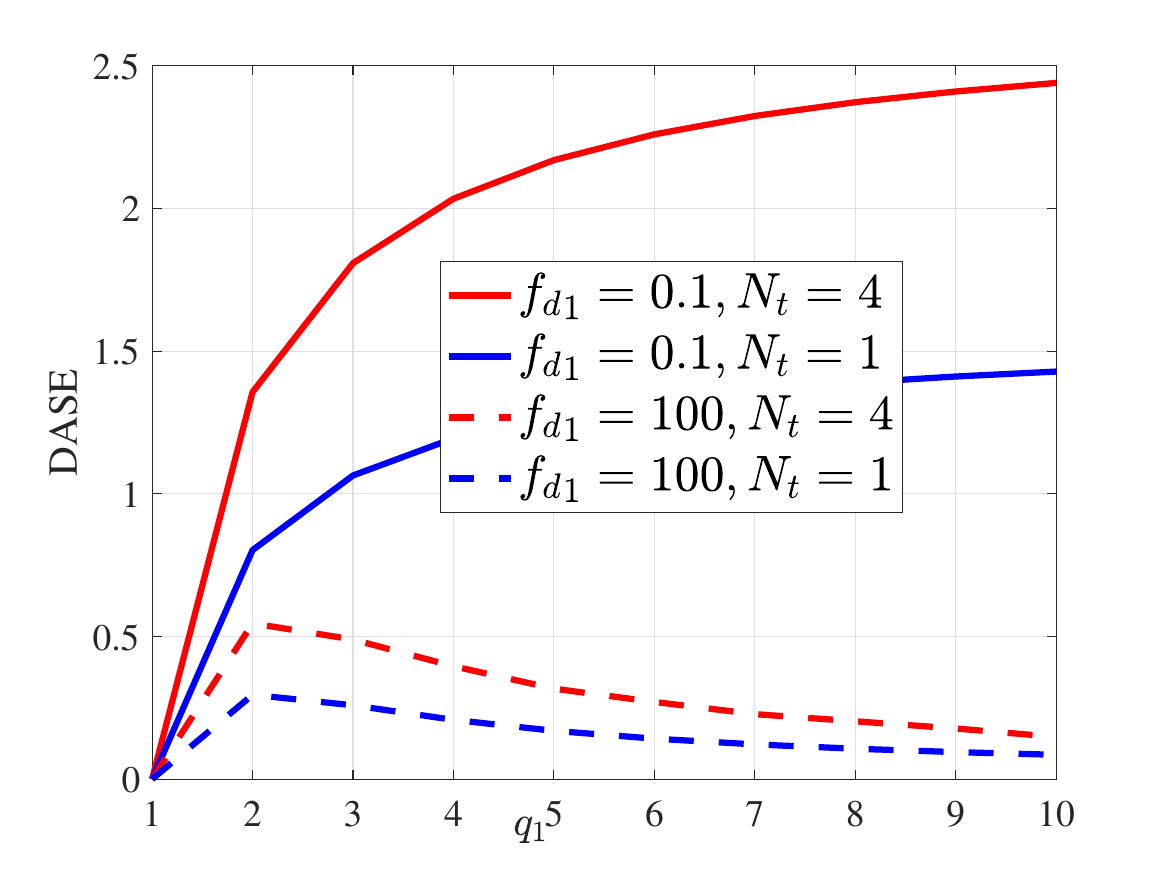}
        \caption{}
        \label{fig:SE(a)}
    \end{subfigure}
       \hfill
    \begin{subfigure}[]{0.24\textwidth}
        \centering
 \includegraphics[scale=.23,trim={2cm 0cm 0cm 0cm}]{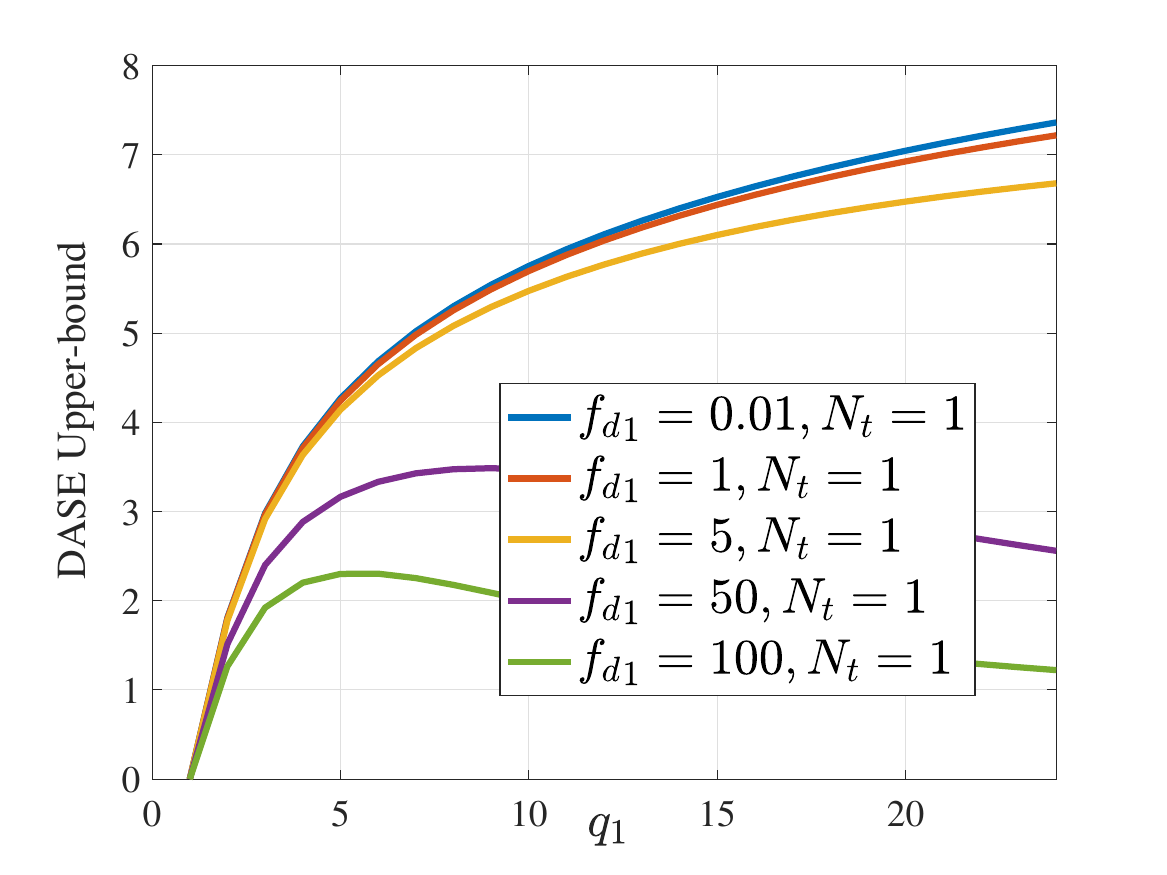}
        \caption{}
      \label{fig:SE(b)}
    \end{subfigure}
     \hfill
     \begin{subfigure}[]{0.24\textwidth}
        \centering
\includegraphics[scale=.23,trim={2cm 0cm 0cm 0cm}]{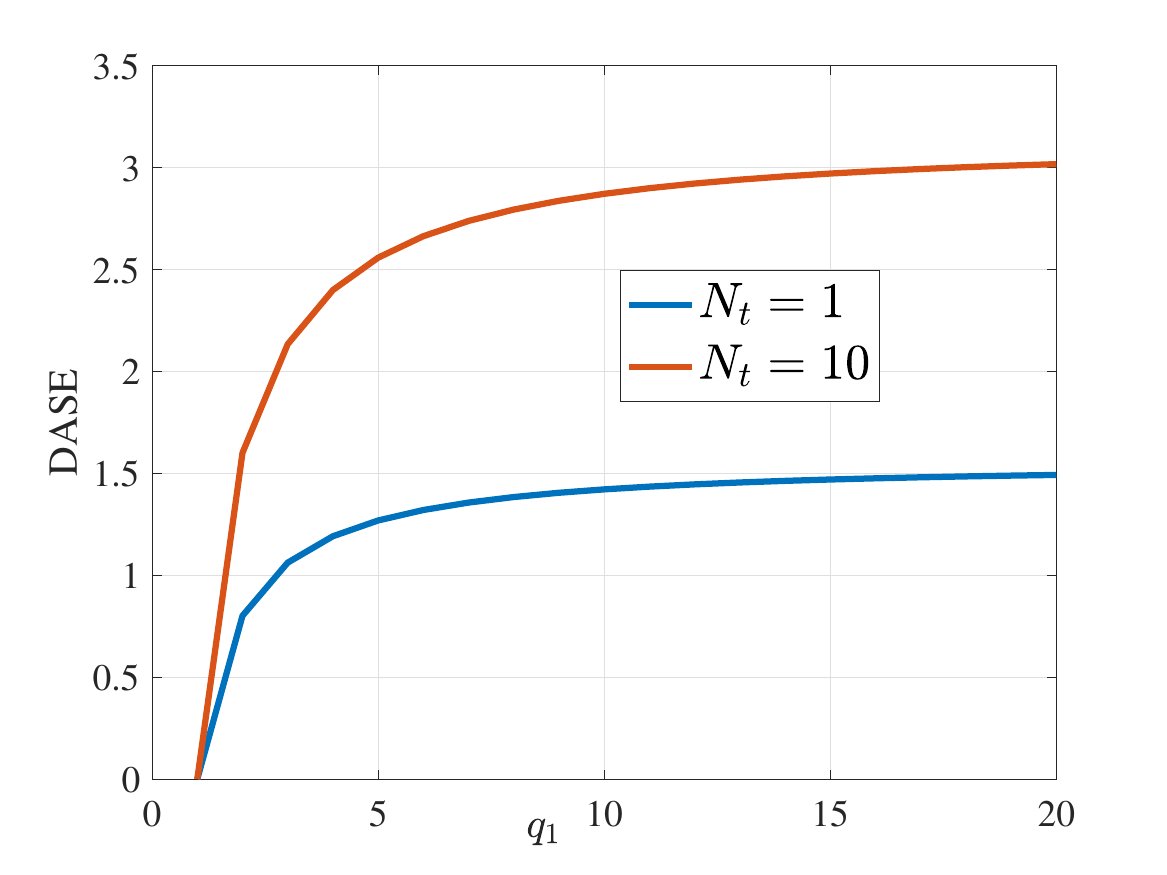}
        \caption{}
        \label{fig:SE(c)}
    \end{subfigure}
    \hfill
    \begin{subfigure}[]{0.24\textwidth}
        \centering
 \includegraphics[scale=.23,trim={2cm 0cm 0cm 0cm}]{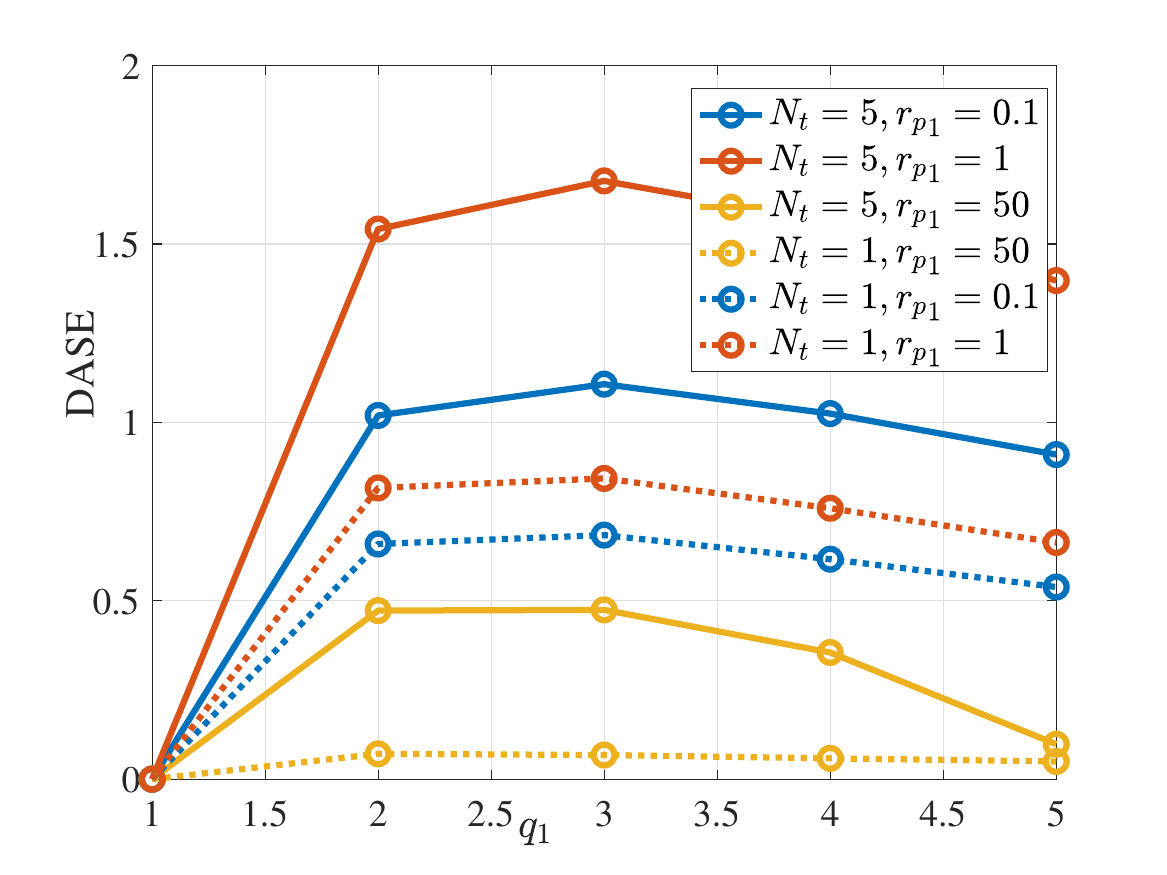}
        \caption{}
        \label{fig:SE(d)}
    \end{subfigure}
  
    \caption{{This figure evaluates the DASE levels in various numbers of transmit antennas, Doppler frequencies, and pilot an data powers. \ref{fig:SE(a)} shows DASE levels in \eqref{eq:DASE} versus Doppler frequency. \ref{fig:SE(b)} shows an upper-bound for DASE in single-antenna case provided in \cite[Eq. 30]{Fodor:23}. \ref{fig:SE(c)} shows the effect of number of transmit antennas on the spectral efficiency. \ref{fig:SE(c)} shows the effect of the power ratio on the DASE levels in the case of having multiple and single transmit antennas.}}
     \label{fig:img1}
\end{figure}
\begin{figure}[h]
    \centering
    
    \begin{subfigure}[]{0.24\textwidth}
        \centering
 \includegraphics[scale=.23,trim={2cm 0cm 0cm 0cm}]{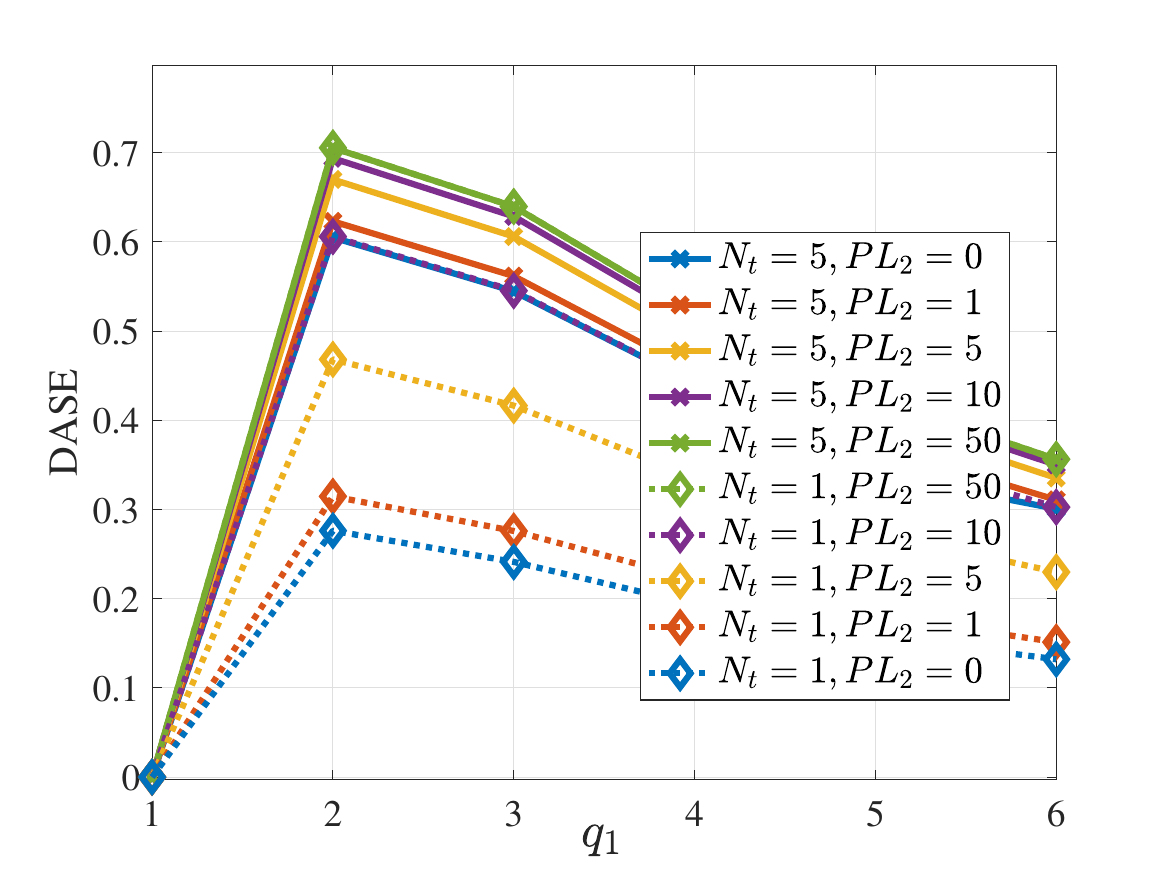}
        \caption{}
         \label{fig:inter(a)}
    \end{subfigure}
       \hfill
    \begin{subfigure}[]{0.24\textwidth}
        \centering
 \includegraphics[scale=.23,trim={2cm 0cm 0cm 0cm}]{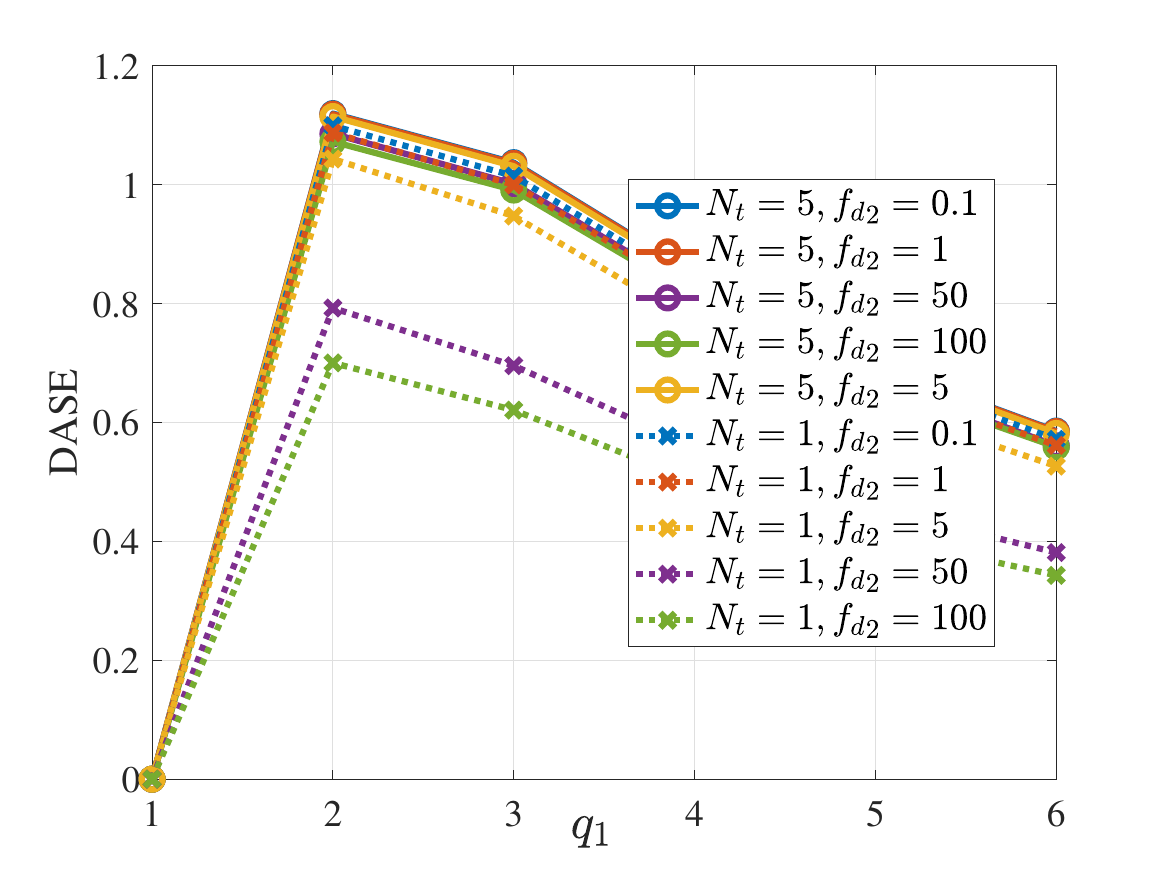}
        \caption{}
        \label{fig:inter(b)}
    \end{subfigure}
     \hfill
     \begin{subfigure}[]{0.24\textwidth}
        \centering
\includegraphics[scale=.23,trim={2cm 0cm 0cm 0cm}]{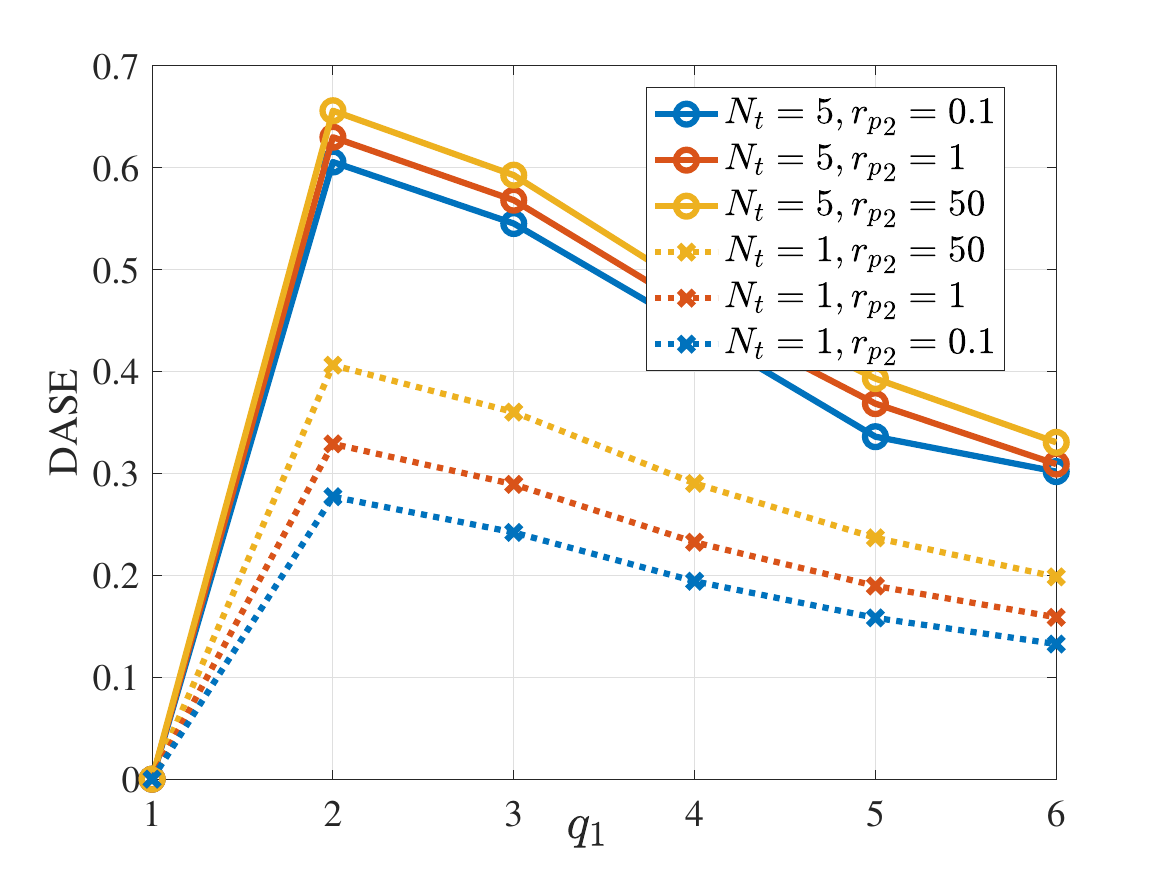}
        \caption{}
        \label{fig:inter(c)}
    \end{subfigure}
    \hfill
    \begin{subfigure}[]{0.24\textwidth}
        \centering
 \includegraphics[scale=.23,trim={2cm 0cm 0cm 0cm}]{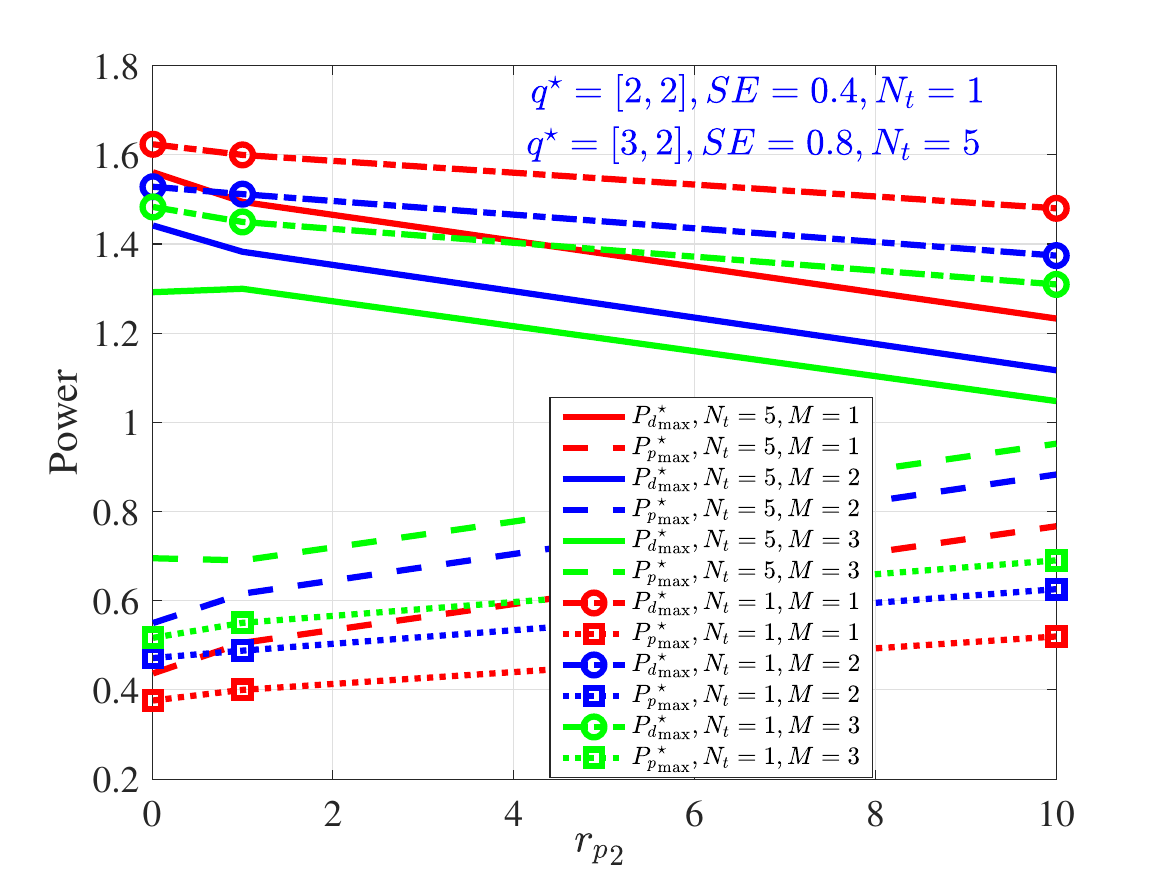}
        \caption{}
        \label{fig:inter(d)}
    \end{subfigure}
  \hfill
    \begin{subfigure}[]{0.24\textwidth}
        \centering
 \includegraphics[scale=.23]{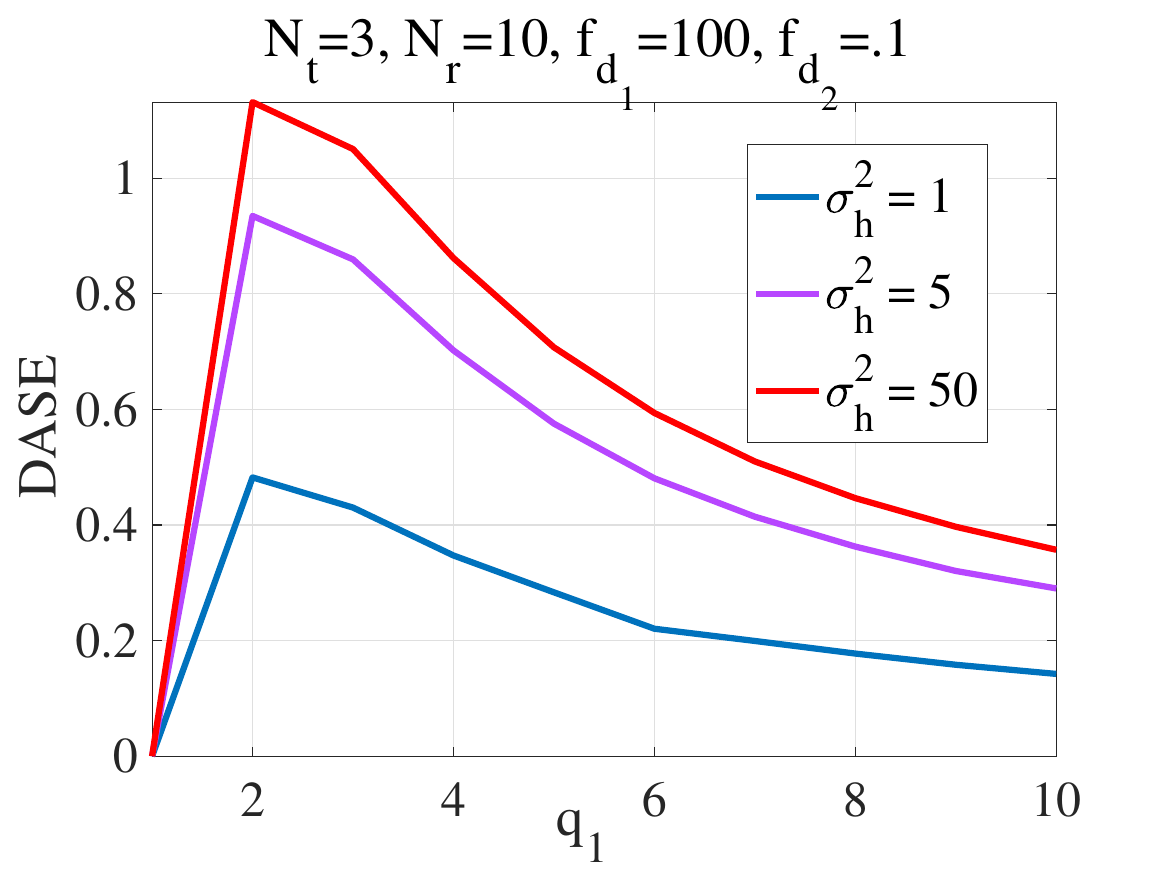}
        \caption{}
        \label{fig:inter(e)}
    \end{subfigure}
    \caption{\ref{fig:inter(a)}: \ac{DASE} versus the interference path loss. \ref{fig:inter(b)}: \ac{DASE} versus the interference maximum Doppler frequency. \ref{fig:inter(c)}: \ac{DASE} versus the interference the power ratio of the interference. \ref{fig:inter(d)}: Pilot and data powers versus the power ratio of the interference .{\color{\change} \ref{fig:inter(e)}: Sensitivity of optimal frame size to variance mismatch.}}
     \label{fig:img2}
\end{figure}

\begin{table}[h]
\caption{Performance of the proposed method in diverse non-stationary and stationary scenarios}
\vspace{0cm}
\label{table:comparison}
\resizebox{.4\textwidth}{!}{

\begin{tabular}{|p{1cm}|p{.5cm}|p{.9cm}|p{.6cm}|p{.2cm}|p{.8cm}|p{.7cm}|p{.3cm}|p{.4cm}|p{.5cm}|p{.5cm}|}
    \hline
    \hline
    ${\mx{q}^{\star}}$ &$M$ &SE& ${\rm SNR}$&$N_t$&${f_d}_1$&${f_d}_2$&${{\rm PL}}_1$&${{\rm PL}}_2$& $\sigma^2_{\mx{h}_1}$&$\sigma^2_{\mx{h}_2}$ \\
    \hline
     $\mx{6}$ &$\mx{1}$&$\mx{2.3646}$&$20$&$5$&$1$&$100$&$10$&$1$&$1$&$1$\\ \hline
    $[3,3]$ &$2$&$2.0586$             &$20$&$5$&$1$&$100$&$10$&$1$&$1$&$1$\\ \hline
    $[2,2,2]$ &$3$&$1.662$            &$20$&$5$&$1$&$100$&$10$&$1$&$1$&$1$\\ \hline\hline
     $\mx{6}$ &$\mx{1}$&$\mx{0.9839}$&$20$&$1$&$1$&$100$&$10$&$1$&$1$&$1$\\ \hline
    $[3,3]$ &$2$&$0.8973$             &$20$&$1$&$1$&$100$&$10$&$1$&$1$&$1$\\ \hline
    $[2,2,2]$ &$3$&$0.7588$            &$20$&$1$&$1$&$100$&$10$&$1$&$1$&$1$\\ \hline\hline
    ${2}$ &${1}$&${1.0722}$&$20$&$5$&$100$&$100$&$1$&$0$&$1$&$1$\\ \hline
    $\mx{[3,2]}$ &$\mx{2}$&$\mx{1.3759}$           &$20$&$5$&$100$&$100$&$1$&$0$&$1$&$1$\\ \hline
    $[2,2,2]$ &$3$&$1.3409$            &$20$&$5$&$100$&$100$&$1$&$0$&$1$&$1$\\ \hline\hline
   ${2}$ &${1}$&${0.7001}$&$20$&$1$&$100$&$100$&$1$&$0$&$1$&$1$\\ \hline
    ${[3,2]}$ &${2}$&${0.9128}$           &$20$&$1$&$100$&$100$&$1$&$0$&$1$&$1$\\ \hline
    $\mx{[2,2,2]}$ &$\mx{3}$&$\mx{0.9272}$            &$20$&$1$&$100$&$100$&$1$&$0$&$1$&$1$\\ \hline\hline
    $\mx{3}$ &$\mx{1}$&$\mx{2.9929}$&$20$&$5$&$5t$&$10$&$1$&$0$&$\tfrac{1}{5t}$&$1$\\ \hline
    ${[3,2]}$ &${2}$&${2.8397}$           &$20$&$5$&$5t$&$10$&$1$&$0$&$\tfrac{1}{5t}$&$1$\\ \hline
    ${[2,2,2]}$ &${3}$&${2.3529}$            &$20$&$5$&$5t$&$10$&$1$&$0$&$\tfrac{1}{5t}$&$1$\\ \hline\hline
     ${2}$ &${1}$&${1.3653}$&$20$&$1$&$5t$&$10$&$1$&$0$&$\tfrac{1}{5t}$&$1$\\ \hline
    $\mx{[3,2]}$ &$\mx{2}$&$\mx{1.4132}$           &$20$&$1$&$5t$&$10$&$1$&$0$&$\tfrac{1}{5t}$&$1$\\ \hline
    ${[2,2,2]}$ &${3}$&${1.1971}$            &$20$&$1$&$5t$&$10$&$1$&$0$&$\tfrac{1}{5t}$&$1$\\ \hline\hline
\end{tabular}}
\end{table}

\section{Numerical Results}
\label{Sec:simulations}
{

In this section, we present numerical experiments to evaluate the performance of the proposed method and compare its effectiveness with several alternative bounds and methods. The analysis focuses on the tightness of these bounds with respect to the expected SE.

\textbf{Experimental Setup}: The first experiment investigates the capacity bound provided in Theorem \ref{thm.stiel}, comparing it with the actual expected spectral efficiency $\mathds{E}[{\rm SE}]$ and four other methods:
\begin{enumerate}
    \item Actual Expected SE ($\mathds{E}[{\rm SE}]$): Calculated as the average of $10000$ Monte Carlo realizations of the instantaneous SE from \eqref{eq:random_SE}.
    \item UTF Bound (Monte Carlo): Derived using \eqref{eq:utf_bound}, with both the numerator and denominator approximated using $10000$ Monte Carlo simulations.
    \item Hoydis Bound (Deterministic Equivalent): Computed using deterministic equivalent expressions for the numerator and denominator, as outlined in \cite{Hoydis:13}.

    \item Ngo Lower-Bound: From \cite[Eq. 25]{ngo2013energy}, calculated using \eqref{eq:ngo_lower_bound}, where $\mathds{E}[\frac{1}{\gamma_{\rm lower}}]$ is approximated via $10000$ Monte Carlo iterations.

    \item Jensen's upper-bound: The Jensen's upper-bound for the expected SE given in \eqref{eq:jensen_upper} is calculated using Monte Carlo approximations of $\mathds{E}[\gamma]$. The deterministic equivalent expressions of Jensen's bound is also presented in several papers such as \cite{couillet2012random,couillet2011deterministic}. 
\end{enumerate}
We consider two distinct channel models:
\begin{enumerate}
    \item Sub-6GHz Channels:  For these types of channels, we consider two scenarios:
Rayleigh fading with widely spaced, independent antennas and uniformly distributed \ac{AoA}.
Rayleigh fading with high spatial correlation among the transmit antennas and independent receive antennas, where the channel may exhibit low-rank covariance matrices.
In both scenarios, the number of scatterers is assumed to be very large.
    \item Millimeter-Wave (mm-wave) Channels: These channels are characterized by limited scattering, with AoA concentrated around a central angle or certain clusters of angles, leading to low-rank channel covariance matrices.
\end{enumerate}
The channels corresponding to the users ($i=1,..., K$) are given by $\mx{C}_{\widehat{\mx{h}}_i}=\sigma_i^2 \mx{B}_i \mx{B}_i^H$ where $\mx{B}_i\in\mathbb{C}^{N\times r}$ is a matrix whose columns form steering vectors and $r$ is the rank of the covariance matrix. The variance of the interfering channels is assumed to be equal to $\sigma_{\rm inter}^2$. The \ac{SNR} is defined as ${\rm SNR}=10\log_{10}(\tfrac{P_{{\rm d},1}}{\sigma_d^2})-{\rm PL}_1$ where ${\rm PL}_1=20\log(\alpha_1)$ is the path loss of the channel corresponding to the tagged user. In the experiments shown in Figure \ref{fig:capacity_bounds}, all path losses are set to $\alpha_i=1, i=1,..., K$ and the data powers are ${P_{{\rm d},i}}=1$.

Figures \ref{fig:SEbound(b)}, \ref{fig:SEbound(e)}, \ref{fig:SEbound(g)}, \ref{fig:SEbound(h)} show results for Rayleigh fading channels with rich scattering environments. In these cases:

\begin{itemize}
    \item For small and moderate numbers of antennas ($N$), the UTF bound from \eqref{eq:utf_bound} provides a loose lower-bound for $\mathds{E}[{\rm SE}]$ and the lower-bound from \eqref{eq:ngo_lower_bound} is similarly non-tight.
    \item As we discussed earlier, the Hoydis bound might be an upper-bound for $\mathds{E}[{\rm SE}]$ which overestimates the actual expected SE.
    \item As $N$ increases, the UTF and Hoydis bounds become tighter and approaches the actual expected SE.
    \item Our proposed bound (highlighted in red) consistently offers a tight estimate of $\mathds{E}[{\rm SE}]$ , even for non-asymptotic values of $N$.
    \item If there are spatial correlations in the transmit antennas, the UTF bound and Jensen's bound fail to reflect the expected spectral efficiency as can be seen from Figures \ref{fig:SEbound(g)}, \ref{fig:SEbound(h)}. 
\end{itemize}
Figures \ref{fig:SEbound(a)}, \ref{fig:SEbound(c)}, \ref{fig:SEbound(d)}, and \ref{fig:SEbound(f)} illustrate results for mm-wave channels with limited scattering, where the \acp{AoA} are non-uniformly distributed around a central angle such as the von Misses distribution. These experiments reveal:
\begin{itemize}
\item As it turns out from Figures \ref{fig:SEbound(a)}, \ref{fig:SEbound(c)}, \ref{fig:SEbound(d)}, \ref{fig:SEbound(f)}, the UTF lower-bound \eqref{eq:utf_bound} is not tight. The gap between the UTF bound and $\mathds{E}[{\rm SE}]$ can be substantial, sometimes exceeding 5 bps/Hz, as seen in Figure \ref{fig:SEbound(a)}.
    \item The Hoydis theoretical bound may act as either an upper or lower bound depending on the specific scenario, e.g., it serves as a upper-bound in Figures \ref{fig:SEbound(c)} and \ref{fig:SEbound(e)}, but as a lower-bound in Figures \ref{fig:SEbound(a)}, \ref{fig:SEbound(a)}, \ref{fig:SEbound(d)} and \ref{fig:SEbound(f)}. However, it is consistently non-tight, even for large $N$. In addition, the deterministic equivalent expressions provided in \cite{Hoydis:13}, derived from random matrix theory, fail to accurately reflect the behavior of the expected spectral efficiency. This failure arises from the inaccuracies in deterministic approximations of the quadratic terms in the numerator and denominator and the fact that passing the expectation operator to both numerator and denominator does not provide necessarily a lower or upper-bound.

    \item The Ngo lower-bound is loose in all scenarios, offering little practical value.
\item Jensen's bound is relatively closer to the actual $\mathds{E}[{\rm SE}]$ , but its tightness does not improve asymptotically as
$N$ increases.
\end{itemize}
In all cases, the proposed bound accurately predicts $\mathds{E}[{\rm SE}]$ , demonstrating exceptional tightness even for small numbers of antennas. This highlights the robustness and reliability of our approach across both Sub-6GHz and mm-wave channel scenarios. 
 }

In the second round of experiments, we will evaluate our approach in pilot spacing and power control, using a Kronecker model for the correlation matrix, where $\mx{h}$ represents a typical channel, including the individual channel $\mx{h}_i$ of each user. The correlation matrix is given by $\mx{P}_{\mx{h}}(t_1,t_2)=\mx{P}_{\mx{h},{\rm T}}(t_1,t_2)\otimes \mx{P}_{\mx{h},{\rm R}}(t_1,t_2)$ where $\mx{P}_{\mx{h},{\rm T}}(t_1,t_2)\in\mathbb{C}^{N_t\times N_t}$ and $\mx{P}_{\mx{h},{\rm R}}(t_1,t_2)\in\mathbb{C}^{N_t\times N_t}$ are the correlation matrices of the transmit and receive antennas, respectively. In the stationary case, we assume in the experiments that $\mx{P}_{\mx{h},{\rm T}}=\rho_1 \mx{1}_{N_t}\mx{1}_{N_t}^H+(1-\rho_1)\mx{I}_{N_t}$ and $\mx{P}_{\rm R}=\rho_2\mx{1}_{N_r}\mx{1}_{N_r}^H+(1-\rho_2)\mx{I}_{N_r}$ with $\rho_1=0.9, \rho_2=0$ which imply that the transmit antennas are close to each other and they are to a great extent correlated and the receive antennas are uncorrelated. 
In the first experiment, we consider $q_{\max}=6$ and $M_{\max}=3$, ${P_{{\rm p},1}}_{\max}={P_{{\rm d},1}}_{\max}={{P_{{\rm p},2}}}_{\max}={{P_{d,2}}}_{\max}=1$ and compare three cases in Table \ref{table:comparison}: 1. using one frame with one pilot time slot and $5$ data time slots. 2. using $M=2$ frames with two pilot time slots at the first of each frame. 3. using $M=3$ frames with three pilots. Each of the three rows in Table \ref{table:comparison} corresponds to one experiment. The first four experiments are related to the stationary settings, while the rest examine the performance of our method in non-stationary scenarios. We consider $K=2$ users, where the first user is the tagged user. The Doppler frequencies, pilot power, data power, path loss and channel variances are shown respectively by ${f_d}_i, {{P_{{\rm p},i}}}_{\max}, {{P_{{\rm d},i}}}_{\max}, {\rm PL}_i, \sigma_{\mx{h}_i}^2$ where $i=1$ corresponds to the tagged user and $i=2$ shows the interference component. The optimal values for the number $M$ of frames, frame size $\mx{q}$ and their corresponding \ac{DASE} are shown by bold numbers. The pilot and data power in this experiment are assumed to be fixed. Note that the values of both \ac{SNR} and ${\rm PL}$ are shown in dB unit. The first two experiments shown in the first six rows of Table \ref{table:comparison} compare the single-antenna with the multi-antenna case in a scenario where the Doppler frequency of the tagged user is low and our proposed method suggests to use the maximum possible data slots with one frame. However, using multiple antennas leads to higher \ac{SE} compared to single-antenna case. In the second experiment, shown in the 7th-12th rows of Table \ref{table:comparison}, we examine a scenario where the user is rapidly moving with a high speed in the two cases of $N_t=5$ and $N_t=1$. We observe that the optimal frame sizes for $N_t=5$ occurs in $\mx{q}^\star=[3,2]$ while in single-antenna case, the optimal frame size is $\mx{q}^\star=[2,2,2]$. This shows that having multiple antennas at the transmit side not only leads to higher \ac{SE} but also spends less pilot time slots compared to the single-antenna case. In the last experiment of Table \ref{table:comparison}, we examine a non-stationary scenario where the tagged user is moving with a time-varying speed. We observe that using multiple antennas ($N_t=5$) and only one frame with frame size $\mx{q}^\star=3$ leads to the highest \ac{SE}, while the optimal design in single-antenna case occurs at $\mx{q}^{\star}=[3,2]$. This suggests that in multi-antenna transmitter with optimized beamforming, one can allocate the whole power to data time slots while in single-antenna, the power is distributed in two frames, which requires 
more pilot time slots to achieve the highest possible \ac{SE}.

{In the third round of experiments, we investigate the impact of the Doppler effect and the number of transmit antennas and power ratio on the optimal frame design, as illustrated in Figure \ref{fig:img1}. Here, the power ratio for user $k$ specifies the distribution between pilot and data power, and is defined as ${r_{p,k}}=\tfrac{{{P_{{\rm p},i}}}_{\max}}{{{P_{{\rm d},k}}}_{\max}}$ for the tagged user ($k=1$) and the interference ($k=2$). The used parameters in Figures \ref{fig:SE(a)} and \ref{fig:SE(b)} are ${f_d}_1=0.1, {f_d}_2=100, f_c=1000, {\rm PL}_1=0, {\rm PL}_2=0, q_{\max}=10, M=1, {P_{{\rm p},1}}_{\max}={P_{{\rm d},1}}_{\max}={{P_{p,2}}}_{\max}={{P_{d,2}}}_{\max}=1, {\rm SNR}=0 dB,  N_r=10, \tau_p=4$. In Figure \ref{fig:SE(c)}, we used the parameters 
${f_d}_1=10, {f_d}_2=1, f_c=1000, {\rm PL}_1=.5, {\rm PL}_2=0, q_{\max}=20, M=1, {P_{{\rm p},1}}_{\max}={P_{{\rm d},1}}_{\max}={{P_{p,2}}}_{\max}={{P_{d,2}}}_{\max}=1, {\rm SNR}=0 dB,  N_r=10, \tau_p=10$ and for Figure \ref{fig:SE(d)}, the parameters as as follows: ${f_d}_1=50, {f_d}_2=100, f_c=1000, {\rm PL}_1=1, {\rm PL}_2=0, q_{\max}=5, M=1, {{P_{p,2}}}_{\max}={{P_{d,2}}}_{\max}=1, P_{\rm tot}=5,  N_r=20, \tau_p=5, \sigma^2_{p,1}=\sigma_{d,1}^2=7.94\times 10^{-5}, {\sigma^2_{p,2}}= 10^{-4} $.
Figures \ref{fig:SE(a)} shows the effect of Doppler and number of transmit antennas on the optimal frame size which is determined by maximizing the DASE expression in \eqref{eq:DASE}. 
For comparison, in the single-antenna case ($N_t=1$), we also depict an upper-bound for the DASE expression provided in \cite[Eq. 30]{Fodor:23}. This is shown in Figure \ref{fig:SE(b)}. 
While the approach in \cite[Eq. 30]{Fodor:23} predicts a higher \ac{SE}, it is important to note that the actual \ac{SE} is lower. Moreover, maximizing the upper bound does not necessarily lead to the maximization of the actual SE. For instance, at ${f_d}_1=100$ in the single-antenna case, \cite[Eq. 30]{Fodor:23} indicates an optimal frame size $q_1^\star=5$ whereas our proposed expression in \eqref{eq:DASE} identifies the optimal frame size as ${q}_1^\star=2$ further reveals that the optimal frame size increases when the speed of the tagged user decreases. This finding suggests that in high-speed scenarios, concentrating the power budget in the initial time slots is advantageous due to the rapidly varying nature of the channel. 

From Figure \ref{fig:SE(c)}, we observe that utilizing multiple transmit antennas with optimal real-time beamforming can increase the \ac{DASE}. The level of improvement is also affected by frame size, Doppler and path loss of the tagged user.  
In Figure \ref{fig:SE(d)}, we analyze the impact of $r_p={r_p}_1$ on our frame design. It is evident that the power distribution between pilot and data, as represented by $r_p={r_p}_1$, influences the optimal frame size. Additionally, employing multiple transmit antennas contributes to an increased \ac{SE}.
Examining Figure \ref{fig:SE(d)}, we find that the power ratio between pilot and data time slots plays a crucial role in determining the optimal frame size. Specifically, a higher pilot power results in an augmentation of data time slots. This effect is more pronounced in scenarios with multiple transmit antennas compared to those with a single transmit antenna.

In the forth experiment, illustrated in Figure \ref{fig:img2}, we examine experiments to find out the effect of the interference factors such as interference path loss, interference Doppler and interference power ratio on the optimal frame design and the resulting \ac{SE} in case of the following parameters ${f_d}_1={f_d}_2=100, f_c=1000, {\rm PL}_1=1, q_{\max}=6, M_{\max}=1, {{P_{{\rm p},1}}}_{\max}={{P_{{\rm d},1}}}_{\max}={{P_{p,2}}}_{\max}={{P_{{\rm d},2}}}_{\max}=1,  N_r=10, \tau_p=5, \sigma^2_{p,1}=\sigma_{d,1}^2=7.94\times 10^{-5}, {\sigma^2_{p,2}}= 10^{-4}$.

Figure \ref{fig:inter(a)} shows that while the interference path loss affects \ac{SE}, it does not impact the optimal frame design. Furthermore, our observations from Figure \ref{fig:inter(a)} suggest that when employing multiple antennas at the transmitter side, the change in \ac{SE} caused by the interference path loss is considerably smaller than in the single-antenna case. The used parameters in this figure are as follows: ${f_d}_1={f_d}_2=100, f_c=1000, {\rm PL}_1=1, q_{\max}=6, M_{\max}=1, {{P_{{\rm p},1}}}_{\max}={{P_{{\rm d},1}}}_{\max}={{P_{p,2}}}_{\max}={{P_{d,2}}}_{\max}=1,  N_r=10, \tau_p=5, \sigma^2_{p,1}=\sigma_{d,1}^2=7.94\times 10^{-5}, {\sigma^2_{p,2}}= 10^{-4}$.

Figure \ref{fig:inter(b)} examines the effect of interference Doppler frequency on the optimal frame design and \ac{SE}. We observe from this figure that while different interference Doppler frequencies lead to different \ac{DASE} levels, the optimal frame size is fixed and not sensitive to changing interference Doppler frequency. We also see that the level of change in the resulting \ac{DASE} is more enhanced in single antenna rather than multi-antenna case. The used parameters in this figure are: ${f_d}_1=100, f_c=1000, {\rm PL}_1=1, {\rm PL}_2=0, q_{\max}=6, M_{\max}=1, {{P_{{\rm p},1}}}_{\max}={{P_{{\rm d},1}}}_{\max}={{P_{p,2}}}_{\max}={{P_{d,2}}}_{\max}=1,  N_r=10, \tau_p=5, \rho_1=0.9, \rho_2=0, \sigma^2_{p,1}=\sigma_{d,1}^2=7.94\times 10^{-5}, {\sigma^2_{p,2}}= 10^{-4} $.

In Figure \ref{fig:inter(c)}, we observe that changing the pilot and data powers of the interference components can change the \ac{DASE} while it does not change the optimal frame size. As depicted in Figures \ref{fig:inter(a)}, \ref{fig:inter(b)}, \ref{fig:inter(c)}, a significant observation emerges: as the number of transmit antennas increases, the disparity in multi-antenna \ac{DASE} among different interference factors diminishes. This outcome underscores that, with our proposed method and in scenarios where $N_t$ is large, the tagged user can experience approximately uniform \ac{SE} across various interference path loss and Doppler. The used parameters in this figure are ${f_d}_1=100, f_c=1000, {\rm PL}_1=1, {\rm PL}_2=0, q_{\max}=6, M_{\max}=1, {{P_{{\rm p},1}}}_{\max}={{P_{{\rm d},1}}}_{\max}=1,  N_r=10, \tau_p=5, \sigma^2_{p,1}=\sigma_{d,1}^2=7.94\times 10^{-5}, {\sigma^2_{p,2}}= 10^{-4} .$

The results of the fifth experiment, displayed in Figure \ref{fig:inter(d)}, depict the outcomes of the joint optimization of frame size and power. This experiment highlights the influence of interference power distribution, as represented by ${r_p}_2$, on the optimal power. Notably, we observe a shift in the optimal data time slot to the right when multiple transmit antennas are employed, indicating the preference for utilizing more data time slots in a multi-antenna scenario compared to a single-antenna configuration. The used parameters are ${f_d}_1={f_d}_2=100, f_c=1000, {\rm PL}_1=10, {\rm PL}_2=1, q_{\max}=6, M_{\max}=3, {P_{\rm tot}}_1={P_{\rm tot}}_2=2,  N_r=10, \tau_p=10, \sigma^2_{p,1}=\sigma_{d,1}^2=7.94\times 10^{-5}, {\sigma^2_{p,2}}= 10^{-4} $.}

{\color{\change}
Figure~\ref{fig:inter(e)} evaluates the sensitivity of the optimal frame size to varying channel variance levels. While different channel variance levels clearly affect the achievable DASE values, it is notable that the optimal frame size--i.e., the frame size maximizing DASE--remains consistent across these variations. This indicates that the optimal frame size is robust to channel variance mismatches arising in non-stationary environments, and depends primarily on the velocity profile of the tagged user.
}

\section{Conclusions}
\label{sec:conclusion}
This work delved into the uplink communications of MIMO multi-user systems featuring multiple transmit and receive antennas, operating over fast-fading non-stationary wireless channels that undergo aging between consecutive pilot signals. A dedicated beamforming framework was developed to address spatial correlations among transmit antennas, complemented by a channel estimation process that capitalizes on the temporal correlations of the channel.
Subsequently, a deterministic expression was introduced to approximate the average spectral efficiency in scenarios involving multi-frame data transmission. To enhance the performance further, an optimization framework was proposed to determine the optimal values for beamforming vectors, the number of frames, frame sizes, and power control, while satisfying some power constraints. Notably, our optimization approach exclusively leverages the knowledge of the temporal dynamics of the channel, eliminating the need for measurements or channel estimates.
More importantly, optimal frame design and pilot spacing were demonstrated to be unaffected by the velocity of interference users and interference path loss. Furthermore, we showed that the impact of interference path loss and velocity diminishes as the number of transmit antennas increases.
Simulation results robustly validated the effectiveness of our methodology, underscoring its profound influence on optimizing beamforming, pilot and data powers, frame sizes, and the number of frames across various practical scenarios.

\appendices
\section{Proof of Proposition \ref{prop.rician}}\label{proof.prop_rician}
{\color{\change}
We first recall the channel model at time $t$:
\begin{align}
{\mx{h}}(t) =\sum_{i=1}^{L(t)} {\rm e}^{\mathrm{j} (2\pi f_d^i(t) t+ \beta_i)} \mx{a}_R(\theta^i_{\rm AoA}(t))\otimes
\mx{a}_T(\theta^i_{\rm AoD}(t)).
\end{align}
The normalized channel at time $t$ is given by:
\begin{align}
  \mx{h}^{\prime}(t)=\mx{C}_{\mx{h}}^{-\tfrac{1}{2}}(t){\mx{h}}(t)=\frac{1}{\sqrt{L(t)}}\mx{R}^{-\frac{1}{2}}(t){\mx{h}}(t), 
\end{align}
where $ \mx{{R}}(t)\triangleq\mx{R}_{\mx{a}_R}(t)\otimes \mx{R}_{\mx{a}_T}(t)$
and 
\begin{align}
    &\mx{R}_{\mx{a}_R}(t)\triangleq\mathds{E}_{\theta(t)}[\mx{a}_R(\theta(t))\mx{a}^H_R(\theta(t))], \nonumber\\
    &\mx{R}_{\mx{a}_T}(t)\triangleq \mathds{E}_{\theta(t)}[\mx{a}_T(\theta(t))\mx{a}^H_T(\theta(t))].
\end{align}
\gf{Notice that} the correlation matrix between times $t_1$ and $t_2$ can be expressed as:
\begin{align}\label{eq:Ph_rel}
 \mx{P}_{\mx{h}}(t_1,t_2)=\frac{1}{\sqrt{L(t_1)L(t_2)}} \mx{R}^{-\frac{1}{2}}(t_1) \mx{C}_{\mx{h}}(t_1,t_2) \mx{R}^{-\frac{1}{2}}(t_2), 
\end{align}
where the cross-covariance matrix $\mx{C}_{\mx{h}}(t_1,t_2)$ between times $t_1$ and $t_2$ is given as:
\begin{align}\label{eq:raw1}
   &\mx{C}_{\mx{h}}(t_1,t_2) =\mathds{E}\Big[\sum_{i=1}^{L(t_1)}\sum_{l=1}^{L(t_2)}{\rm e}^{\mathrm{j} \Big(2\pi \big(f_d^i(t_1) t_1-f_d^l(t_2) t_2\big)+ \beta_i-\beta_l\Big)}\nonumber\\
  & \scalebox{.9}{$\left(\mx{a}_R(\theta_{\rm AoA}^i(t_1))\otimes \mx{a}_T(\theta_{\rm AoD}^i(t_1))\right)\left(\mx{a}_R^H(\theta_{\rm AoA}^l(t_2))\otimes \mx{a}_T^H(\theta_{\rm AoD}^l(t_2))\right)\Big]$}.
\end{align}
Since the phases $(\beta_i)$ are i.i.d. uniform on $[-\pi,\pi]$ and
independent of the \acp{AoA} and \acp{AoD} based on Assumption \eqref{A1},  
$\mathbb E[{\rm e}^{\mathrm{j}(\beta_i-\beta_l)}]=0$ whenever $i\neq l$.  Hence only the
$\min\!\bigl(L(t_1),L(t_2)\bigr)$ terms with $i=l$ survive in
\eqref{eq:raw1}, leading to the following relation:
\begin{align}
    &\mx{C}_{\mx{h}}(t_1,t_2)=\mathds{E}\Big[\sum_{i=1}^{\min(L(t_1),L(t_2))}{\rm e}^{\mathrm{j} \Big(2\pi \big(f_d^i(t_1) t_1-f_d^i(t_2) t_2\big)\Big)}
  \nonumber\\
   & \mx{a}_R(\theta_{\rm AoA}^i(t_1))\mx{a}_R^{\mathsf{H}}(\theta_{\rm AoA}^i(t_2))\otimes \mx{a}_T(\theta_{\rm AoD}^i(t_1))\mx{a}_T^{\mathsf{H}}(\theta_{\rm AoD}^i(t_2))\Big],
\end{align}
It is worth mentioning that path index $i$ refers to the same physical ray at both $t_1$ and $t_2$.  Rays with $i>\min\{L(t_1),L(t_2)\}$ exist only at one time and do not contribute to the matched-index sum.

Since those surviving terms are i.i.d.\ by Assumption \eqref{A2}, the sum can be further approximated as follows:

\begin{align}\label{eq:cov_elem}
    &\varsigma(t_1,t_2,p_1,p_2,q_1,q_2)\triangleq[\mx{C}_{\mx{h}}(t_1,t_2)]_{{N_t(p_1-1)+q_1,N_t(p_2-1)+q_2}} \approx\nonumber\\
    &\min(L(t_1),L(t_2)) I_1 I_2,\nonumber\\
    &p_1,p_2=0,..., N_r-1, q_1,q_2=0,..., N_t-1,
\end{align}
{\color{\change}where $I_1, I_2$ are defined in \eqref{eq:I1I2}}.

\begin{figure*}[h]
\begin{align}\label{eq:I1I2}
 &I_1\triangleq \mathds{E}\Big[\scalebox{1}{$
   {\rm e}^{\mathrm{j}\tfrac{2\pi}{\lambda}\big[(t_1\nu(t_1)+d_T q_1)\cos(\eta(t_1)-\theta_{\rm AoD}(t_1))-(t_2\nu(t_2)+d_T q_2)\cos(\eta(t_2)-\theta_{\rm AoD}(t_2))\big]} $}\Big],\nonumber\\
   &I_2\triangleq\mathds{E}\Big[{\rm e}^{\mathrm{j}\tfrac{2\pi d_R}{\lambda}\big[p_1\cos(\alpha-\theta_{\rm AoA}(t_1))-p_2\cos(\alpha-\theta_{\rm AoA}(t_2))\big]}
   \Big]
\end{align}
\end{figure*}

This is the case in environments with many comparably weak \ac{NLoS} components (e.g. urban macro/micro or dense indoor), where each path's AoA/AoD can be treated as an i.i.d. draw from a common PDF at each $t$. 
In the matrix-vector format, the cross-covariance matrix $\mx{C}_{\mx{h}}(t_1,t_2)$ can be equivalently written as
\begin{align}\label{eq:cov_mat}
 \mx{C}_{\mx{h}}(t_1,t_2)\approx L_{\min} \mx{R}_{\mx{a}_R}(t_1,t_2)\otimes \mx{T}_T(t_1,t_2),   
\end{align}
where $R_{a_R}(t_1,t_2)[p_1,p_2]\triangleq I_2$, $T_T(t_1,t_2)[q_1,q_2]\triangleq I_1$ and $L_{\min}\triangleq \min\{L(t_1),L(t_2)\}$.

The channel correlation matrix in \eqref{eq:Ph_rel} can then be obtained using \eqref{eq:cov_mat} and \eqref{eq:cor1}.

Computing the expressions $\mx{R}_{\mx{a}_R}(t_1,t_2)$, $\mx{R}_{\mx{a}_T}(t_1,t_2)$, and $\mx{T}_T(t_1,t_2)$ requires to know the joint \ac{PDF} of $\theta_{\rm AoD}(t_1)$ and $\theta_{\rm AoD}(t_2)$, joint \ac{PDF} of $\theta_{\rm AoA}(t_1)$ and $\theta_{\rm AoA}(t_2)$ and the joint \ac{PDF} for $\theta_{\rm AoA}(t_1)$, $\theta_{\rm AoA}(t_2)$, respectively. 

In the special case of Corollary \ref{corr.von_miss} where the statistics of user and the environment do not change with time, then the parameters $\nu(t)=\nu, \eta(t)=\eta, \theta_{\rm AoA}(t)=\theta_{\rm AoA}, \theta_{\rm AoD}(t)=\theta_{\rm AoD}$ are fixed. Define $\tau=t_2-t_1$ and $n_R=p_2-p_1, n_T=q_2-q_1$. When the \ac{AoA} and \ac{AoD} follow von Mises distribution with \acp{PDF} provided in \eqref{eq:pdfs_von}, the expressions $\mx{R}_{\mx{a}_R}$ and $\mx{R}_{\mx{a}_T}$ become constant and $\mx{T}_{T}(\tau)$ depends on $\tau$. These expressions can be calculated in closed-form as follows:
\begin{equation}\label{eq:expppr1}
\begin{aligned}
R_{a_T}[q_1,q_2]
&= \frac{I_0\Bigl(
     \sqrt{%
       \begin{aligned}
         &\kappa_{\rm AoD}^2
          - k_T^2 n_T^2 \\[-0.5ex]
         &\quad
          + 2\,\mathrm{j}\,\kappa_{\rm AoD}\,k_T n_T\,
            \cos\!\bigl(\eta - \theta_{\rm AoD}^c\bigr)
       \end{aligned}
     }
   \Bigr)}
   {I_0(\kappa_{\rm AoD})},\\[1ex]
R_{a_R}[q_1,q_2]
&= \frac{I_0\Bigl(
     \sqrt{%
       \begin{aligned}
         &\kappa_{\rm AoA}^2
          - k_R^2 n_R^2 \\[-0.5ex]
         &\quad
          + 2\,\mathrm{j}\,\kappa_{\rm AoA}\,k_R n_R\,
            \cos\!\bigl(\eta - \theta_{\rm AoA}^c\bigr)
       \end{aligned}
     }
   \Bigr)}
   {I_0(\kappa_{\rm AoA})},\\[1ex]
T_T(\tau)[q_1,q_2]
&= \frac{I_0\Bigl(
     \sqrt{%
       \begin{aligned}
         &\kappa_{\rm AoD}^2
          - \bigl(k_T n_T + \psi_{\tau}\bigr)^2+ \\
         &
           2\,\mathrm{j}\,\kappa_{\rm AoD}\,
            \bigl(k_T n_T + \psi_{\tau}\bigr)\,
            \cos\!\bigl(\eta - \theta_{\rm AoD}^c\bigr)
       \end{aligned}
     }
   \Bigr)}
   {I_0(\kappa_{\rm AoD})}\,.
\end{aligned}
\end{equation}
Here, we used the trigonometric property from \cite{harmonic2017} given by
\begin{align}
  &\tfrac{1}{2\pi} \int_{-\pi}^{\pi} {\rm e}^{\kappa\cos(\phi-b)+jc\cos(\phi-a)}{\rm d}\phi=\tfrac{I_0(\sqrt{\kappa^2-c^2+2j\kappa c\cos(a-b)})}{I_0(\kappa)} \nonumber\\
  &\forall c,\kappa,a,b \in\mathbb{R}.  
\end{align}
By using the expressions provided in \eqref{eq:expppr1}, the covariance matrix reads as
\begin{align}
 \mx{C}_{\mx{h}}(\tau)=  L_{\min} \mx{R}_{\mx{a}_R}\otimes \mx{T}_T(\tau). 
\end{align}
By inserting the covariance matrix in the equation \eqref{eq:Ph_rel}, we have

\begin{align}
 \mx{P}_{\mx{h}}(\tau) =\mx{I}_{N_r}\otimes \mx{R}_{a_T}^{-\frac12}\mx{T}(\tau)\mx{R}_{a_T}^{-\frac12},  
\end{align}
 which proves the Corollary \ref{corr.von_miss}. 
In the special case of Corollary \ref{corr.uniform} where $\kappa_{\rm AoD}=\kappa_{\rm AoA}=0$, we have a uniform distribution for \ac{AoD} and \ac{AoA}. By replacing both $\kappa_{\rm AoD}$, $\kappa_{\rm AoA}$ with zero, it holds that
\begin{align}
 &{R}_{a_T}[q_1,q_2]=J_0(k_T(q_2-q_1)),\nonumber\\
 &{R}_{a_T}[p_1,p_2]=J_0(k_R(p_2-p_1)),\nonumber\\
 &T_T(\tau)[q_1,q_2]=J_0(k_T(q_2-q_1)+\psi_{\tau}), 
\end{align}
and the correlation matrix becomes in the form of: 
\begin{align}
 \mx{P}_{\mx{h}}(\tau) =\mx{I}_{N_r}\otimes \mx{R}_{a_T}^{-\frac12}\mx{T}(\tau)\mx{R}_{a_T}^{-\frac12}.  
\end{align}
By single-transmit antenna assumptions, i.e., $N_t=1$, this leads to
\begin{align}
    R_{a_T}=1, T(\tau)= J_0(\psi_{\tau}),
 \end{align}
and 
\begin{align}
  \mx{P}_{\mx{h}}(\tau)=J_0(\psi_{\tau}) \mx{I}_{N_r},  
\end{align}
which exactly matches the well-known Jakes model \cite{jakes1974mobile,baddour2004accurate}.}

\section{Proof of Theorem \ref{thm.Instantanous_SINR}}\label{proof.thm.instantnaous_sinr}
{
Based on the received measurements at data time slots, the BS employs the optimal \ac{MMSE} receiver to estimate the transmitted data symbol of the tagged user at time slot $t$, i.e., $\widehat{{s}}_1(t)={\mx{g}_1^{\star}}^H(t)\mx{y}_{\rm d}$. Given the channel estimate, the received vector for all users after applying the linear \ac{MMSE} detector is given by $\mx{r}=\mx{G} \mx{y}_d$ where $\mx{G}=[\mx{g}_1,...,\mx{g}_K]^T\in\mathbb{C}^{K\times N_r}$. The first element of the latter vector is related to the tagged user (first user) which is as follows:
\begin{align}
 &r_1=\underbrace{\alpha_1 \mx{g}_1^H \widehat{\mx{H}}_1\mx{w}_1 s_1}_{\text{Desired signal}}+\underbrace{\sum_{l\neq 1}  \alpha_l \mx{g}_1^H \widehat{\mx{H}}_l\mx{w}_l s_l+\sum_{l= 1}^K  \alpha_l \mx{g}_1^H \widetilde{\mx{H}}_l\mx{w}_l s_l}_{{\text{Interference and channel estimation errors}}}\nonumber\\
 &+\underbrace{\mx{g}_1^H\mx{n}_d}_{\text{noise}}
\end{align}
where we have used the channel as the summation of the channel estimate and the channel estimate error, i.e, $\mx{H}_k=\widehat{\mx{H}}_k+\widetilde{\mx{H}}_k, k=1,..., K$. Given the channel estimates of all users, the first term is considered to be the desired signal and the noise+interference+channel errors are considered to be noise random variables. Given the channel estimates of all users, the instantaneous \ac{SINR} formula for the tagged user is obtained by dividing the power of desired signal to the power of interference and noise \cite[Eq. 2.46]{marzetta2016fundamentals}:
\begin{align}\label{eq:sinr1}
   \scalebox{.8}{$ \gamma(\mx{q},t,\widehat{\mx{h}}_1(t))=\frac{|\alpha_1 \mx{g}_1^H \widehat{\mx{H}}_1 \mx{w}_1|^2P_{{\rm d},1}}{ \sum_{l\neq 1}|\alpha_l \mx{g}_l^H \widehat{\mx{H}}_l\mx{w}_l |^2P_{d,l}+\sum_{l=1}^K\mathds{E}[|\alpha_l \mx{g}_l^H \widetilde{\mx{H}}_l\mx{w}_l |^2]P_{d,l}+\sigma_d^2 \|\mx{g}_1\|_2^2}$}
\end{align}
The numerator can be simplified as
\begin{align}\label{eq:numerator1}
    &|\alpha_1 \mx{g}_1^H \widehat{\mx{H}}_1 \mx{w}_1|^2P_{{\rm d},1}=\alpha_1^2 \langle \mx{g}_1 \mx{g}_1^H,\widehat{\mx{H}}_1\mx{C}_{\mx{x}_1}\widehat{\mx{H}}_1^H \rangle =\nonumber\\
    &\alpha_1^2\langle \mx{g}_1 \mx{g}_1^H,\mathcal{A}_1(\widehat{\mx{h}}_1\widehat{\mx{h}}_1^H) \rangle=\alpha_1^2\langle \widehat{\mx{h}}_1\widehat{\mx{h}}_1^H, \mx{C}_{\mx{x}_1}\otimes \mx{g}_1 \mx{g}_1^H \rangle
\end{align}
where we used Proposition \ref{prop.A(zz^T)} and Lemma \ref{lem.adjoint} in the following.
\begin{prop}\label{prop.A(zz^T)}
   For any operator $\mathcal{A}_k(\cdot): \mathbb{C}^{N_tN_r\times N_tN_r}\rightarrow \mathbb{C}^{N_r\times N_r}$, the following relationship holds: 
\begin{align}\label{eq:A(zz^T)}
    \mathcal{A}_k({\rm vec}(\mx{H}_k) {\rm vec}^H(\mx{H}_k))=\mx{H}_k \mx{C}_{\mx{x}_k} \mx{H}_k^{\mathsf{H}}.
\end{align}
\end{prop}
Proof: See Appendix \ref{proof.prop.A(zz^T)}.
\begin{lem}\label{lem.adjoint}
The ad-joint of $\mathcal{A}_k(\cdot)$ is obtained as: $\mathcal{A}_k^{Adj}(\mx{Z})=  \mx{C}_{\mx{x}_k} \otimes \mx{Z}$. 
\end{lem}
Proof: See Appendix \ref{proof.lem.adjoint}.

The middle term in the denominator of \eqref{eq:sinr1} can be simplified as follows:
\begin{align}\label{eq:denum1}
    &\mathds{E}[|\alpha_l \mx{g}_l^H \widetilde{\mx{H}}_l\mx{w}_l |^2]P_{{\rm d},l}=
    \alpha_l^2P_{{\rm d},l} (\mx{w}_l^H\otimes \mx{g}_l^H)\widetilde{\mx{h}}_l \widetilde{\mx{h}}_l^H(\mx{w}_l \otimes \mx{g}_l) \nonumber\\
    &\langle \mx{Q}_l, \mx{C}_{\mx{x}_l}\otimes \mx{g}_l\mx{g}_l^H\rangle,
\end{align}
where $\mx{Q}_l\triangleq \mathds{E}[\widetilde{\mx{h}}_l\widetilde{\mx{h}}_l^H]$ is the covariance of the channel estimate for user $l$.
The equation \eqref{eq:numerator1} together with \eqref{eq:denum1} leads to the following expression for the instantaneous \ac{SINR}:
\begin{align}\label{eq:SINR1}
 & \scalebox{.8}{$\gamma(\mx{q},t,\widehat{\mx{h}}_1(t))\triangleq \tfrac{\alpha_1^2  \left\langle \widehat{\mx{h}}_1\widehat{\mx{h}}_1^{\mathsf{H}}, \mx{C}_{\mx{x}_k}\otimes \mx{g}_1\mx{g}_1^H \right\rangle}
  {\sum_{k=2}^K\alpha_k^2
   \left\langle \widehat{\mx{h}}_k\widehat{\mx{h}}_k^{\mathsf{H}}, \mx{C}_{\mx{x}_k}\otimes  \mx{g}_1 \mx{g}_1^H   \right\rangle+\sum_{k=1}^K \alpha_k^2 \left\langle \mx{Q}_k, \mx{C}_{\mx{x}_k}\otimes\mx{g}_1\mx{g}_1^H   \right\rangle+\sigma^2_d \|\mx{g}_1\|_2^2}.$}  
\end{align}

    By having the definition of $\mx{F}_1$ in \eqref{eq:F_1_def}, the latter instantaneous \ac{SINR} can be rewritten as
\begin{align}\label{eq:SINR2}
    \gamma(\mx{q},t,\widehat{\mx{h}}_1(t))=\tfrac{\left\langle \mx{g}_1\mx{g}_1^{\mathsf{H}} , \mx{F}-\mx{F}_1 \right\rangle}{\left\langle \mx{g}_1\mx{g}_1^{\mathsf{H}} , \mx{F}_1 \right\rangle}.
\end{align}
It then follows that based on \eqref{eq:G_star}, $\mx{g}_1\mx{g}_1^{\mathsf{H}} $ can be stated as:
\begin{align}\label{eq:G^HG}
\mx{g}_1\mx{g}_1^{\mathsf{H}} =\alpha_1^2 P_{{\rm d},1} \mx{F}^{-1} \widehat{\mx{H}}_1 \mx{w}_1\mx{w}_1^{\mathsf{H}} \widehat{\mx{H}}_1^{\mathsf{H}} \mx{F}^{-1}.  
\end{align}

By incorporating \eqref{eq:A(zz^T)} and \eqref{eq:G^HG} into \eqref{eq:SINR2}, we reach to the following relation: 
    \begin{align}\label{eq:SINR3}
&\gamma(\mx{q},t,\widehat{\mx{h}}_1(t))=\tfrac{\alpha_1^2 P_{{\rm d},1} \left\langle  \mx{F}^{-1} \widehat{\mx{H}}_1 \mx{w}_1\mx{w}_1^{\mathsf{H}} \widehat{\mx{H}}_1^{\mathsf{H}} \mx{F}^{-1} , \alpha^2 \widehat{\mx{H}}_1\mx{C}_{\mx{x}_1} \widehat{\mx{H}}_1^{\mathsf{H}} \right\rangle}{\alpha_1^2 P_1\left\langle \mx{F}^{-1} \widehat{\mx{H}}_1 \mx{w}_1\mx{w}_1^{\mathsf{H}} \widehat{\mx{H}}_1^{\mathsf{H}} \mx{F}^{-1}, \mx{F} -\alpha_1^2 \widehat{\mx{H}}_1 \mx{w}_1\mx{w}_1^{\mathsf{H}} \widehat{\mx{H}}_1^{\mathsf{H}} \right\rangle}=\nonumber\\
&\tfrac{\alpha_1^2 P_{{\rm d},1} \left\langle  \mx{F}^{-1} \widehat{\mx{H}}_1 \mx{w}_1\mx{w}_1^{\mathsf{H}} \widehat{\mx{H}}_1^{\mathsf{H}} \mx{F}^{-1} , \widehat{\mx{H}}_1\mx{C}_{\mx{x}_1} \widehat{\mx{H}}_1^{\mathsf{H}} \right\rangle}
{\left\langle \mx{F}^{-1}\widehat{\mx{H}}_1 \mx{w}_1\mx{w}_1^{\mathsf{H}} \widehat{\mx{H}}_1^{\mathsf{H}} \mx{F}^{-1}, F\right\rangle-\alpha_1^2\left\langle  \mx{F}^{-1} \widehat{\mx{H}}_1 \mx{w}_1\mx{w}_1^{\mathsf{H}} \widehat{\mx{H}}_1^{\mathsf{H}} \mx{F}^{-1} , \widehat{\mx{H}}_1\mx{C}_{\mx{x}_1} \widehat{\mx{H}}_1^{\mathsf{H}} \right\rangle}.
\end{align}
When a single symbol is transmitted through multiple antennas at the user side, the covariance matrix satisfy $\mx{C}_{\mx{x}_1}=P_{{\rm d},1}\mx{w}_1\mx{w}_1^{\mathsf{H}}$ where $\|\mx{w}_1\|_2=1$. By using the rotational invariance property of the Frobenius inner product in the numerator of \eqref{eq:SINR3}, we can rewrite the numerator of \eqref{eq:SINR3} as follows
{
\begin{align}\label{eq:num_rel1}
\scalebox{.8}{$\alpha_1^2 P_{{\rm d},1} {\rm tr}(  \mx{F}^{-1} \widehat{\mx{H}}_1 \mx{w}_1 \mx{w}_1^{\mathsf{H}} \widehat{\mx{H}}_1^{\mathsf{H}} \mx{F}^{-1} \widehat{\mx{H}}_1  \mx{w}_1 \mx{w}_1^{\mathsf{H}} \widehat{\mx{H}}_1^{\mathsf{H}})=\alpha_1^2 P_{{\rm d},1}(\mx{w}_1^{\mathsf{H}} \widehat{\mx{H}}_1^{\mathsf{H}} \mx{F}^{-1}\widehat{\mx{H}}_1 \mx{w}_1)^2.$}    
\end{align}}
Also, the first term in the denominator of \eqref{eq:SINR3} is simplified as 
\begin{align}\label{eq:denum_rel1}
   {\rm tr} (\mx{F}^{-1} \widehat{\mx{H}}_1 \mx{w}_1 \mx{w}_1^{\mathsf{H}} \widehat{\mx{H}}_1^{\mathsf{H}} \mx{F}^{-1} \mx{F})=\mx{w}_1^{\mathsf{H}} \widehat{\mx{H}}_1^{\mathsf{H}} \mx{F}^{-1}\widehat{\mx{H}}_1 \mx{w}_1.
\end{align}
By having \eqref{eq:num_rel1} and \eqref{eq:denum_rel1}, the SINR expression in \eqref{eq:SINR3} simplifies to 
\begin{align}
    &\gamma(\mx{q},t,\widehat{\mx{h}}_1(t))=\tfrac{\alpha_1^2 P_{{\rm d},1} \mx{w}_1^{\mathsf{H}} \widehat{\mx{H}}_1^{\mathsf{H}} \mx{F}^{-1}\widehat{\mx{H}}_1 \mx{w}_1}{1-\alpha_1^2 P_{{\rm d},1} \mx{w}_1^{\mathsf{H}} \widehat{\mx{H}}_1^{\mathsf{H}} \mx{F}^{-1}\widehat{\mx{H}}_1 \mx{w}_1}.
\end{align}
Now, we can use the following lemma whose proof is provided in Appendix  \ref{proof.lem.SINR_equivalency}.
\begin{lem}\label{lem.SINR_equivalency}
Let $\mx{A}\in\mathbb{C}^{N_r\times N_r}$ be an invertible matrix and $\beta$ is a scalar. For any $\mx{z}\in\mathbb{C}^{N_r\times 1}$, it holds that
    \begin{align}\label{eq:sinr_rel}
        \tfrac{\mx{z}^{\mathsf{H}}\mx{A} \mx{z}}{\beta-\mx{z}^{\mathsf{H}}\mx{A} \mx{z}}=\mx{z}^{\mathsf{H}}\left(\beta \mx{A}^{-1}-\mx{z}\mx{z}^{\mathsf{H}}\right)^{-1}\mx{z}.
    \end{align}
\end{lem}
\noindent By  replacing $\mx{z}=\widehat{\mx{H}}_1 \mx{w}_1$ and $\mx{A}=\mx{F}^{-1}$ in Lemma \ref{lem.SINR_equivalency} and using the definition of $\mx{F}_1$ in Theorem \ref{thm.Instantanous_SINR} and Proposition \ref{prop.A(zz^T)}, we have
\begin{align}
     &\gamma(\mx{q},t,\widehat{\mx{h}}_1(t))= \alpha_1^2 P_{{\rm d},1} \mx{w}_1^{\mathsf{H}} \widehat{\mx{H}}_1^{\mathsf{H}}\left(\mx{F}- \alpha_1^2 \widehat{\mx{H}}_1 \mx{w}_1\mx{w}_1^{\mathsf{H}} \widehat{\mx{H}}_1^{\mathsf{H}}\right)^{-1} \widehat{\mx{H}}_1 \mx{w}_1\nonumber\\
     &=\alpha_1^2 P_{{\rm d},1} \mx{w}_1^{\mathsf{H}} \widehat{\mx{H}}_1^{\mathsf{H}} \mx{F}_1^{-1}\widehat{\mx{H}}_1 \mx{w}_1=\alpha_1^2 P_{{\rm d},1} \left\langle \widehat{\mx{H}}_1 \mx{w}_1\mx{w}_1^{\mathsf{H}} \widehat{\mx{H}}_1^{\mathsf{H}}, \mx{F}_1^{-1}\right\rangle=\nonumber\\
     &
     \alpha_1^2 \left\langle \mathcal{A}_1(\widehat{\mx{h}}_1 \widehat{\mx{h}}_1^{\mathsf{H}}), \mx{F}_1^{-1}\right\rangle
     =\alpha_1^2 \left\langle \widehat{\mx{h}}_1 \widehat{\mx{h}}_1^{\mathsf{H}}, \mx{C}_{\mx{x}_1}\otimes \mx{F}_1^{-1}\right\rangle=\nonumber\\
     &
     \alpha_1^2  \widehat{\mx{h}}_1^{\mathsf{H}} \left( \mx{C}_{\mx{x}_1}\otimes \mx{F}_1^{-1}\right) \widehat{\mx{h}}_1,
     \end{align}
      which proves the result.}
\section{Proof of Theorem \ref{thm.stiel}}\label{proof.thm.stiel}
Define the expectation of \ac{SE} and \ac{SINR},  by
\begin{align}
&\overline{\textup{SE}}(\mx{q},t)\triangleq \mathds{E}\left[\textup{SE}(\mx{q},t,
\widehat{\mx{h}}_1)\right], \overline{\gamma}(\mx{q},t)\triangleq  \mathds{E}\left[\gamma(\mx{q},t,\widehat{\mx{h}}_1)\right].
\end{align}
The aim is to find a deterministic equivalent expression for the instantaneous SE. First, we resort the second-order delta method to find the asymptotic distribution of the function ${\rm SE}=\log_2(1+\gamma)\triangleq f(\gamma)$. Let $\gamma^\circ$ be the deterministic equivalent expression such that $\gamma \xrightarrow{\mathds{P}} \gamma^\circ$ in the asymptotic limit.
First, we write the Taylor expansion of $f(\cdot)$ around the ${\gamma}^\circ$ as follows:
\begin{align}\label{eq:taylor_exan}
f(\gamma)=f({\gamma}^\circ)+(\gamma-{\gamma}^\circ)\frac{\partial f}{\partial \gamma}\Big|_{\gamma={\gamma}^\circ}+\frac{(\gamma-{\gamma}^\circ)^2}{2} \frac{\partial^2 f}{\partial^2 \gamma}\Big|_{\gamma=\widetilde{\gamma}}   
\end{align}
where $\widetilde{\gamma}$ lies between $\gamma$ and ${\gamma}^\circ$. We can rewrite the above relation as follows:
\begin{align}\label{eq:taylor_exan1}
&f(\gamma)=f({\gamma}^\circ)+(\gamma-\overline{\gamma}+\overline{\gamma}-{\gamma}^\circ)\frac{\partial f}{\partial \gamma}\Big|_{\gamma={\gamma}^\circ}+\nonumber\\
&\frac{(\gamma-\overline{\gamma}+\overline{\gamma}-{\gamma}^\circ)^2}{2} \frac{\partial^2 f}{\partial^2 \gamma}\Big|_{\gamma=\widetilde{\gamma}}   
\end{align}
If $\gamma\xrightarrow{\mathds{P}} \gamma^\circ$, then, because $|\widetilde{\gamma}-\gamma|\le |\gamma^\circ-\gamma|$ and $f''(\cdot)$ is continuous, we conclude that $f''(\widetilde{\gamma})\xrightarrow{\mathds{P}} f''(\gamma^\circ) $ where we used the continuous mapping theorem \cite{continous_mapp}.
By taking the expectation with respect to the random variable $\gamma$, we have:
\begin{align}\label{eq:E[f(gamma)]}
&\mathds{E}[f(\gamma)]=f({\gamma}^\circ)-\frac{{\rm var}(\gamma)}{2{\rm ln}(2)(1+\gamma^\circ)^2}+\frac{\overline{\gamma}-{\gamma}^\circ}{{\rm ln}(2)(1+\gamma^\circ)}\nonumber\\
&-\frac{(\overline{\gamma}-\gamma^\circ)^2}{2{\rm ln}(2)(1+\gamma^\circ)^2},
\end{align}
where we used the second-order delta method \cite{second_order_delta} to replace $f''(\widetilde{\gamma})$ by $f''(\gamma^\circ)$.


In what follows, we show that expected \ac{SINR} $\overline{\gamma}$ is well approximated by the \ac{SINR} expression ${\gamma}^{\circ}(\mx{q},t)$. This implies that 
\begin{align}
\mathds{E}[f(\gamma)]\xrightarrow{a.s.}    f({\gamma}^\circ)-\frac{{\rm var}(\gamma)}{2{\rm ln}(2)(1+\gamma^\circ)^2}
\end{align}
which is our final result.
Define $\mx{B}=\sum_{k=2}^K \alpha_k^2 \mathcal{A}_k(\widehat{\mx{h}}_k \widehat{\mx{h}}_k^{\rm H})+\bs{\Theta}$, we can write $\mx{F}_1$ in Theorem \ref{thm.Instantanous_SINR} as
$\mx{F}_1=\mx{B}+\bs{\Theta}+\rho_{\rm d} \mx{I}_{N_r}$.
\begin{align}\label{eq:t1}
&\tfrac{\overline{\gamma}(\mx{q},t)}{N_r}= \tfrac{1}{N_r} \alpha_1^2\left\langle \mx{C}_{\widehat{\mx{h}}_1}, \mx{C}_{\mx{x}_1}\otimes \mathds{E}[{\mx{F}_1}^{-1}]  \right\rangle 
 \stackrel{N_r\rightarrow \infty}{=} \nonumber\\
 &\tfrac{1}{N_r} \alpha_1^2\left\langle \mx{C}_{\widehat{\mx{h}}_1},\mx{C}_{\mx{x}_1}\otimes \bs{\Xi}^{-1}  \right\rangle\triangleq \tfrac{{\gamma}^{\circ}(\mx{q},t)}{N_r}
\end{align}
where $\bs{\Xi}\triangleq \mx{\Gamma}+\bs{\Theta}+\rho_{\rm d}  \mx{I}_{N_r}$ and $\mx{\Gamma}$ is a matrix depending on the Stieltjes transform corresponding to the empirical distribution of $\mx{B}$ and is determined later.

The approximation error in \eqref{eq:t1} is defined as the difference between $\overline{\gamma}(\mx{q},t)$ and $\gamma^{\circ}(\mx{q},t)$ given by:

\begin{align}\label{eq:t3}
   e_N=\tfrac{\alpha_1^2}{N_r}\left\langle \mathcal{A}_1(\mx{C}_{\widehat{\mx{h}}_1}),(\mx{B}+\rho_{\rm d}  \mx{I}_{N_r})^{-1}  \right\rangle-
   \tfrac{\alpha_1^2}{N_r} \left\langle \mathcal{A}_1(\mx{C}_{\widehat{\mx{h}}_1}), \bs{\Xi}^{-1} \right\rangle .
\end{align}
The aim is to show that $e_N\xrightarrow[]{N_r\rightarrow \infty} 0$ with high probability.
First, we state the resolvent matrix identity between two matrices i.e. 
\begin{align}\label{eq:resolvent_identity}
   \mx{U}^{-1}-\mx{V}^{-1}=\mx{V}^{-1}(\mx{V}-\mx{U})\mx{U}^{-1}, \forall \mx{U},\mx{V} 
\end{align}
to have
\begin{align}\label{eq:t2}
\scalebox{.9}{$(\mx{B}+\rho_{\rm d}  \mx{I}_{N_r})^{-1}- \bs{\Xi}^{-1}=\bs{\Xi}^{-1}(\bs{\Xi}-\mx{B}-\rho_{\rm d}  \mx{I}_{N_r})) (\mx{B}+\rho_{\rm d}\mx{I}_{N_r})^{-1}.$}    
\end{align}
We can proceed with \eqref{eq:t2} as follows:
\begin{align}
 &(\mx{B}+\rho_{\rm d} \mx{I}_{N_r})^{-1}- \bs{\Xi}^{-1}=\bs{\Xi}^{-1} \mx{\Gamma} (\mx{B}+\rho_{\rm d} \mx{I}_{N_r})^{-1} \nonumber\\
 &-\sum_{k=2}^K \alpha_k^2  \bs{\Xi}^{-1} \mathcal{A}_k(\widehat{\mx{h}}_k\widehat{\mx{h}}_k^H)(\mx{B}+\rho_{\rm d}  \mx{I}_{N_r})^{-1}.
\end{align}
Now, the residual $e_N$ in \eqref{eq:t3} becomes as
\begin{align}\label{eq:t4}
   &e_N=\tfrac{\alpha_1^2}{N_r}\left\langle \mathcal{A}_1(\mx{C}_{\widehat{\mx{h}}_1}),\bs{\Xi}^{-1}\mx{\Gamma} (\mx{B}+\rho_{\rm d}  \mx{I}_{N_r})^{-1}  \right\rangle-\nonumber\\
   &
   \tfrac{\alpha_1^2}{N_r} \sum_{k=2}^K \alpha_k^2  \left\langle \mathcal{A}_1(\mx{C}_{\mx{z}_1}), \bs{\Xi}^{-1} \mathcal{A}_k (\widehat{\mx{h}}_k {\widehat{\mx{h}}_k}^{\rm H} )(\mx{B}+\rho_{\rm d}  \mx{I}_{N_r})^{-1} \right\rangle.    
\end{align}
Due to the relation \ref{eq:A(zz^T)}, it follows that
\begin{align}\label{t5}
&\langle \mathcal{A}_{k}({\widehat{\mx{h}}}_k{{\widehat{\mx{h}}}_k}^{\rm H}), \bs{\Xi}^{-1} \mathcal{A}_1(\mx{C}_{{\widehat{\mx{h}}}_1}) (\mx{B}+\rho_{\rm d}  \mx{I}_{N_r})^{-1}\rangle\nonumber\\
&\langle \mx{J}_k \mx{C}_{\mx{x}_k} {\mx{J}_k}^{\rm H}, \bs{\Xi}^{-1} \mathcal{A}_1(\mx{C}_{{\widehat{\mx{h}}}_1}) (\mx{B}+\rho_{\rm d}  \mx{I}_{N_r})^{-1} \rangle=\nonumber\\
&P_{d_k}\mx{w}_k^{\rm H} {\mx{J}_k}^{\rm H} (\mx{B}+\rho_{\rm d}  \mx{I}_{N_r})^{-1} \mathcal{A}_1(\mx{C}_{{\widehat{\mx{h}}}_1}) \bs{\Xi}^{-1}  \mx{J}_k \mx{w}_k,
\end{align}
where in the last step, we use the definition of the covariance matrix $\mx{C}_{\mx{x}_k}= P_{{\rm d},k}\mx{w}_k \mx{w}_k^{\rm H}$ and trace rotational invariance property. 
Now, by using \cite[Equation 2.2]{silverstein1995empirical}, it follows that
\begin{align}\label{eq:t6}
    \mx{w}_k^{\rm H} {\mx{J}_k}^{\rm H} (\mx{B}+\rho_{\rm d}  \mx{I}_{N_r})^{-1} =\tfrac{\mx{w}_k^{\rm H} {\mx{J}_k}^{\rm H} (\mx{B}_{(k)}+\rho_{\rm d}  \mx{I}_{N_r})^{-1} }{1+\alpha_k^2P_{{\rm d},k}\mx{w}_k^{\rm H} {\mx{J}_k}^{\rm H} (\mx{B}_{(k)}+\rho_{\rm d}  \mx{I}_{N_r})^{-1}  \mx{J}_k \mx{w}_k},
\end{align}
in which $\mx{B}_{(k)}\triangleq \mx{B}- \alpha_k^2\mx{J}_k \mx{C}_{\mx{x}_k} {\mx{J}_k}^{\rm H}$.
We also have:
\begin{align}\label{eq:t7}
    &\scalebox{.85}{$P_{{\rm d},k} \mx{w}_k^{\rm H} {\mx{J}_k}^{\rm H} (\mx{B}+\rho_{\rm d}  \mx{I}_{N_r})^{-1}\mx{J}_k \mx{w}_k={{\widehat{\mx{h}}}_k}^{\rm H} \left(\mx{C}_{\mx{x}_k}\otimes  (\mx{B}+\rho_{\rm d}  \mx{I}_{N_r})^{-1}\right) {\widehat{\mx{h}}}_k .$}
\end{align}
By incorporating \eqref{eq:t6} and \eqref{eq:t7} into \eqref{eq:t4}, the approximation error reads as:
\begin{align}\label{eq:t8}
   &e_N=\tfrac{\alpha_1^2}{N_r} \sum_{k=2}^K \alpha_k^2  \tfrac{{{\widehat{\mx{h}}}_k}^{\rm H} \left( \mx{C}_{\mx{x}_k}\otimes  (\mx{B}_{(k)}+\rho_{\rm d}  \mx{I}_{N_r})^{-1} \mathcal{A}_1(\mx{C}_{{\widehat{\mx{h}}}_1})\bs{\Xi}^{-1}  \right) {\widehat{\mx{h}}}_k}{1+\alpha_k^2{{\widehat{\mx{h}}}_k}^H \mx{C}_{\mx{x}_k}\otimes (\mx{B}_{(k)}+\rho_{\rm d}  \mx{I}_{N_r})^{-1} {\widehat{\mx{h}}}_k}-\nonumber\\
   &\tfrac{\alpha_1^2}{N_r}\left\langle \mathcal{A}_1(\mx{C}_{{\widehat{\mx{h}}}_1}),\bs{\Xi}^{-1}\mx{\Gamma} (\mx{B}+\rho_{\rm d}  \mx{I}_{N_r})^{-1}  \right\rangle.
\end{align}
We next prove that with the following choice of $\mx{\Gamma}$ as
\begin{align}
    \mx{\Gamma}=\sum_{k=2}^K \tfrac{\alpha_k^2  \mathcal{A}_k(\mx{C}_{{\widehat{\mx{h}}}_k})}{1+ \alpha_k^2 N_r m_{{\mx{C}_{{\widehat{\mx{h}}}_k},\mx{B}}}(\rho_{\rm d})},
\end{align}
where $m_{{\mx{C}_{{\widehat{\mx{h}}}_k},\mx{B}}}(\rho_{\rm d})\triangleq\tfrac{1}{N_r}\langle  \mx{C}_{{\widehat{\mx{h}}}_k}, \mx{C}_{\mx{x}_k}\otimes (\mx{B}+\rho_{\rm d} \mx{I}_{N_r})^{-1} \rangle$, the residual $e_N$ goes to zero as $N_r$ goes to infinity. To see this, first, we replace this specific choice of $\mx{\Gamma}$ into \eqref{eq:t8} to form $e_N$ as follows:

\begin{align}\label{eq:t9}
  &e_N=\tfrac{\alpha_1^2}{N_r} \sum_{k=2}^K \alpha_k^2  \tfrac{{{\widehat{\mx{h}}}_k}^H \left( \mx{C}_{\mx{x}_k}\otimes  (\mx{B}_{(k)}+\rho_{\rm d}  \mx{I}_{N_r})^{-1} \mathcal{A}_1(\mx{C}_{{\widehat{\mx{h}}}_1})\bs{\Xi}^{-1}  \right) {{\widehat{\mx{h}}}_k}}{1+\alpha_k^2{{\widehat{\mx{h}}}_k}^H \mx{C}_{\mx{x}_k}\otimes (\mx{B}_{(k)}+\rho_{\rm d}  \mx{I}_{N_r})^{-1} {{\widehat{\mx{h}}}_k}} \nonumber\\
  &\scalebox{.9}{$-\tfrac{\alpha_1^2}{N_r}\sum_{k=2}^K\alpha_k^2 \tfrac{\langle \mx{C}_{{\widehat{\mx{h}}}_k}, \mx{C}_{\mx{x}_k}\otimes\bs{\Xi}^{-1}\mathcal{A}_1(\mx{C}_{{\widehat{\mx{h}}}_1})  (\mx{B}+\rho_{\rm d}  \mx{I}_{N_r})^{-1} \rangle }{1+\alpha_k^2N_r m_{{\mx{C}_{{\widehat{\mx{h}}}_k},\mx{B}}}(\rho_{\rm d})}\triangleq\frac{1}{K-1}\sum_{k=2}^K\alpha_k^2d_k,$}
\end{align}
where 
\begin{align}
 &d_k=\tfrac{\alpha_1^2}{N_r}  \tfrac{{{\widehat{\mx{h}}}_k}^H \left( \mx{C}_{\mx{x}_k}\otimes(\mx{B}_{(k)}+\rho_{\rm d}  \mx{I}_{N_r})^{-1} \mathcal{A}_1(\mx{C}_{{\widehat{\mx{h}}}_1}) {{\bs{\Xi}}^\prime}^{-1} \right) {\widehat{\mx{h}}}_k}{1+\alpha_k^2{{\widehat{\mx{h}}}_k}^H \mx{C}_{\mx{x}_k}\otimes (\mx{B}_{(k)}+\rho_{\rm d}  \mx{I}_{N_r})^{-1} {\widehat{\mx{h}}}_k}\nonumber\\
 &-\tfrac{\alpha_1^2}{N_r}\tfrac{\left\langle \mathcal{A}_k(\mx{C}_{{\widehat{\mx{h}}}_k}),{{\bs{\Xi}}^\prime}^{-1} \mathcal{A}_1(\mx{C}_{{\widehat{\mx{h}}}_1})(\mx{B}+\rho_{\rm d}  \mx{I}_{N_r})^{-1}  \right\rangle}{1+\alpha_k^2 N_r m_{{\mx{C}_{{\widehat{\mx{h}}}_k},\mx{B}}}(\rho_{\rm d})}, 
\end{align}
where $\bs{\Xi}^\prime\triangleq \tfrac{1}{K-1}\bs{\Xi}$.
Define 
\begin{align}
 &{\bs{\Xi}^\prime}_{(k)}\triangleq\frac{1}{K-1} \sum_{k=2}^K \tfrac{\alpha_k^2 \mathcal{A}_k(\mx{C}_{{\widehat{\mx{h}}}_k})}{1+\alpha_k^2N_rm_{{\mx{C}_{{\widehat{\mx{h}}}_k},\mx{B}_{(k)}}}(\rho_{\rm d})} +\tfrac{\bs{\Theta}}{K-1}+\tfrac{\rho_{\rm d}}{K-1}\mx{I}_{N_r}.
\end{align}
$d_k$ in the relation \eqref{eq:t9} can be stated as the sum of $4$ terms as below:
\begin{align}\label{eq:error_term}
d_k=d_k^{(1)}+d_k^{(2)}+d_k^{(3)}+d_k^{(4)}, 
\end{align}
where
\begin{align}\label{eq:d1}
&d_k^{(1)}= \tfrac{\alpha_1^2}{N_r} \tfrac{{\widehat{\mx{h}}}_k^H \left( \mx{C}_{\mx{x}_k}\otimes (\mx{B}_{(k)}+\rho_{\rm d}  \mx{I}_{N_r})^{-1} \mathcal{A}_1(\mx{C}_{{\widehat{\mx{h}}}_1}) {\bs{\Xi}^\prime}^{-1} \right) {\widehat{\mx{h}}}_k}{1+\alpha_k^2{{\widehat{\mx{h}}}_k}^H \mx{C}_{\mx{x}_k}\otimes (\mx{B}_{(k)}+\rho_{\rm d}  \mx{I}_{N_r})^{-1} {\widehat{\mx{h}}}_k} -\tfrac{\alpha_1^2}{N_r},\nonumber\\
&\tfrac{{\widehat{\mx{h}}}_k^H \left(\mx{C}_{\mx{x}_k}\otimes  (\mx{B}_{(k)}+\rho_{\rm d}  \mx{I}_{N_r})^{-1} \mathcal{A}_1(\mx{C}_{{\widehat{\mx{h}}}_1}) {\bs{\Xi}^\prime}_{(k)}^{-1}\right) {\widehat{\mx{h}}}_k}
{1+\alpha_k^2{{\widehat{\mx{h}}}_k}^H \mx{C}_{\mx{x}_k}\otimes (\mx{B}_{(k)}+\rho_{\rm d}  \mx{I}_{N_r})^{-1} {\widehat{\mx{h}}}_k},
\end{align}

\begin{align}\label{eq:d2}
&d_k^{(2)}=  \tfrac{\alpha_1^2}{N_r}\tfrac{{\widehat{\mx{h}}}_k^H \left( \mx{C}_{\mx{x}_k}\otimes  (\mx{B}_{(k)}+\rho_{\rm d}  \mx{I}_{N_r})^{-1} \mathcal{A}_1(\mx{C}_{{\widehat{\mx{h}}}_1}) {\bs{\Xi}^\prime}_{(k)}^{-1}  \right) {\widehat{\mx{h}}}_k}{1+\alpha_k^2{{\widehat{\mx{h}}}_k}^H \mx{C}_{\mx{x}_k}\otimes (\mx{B}_{(k)}+\rho_{\rm d}  \mx{I}_{N_r})^{-1} {\widehat{\mx{h}}}_k}-\nonumber\\
&
\tfrac{\alpha_1^2}{N_r}\tfrac{\langle \mx{C}_{{\widehat{\mx{h}}}_k}, {\bs{\Xi}^\prime}_{(k)}^{-1}\mathcal{A}_1(\mx{C}_{{\widehat{\mx{h}}}_1}) \mx{C}_{\mx{x}_k}\otimes (\mx{B}_{(k)}+\rho_{\rm d}  \mx{I}_{N_r})^{-1}  \rangle }
{1+\alpha_k^2{{\widehat{\mx{h}}}_k}^H \mx{C}_{\mx{x}_k}\otimes (\mx{B}_{(k)}+\rho_{\rm d}  \mx{I}_{N_r})^{-1} {\widehat{\mx{h}}}_k}, 
&
\end{align}
\begin{align}\label{eq:d3}
&d_k^{(3)}= \tfrac{\alpha_1^2}{N_r}\tfrac{\langle \mx{C}_{{\widehat{\mx{h}}}_k}, \mx{C}_{\mx{x}_k}\otimes {\bs{\Xi}^\prime}_{(k)}^{-1}\mathcal{A}_1(\mx{C}_{{\widehat{\mx{h}}}_1}) (\mx{B}_{(k)}+\rho_{\rm d}  \mx{I}_{N_r})^{-1}\rangle }
{1+\alpha_k^2{{\widehat{\mx{h}}}_k}^H \mx{C}_{\mx{x}_k}\otimes (\mx{B}_{(k)}+\rho_{\rm d}  \mx{I}_{N_r})^{-1} {\widehat{\mx{h}}}_k} - \nonumber\\
&\tfrac{\alpha_1^2}{N_r}\tfrac{\langle \mx{C}_{{\widehat{\mx{h}}}_k}, \mx{C}_{\mx{x}_k}\otimes {\bs{\Xi}^\prime}^{-1}\mathcal{A}_1(\mx{C}_{{\widehat{\mx{h}}}_1}) (\mx{B}+\rho_{\rm d}  \mx{I}_{N_r})^{-1} \rangle }
{1+\alpha_k^2{{\widehat{\mx{h}}}_k}^H \mx{C}_{\mx{x}_k}\otimes (\mx{B}_{(k)}+\rho_{\rm d}  \mx{I}_{N_r})^{-1} {\widehat{\mx{h}}}_k},
\end{align}
\begin{align}\label{eq:d4}
&d_k^{(4)}=\tfrac{\alpha_1^2}{N_r}\tfrac{\langle \mx{C}_{{\widehat{\mx{h}}}_k}, \mx{C}_{\mx{x}_k}\otimes {\bs{\Xi}^\prime}^{-1}\mathcal{A}_1(\mx{C}_{{\widehat{\mx{h}}}_1}) (\mx{B}+\rho_{\rm d}  \mx{I}_{N_r})^{-1} \rangle }
{1+\alpha_k^2{{\widehat{\mx{h}}}_k}^H \mx{C}_{\mx{x}_k}\otimes (\mx{B}+\rho_{\rm d}  \mx{I}_{N_r})^{-1} {\widehat{\mx{h}}}_k} -  \nonumber\\
&\tfrac{\alpha_1^2}{N_r}\tfrac{\langle \mx{C}_{{\widehat{\mx{h}}}_k}, \mx{C}_{\mx{x}_k}\otimes{\bs{\Xi}^\prime}^{-1}\mathcal{A}_1(\mx{C}_{{\widehat{\mx{h}}}_1})  (\mx{B}+\rho_{\rm d}  \mx{I}_{N_r})^{-1} \rangle }
{1+\alpha_k^2N m_{{\mx{C}_{{\widehat{\mx{h}}}_k},\mx{B}_{(k)}}}(\rho_{\rm d})}.
\end{align}
To bound $d_k^{(1)}$, we use Lemma \ref{lem.stiel} whose proof is given in Appendix \ref{proof.lem.stiel1}.

\begin{lem}\label{lem.stiel}
Let $\mx{J}\in\mathbb{C}^{N_r\times N_t}$, $\mx{B}$ be a Hermitian non-negative definite matrix and $\mx{C}$ is of size $N_t\times N_t$. Then for $t>0$ and $\rho_{\rm d}  >0$,  we have: $    \left|\tfrac{1}{1+t\langle\mx{J}\mx{C}\mx{J}^H, \left(\mx{B}+\rho_{\rm d}  \mx{I}\right)^{-1}\rangle}\right|\le 1$.
\end{lem}

\noindent It then follows that
\begin{align}\label{eq:d1bound1}
  &\scalebox{.9}{$|d_{k}^{(1)}|\le \tfrac{\alpha_1^2 P_{d,k}}{\alpha_k^2 N_rN_t}\|{\widehat{\mx{h}}}_k\|_2^2\|(\mx{B}_{(k)}+\rho_{\rm d}  \mx{I}_{N_r})^{-1}\|\|{\bs{\Xi}^\prime}^{-1}-{\bs{\Xi}^\prime}_{(k)}^{-1}\|\|\mathcal{A}_1(\mx{C}_{{\widehat{\mx{h}}}_1})\|$}, 
\end{align}
where we used Lemma \ref{lem.stiel}. $(\mx{B}_{(k)}+\rho_{\rm d}  \mx{I}_{N_r})^{-1}$ is the Stieltjes transform of a finite measure in $\mathbb{C}^{-}$, so by \cite[Theorem 3.2]{couillet2011random}, we have 
\begin{align}\label{eq:stiel_bound}
    \|\left(\mx{B}_{(k)}+\rho_{\rm d}  \mx{I}_{N_r}\right)^{-1}\|\le \tfrac{1}{|\rho_{\rm d} |}
\end{align}
To bound $\|{\bs{\Xi}^\prime}^{-1}-{\bs{\Xi}^\prime}_{(k)}^{-1}\|$, we can use the resolvent matrix identity \eqref{eq:resolvent_identity} which leads to
\begin{align}\label{eq:t14}
  &\|{\bs{\Xi}^\prime}^{-1}-{\bs{\Xi}^\prime}_{(k)}^{-1}\|\le \|{\bs{\Xi}^\prime}_{(k)}^{-1}\|\|{\bs{\Xi}^\prime}_{(k)}-{\bs{\Xi}^\prime}\|\|{\bs{\Xi}^\prime}^{-1}\|. 
\end{align}
By \cite[Lemma 8]{couillet2011deterministic}, we have that $\|{\bs{\Xi}^\prime}_{(k)}^{-1}\|$ and $\|{\bs{\Xi}^\prime}^{-1}\|$ are less than $\tfrac{1}{|\rho_{\rm d} |}$. Also, it follows that
\begin{align}\label{eq: t12}
    &\|{\bs{\Xi}^\prime}_{(k)}-{\bs{\Xi}^\prime}\|\le \tfrac{\mu_k^\prime}{K-1}\nonumber\\
    &\sum_{k=2}^K \alpha_k^4 N_r  \tfrac{|m_{\mx{C}_{{\widehat{\mx{h}}}_k},\mx{B}_{(k)}}(\rho_{\rm d} )-m_{\mx{C}_{{\widehat{\mx{h}}}_k},\mx{B}}(\rho_{\rm d} )|}{(1+\alpha_k^2 N_r m_{\mx{C}_{{\widehat{\mx{h}}}_k},\mx{B}_{(k)}}(\rho_{\rm d} ))(1+\alpha_k^2 N_rm_{\mx{C}_{{\widehat{\mx{h}}}_k},\mx{B}}(\rho_{\rm d} ))} \le \tfrac{{\mu_k^\prime}^4}{|\rho_{\rm d}|},
\end{align}
where we used \cite[Lemma 2.1]{bai2007signal}, and
\begin{align}\label{eq:A(Rz)_upper bound}
&\|\mathcal{A}_k(\mx{C}_{{\widehat{\mx{h}}}_k})\| =\sup_{\substack{\mx{x}\neq \mx{0}\\\mx{y}\neq \mx{0}} }\tfrac{\langle \mathcal{A}_k(\mx{C}_{{\widehat{\mx{h}}}_k})\mx{y}, \mx{x} \rangle}{\|\mx{x}\|_2\|\mx{y}\|_2} = \sup_{\substack{\mx{x}\neq \mx{0}\\\mx{y}\neq \mx{0}} }\tfrac{\langle \mathcal{A}_k(\mx{C}_{{\widehat{\mx{h}}}_k}), \mx{x}\mx{y}^H \rangle}{\|\mx{x}\|_2\|\mx{y}\|_2}\nonumber\\
&=
\sup_{\substack{\mx{x}\neq \mx{0}\\\mx{y}\neq \mx{0}} }\tfrac{\langle \mx{C}_{{\widehat{\mx{h}}}_k}, \mx{x}\mx{y}^H \otimes \mx{C}_{\mx{x}_k}  \rangle}{\|\mx{x}\|_2\|\mx{y}\|_2}
\le {\rm tr}(\mx{C}_{{\widehat{\mx{h}}}_k})\|\mx{C}_{\mx{x}_k}\|\le\tfrac{\Omega_1^2\Omega_2L_k}{N}\tfrac{{P_d}_k}{N_t}\triangleq \mu_k^\prime,
\end{align}
in which $\mx{C}_{{\widehat{\mx{h}}}_k}=\bs{\Psi}_{{\widehat{\mx{h}}}_k}\mx{C}_{\bs{\zeta}_k}\bs{\Psi}_{{\widehat{\mx{h}}}_k}^H$ and we made the following assumptions:
\begin{align}\label{eq:R_upperbound}
 &\|\mx{\Psi}_{{\widehat{\mx{h}}}_k}\|\le \Omega_1, ~\|\mx{C}_{\bs{\zeta}_k}\|\le \tfrac{\Omega_2}{N}, {\rm tr}(\mx{C}_{\bs{\zeta}_k})\le \tfrac{L_k\Omega_2}{N} ~~\forall k .
\end{align}
Here, $L_k$ is the number of paths between user $k$ and the \ac{BS}. Furthermore,
it holds that
\begin{align}\label{eq:t11}
   |m_{\mx{C}_{{\widehat{\mx{h}}}_k},\mx{B}_{(k)}}(\rho_{\rm d} )-m_{\mx{C}_{{\widehat{\mx{h}}}_k},\mx{B}}(\rho_{\rm d} )|\le  \tfrac{\mu_k^\prime}{N_r|\rho_{\rm d} |},
\end{align}
 and by \cite[Corollary 3.1]{couillet2011random} when $K\rightarrow \infty$,
 \begin{align}\label{eq:t10}
    |\tfrac{1}{K-1}\sum_{k=2}^K\tfrac{\alpha_k^2}{1+\alpha_k^2Nm_{\mx{C}_{{\widehat{\mx{h}}}_k},\mx{B}_{(k)}}(\rho_{\rm d} )}|\le \mu_k^\prime .
 \end{align}
Thus, this leads to finding an upper-bound in \eqref{eq:t14} which finally reads as
\begin{align}
    &\|{\bs{\Xi}^\prime}^{-1}-{\bs{\Xi}^\prime}_{(k)}^{-1}\|\le {\mu_k^\prime}^4 \tfrac{1}{|\rho_{\rm d}|^3}.
\end{align}
Now by having this in mind, together with \eqref{eq:stiel_bound}, \eqref{eq:d1bound1} reads as follows:
\begin{align}\label{eq:d1_main}
   &|d_k^{(1)}|\le \tfrac{ \alpha_1^2\mathds{E}\|{\widehat{\mx{h}}}_k\|_2^2}{\alpha_k^2} \tfrac{P_{{\rm d},k}{\mu_1^\prime}{\mu_k^\prime}^4}{N_tN_r|\rho_{\rm d}|^4}.
\end{align}
To bound $\mathds{E}\|{\widehat{\mx{h}}}_k\|_2^2$ in the above relation, we use the following lemma which provides an adapted version of \cite[Lemma B.26]{bai2010spectral}, showing that non-centered quadratic matrix forms concentrate around their expectation.
\begin{lem}\label{lem.concentration_lemma}
    Let $\mx{A}\in\mathbb{C}^{N\times N}$ be a deterministic matrix and ${\widehat{\mx{h}}}=\overline{\mx{h}}+\bs{\Delta}\mx{a}\in\mathbb{C}^{N\times 1}$ where $\mx{a}\in\mathbb{C}^{r}$ is a random vector of zero mean with i.i.d. elements and variance $\tfrac{1}{N}$ and $\bs{\Delta}\in\mathbb{C}^{N\times r}$ is a deterministic matrix with bounded spectral norm that is 
    $\|\bs{\Delta}\|\le  Q$. Also, assume that the cross-correlations and auto-correlations satisfy $|\mathds{E}\{|a(t)a(j)|^2\}|\le \nu_4~\forall i,j$ and $|\mathds{E}\{|a(t)|^2\}|\le \nu_2~ \forall i$. Then, we have:
\begin{align}
  \mathds{E}|{\widehat{\mx{h}}}^H \mx{A} {\widehat{\mx{h}}}-\langle \mx{R}_{{\widehat{\mx{h}}}}, \mx{A}\rangle | \le C  \tfrac{1}{N}(\nu_2+\sqrt{\nu_4}) Q^2 \|\mx{A}\|_F,
\end{align}    
for some constant $C$.
\end{lem}
Now, by using Lemma \ref{lem.concentration_lemma}, \eqref{eq:d1_main} reads as follows:
\begin{align}
  &\mathds{E}|d_k^{(1)}|\le  \tfrac{\alpha_1^2P_{{\rm d},k}{\mu_1^\prime}{\mu_k^\prime}^4}{\alpha_k^2 N_tN_r|\rho_{\rm d}|^4} 
  \left(\mathds{E}|\|{\widehat{\mx{h}}}_k\|_2^2-{\rm tr}(\mx{C}_{{\widehat{\mx{h}}}_k})|+{\rm tr}(\mx{C}_{{\widehat{\mx{h}}}_k})\right)\le\nonumber\\
  &\tfrac{\alpha_1^2P_{{\rm d},k}{\mu_1^\prime}{\mu_k^\prime}^4}{\alpha_k^2 N_tN_r|\rho_{\rm d}|^4}\left(c^\prime\tfrac{\Omega_1^2\Omega_2}{\sqrt{N}}+\tfrac{\Omega_1^2\Omega_2 L_k}{N}\right)\xrightarrow[]{N_r\rightarrow \infty} 0,
\end{align}
where we used \eqref{eq:R_upperbound} and Lemma \ref{lem.concentration_lemma} and
\begin{align}
  &\mathds{E}|\|{\widehat{\mx{h}}}_k\|_2^2-\mx{C}_{{\widehat{\mx{h}}}_k}|\le c^\prime\tfrac{\Omega_1^2\Omega_2}{\sqrt{N}}.
\end{align}
For $d^{(2)}$ bound based on Lemmas \ref{lem.stiel} and \ref{lem.concentration_lemma} , we have:
{
\begin{align}
   &\scalebox{.9}{$\mathds{E} |d_k^{(2)}|\le 
   \tfrac{c^\prime \alpha_1^2}{\alpha_k^2 N_r N} \Omega_1^2\Omega_2 \|\mx{C}_{\mx{x}_k}\otimes  (\mx{B}_{(k)}+\rho_{\rm d}  \mx{I}_{N_r})^{-1} \mathcal{A}_1(\mx{C}_{{\widehat{\mx{h}}}_1}) {\bs{\Xi}^\prime}_{(k)}^{-1}\|_F$} \nonumber\\
   &\le \tfrac{c^\prime \alpha_1^2}{\alpha_k^2 N_r N} \Omega_1^2\Omega_2 P_{{\rm d},k} \|  (\mx{B}_{(k)}+\rho_{\rm d}  \mx{I}_{N_r})^{-1}\|\|{\bs{\Xi}^\prime}_{(k)}^{-1}\|\|\mathcal{A}(\mx{C}_{{\widehat{\mx{h}}}})\|_F\nonumber\\
   &\le \tfrac{c^\prime \alpha_1^2 \Omega_1^2\Omega_2}{\alpha_k^2\sqrt{N_r}N |\rho_{\rm d} |^2} {\mu_1^\prime}\xrightarrow[]{N_r\rightarrow \infty} 0.
\end{align}}
\noindent for some constant $c^\prime>0$.
To obtain a bound for $d_k^{(3)}$, we use Lemmas \ref{lem.stiel} and \ref{lem.concentration_lemma} which leads to
\begin{align}
   &\mathds{E} d_k^{(3)}\le \tfrac{\alpha_1^2}{N_r\alpha_k^2}\Big\langle \mx{C}_{{\widehat{\mx{h}}}_k}, \mx{C}_{\mx{x}_k}\otimes \nonumber\\
   &\Big[({\bs{\Xi}^\prime}_{(k)}^{-1}-{\bs{\Xi}^\prime}^{-1})\mathcal{A}_1(\mx{C}_{{\widehat{\mx{h}}}_1}) (\mx{B}_{(k)}+\rho_{\rm d}  \mx{I}_{N_r})^{-1}+{\bs{\Xi}^\prime}^{-1}\mathcal{A}_1(\mx{C}_{{\widehat{\mx{h}}}_1}) \nonumber\\
   &\left((\mx{B}_{(k)}+\rho_{\rm d}  \mx{I}_{N_r})^{-1}-(\mx{B}+\rho_{\rm d}  \mx{I}_{N_r})^{-1})\right)\Big]\Big\rangle\le\nonumber\\ 
&\tfrac{\alpha_1^2}{N_r\alpha_k^2}{\mu_k^\prime}\Big(\|({\bs{\Xi}^\prime}_{(k)}^{-1}-{\bs{\Xi}^\prime}^{-1})\|\|(\mx{B}_{(k)}+\rho_{\rm d}  \mx{I}_{N_r})^{-1}\| {\rm tr}(\mathcal{A}(\mx{C}_{{\widehat{\mx{h}}}}))+\nonumber\\
&\|{\bs{\Xi}^\prime}^{-1}{\mu_1^\prime}+{\rm tr}((\mx{B}_{(k)}+\rho_{\rm d}  \mx{I}_{N_r})^{-1}-(\mx{B}+\rho_{\rm d}  \mx{I}_{N_r})^{-1}))\|\Big)\le\nonumber\\
&\tfrac{\alpha_1^2{\mu_k^\prime}}{N_r\alpha_k^2} \Big({\mu_k^\prime}^4 \tfrac{1}{|\rho_{\rm d}|^4}\tfrac{P_d}{N_t}(\tfrac{\Omega_1^2\Omega_2 L_1}{N})+ \tfrac{{\mu_1^\prime}}{|\rho_{\rm d}|^2}\Big)\xrightarrow[]{N_r\rightarrow \infty} 0.
\end{align}
Lastly, $d_k^{(4)}$ can be bounded as:
\begin{align}
     &\mathds{E} d_k^{(4)}\le \tfrac{\alpha_1^2\alpha_k^2}{\alpha_k^4}
|\langle \mx{C}_{{\widehat{\mx{h}}}_k}, \mx{C}_{\mx{x}_k}\otimes {\bs{\Xi}^\prime}^{-1}\mathcal{A}_1(\mx{C}_{{\widehat{\mx{h}}}_1}) (\mx{B}+\rho_{\rm d}  \mx{I}_{N_r})^{-1} \rangle|\nonumber\\
&|{{\widehat{\mx{h}}}_k}^H \mx{C}_{\mx{x}_k}\otimes (\mx{B}+\rho_{\rm d}  \mx{I}_{N_r})^{-1} {\widehat{\mx{h}}}_k-N_r m_{{\mx{C}_{{\widehat{\mx{h}}}_k},\mx{B}_{(k)}}}(\rho_{\rm d})|\le
   \nonumber\\  
     &\tfrac{{\mu_k^\prime} \alpha_1^2}{N_r|\rho_{\rm d}|^2}\tfrac{P_d}{N_t}(\tfrac{\Omega_1^2\Omega_2 L_1}{N})
     c^\prime \tfrac{\Omega_1^2\Omega_2\sqrt{P_{{\rm d},k}}\sqrt{N_r}}{N\alpha_k^2|\rho_{\rm d}|}\xrightarrow[]{N_r\rightarrow \infty} 0,
\end{align}
where we used Lemma \ref{lem.concentration_lemma}. As $N_r$ tends to infinity, $e_N$ in \eqref{eq:error_term} vanishes with high probability. Let $\omega^{\star}$ represent the final solution of the fixed-point equation \eqref{eq:fixedpoint_iter} to which it converges. Consequently, as the number of receive antennas $N_r$ approaches infinity, the expression $|N_r m_{\mx{C}_{{\widehat{\mx{h}}}},\mx{B}}(\rho_{\rm d})-\omega^{\star}|$ tends to zero with high probability.

For obtaining the deterministic equivalent of the variance term i.e. ${\rm var}(\gamma)$, we have:
\begin{align}
& {\rm var}(\gamma)= {\rm var}(\alpha_1^2\widehat{\mx{h}}_1^H(\mx{C}_{\mx{x}_1}\otimes\mx{F}_1^{-1})\widehat{\mx{h}}_1)
 =\nonumber\\
 &2\alpha_1^4 tr((\mx{C}_{\mx{x}_1}\otimes\mx{F}_1^{-1})\mx{C}_{\widehat{\mx{h}}_1}(\mx{C}_{\mx{x}_1}\otimes\mx{F}_1^{-1})\mx{C}_{\widehat{\mx{h}}_1})=\nonumber\\
 &2\alpha_1^4 {\rm tr}(\mx{F}_1^{-1}\mathcal{A}_1(\mx{C}_{\widehat{\mx{h}}_1})\mx{F}_1^{-1}\mathcal{A}_1(\mx{C}_{\widehat{\mx{h}}_1}))
\end{align}
where we used \cite[Eq. (379)]{matrix_codebook} in the last equation. Now, by leveraging the result of \cite[Proof of Theorem 2]{wagner2012large}, the deterministic equivalent expression for variance of the SINR is obtained as in \eqref{eq:var_term}.

\section{Proof of Proposition \ref{prop.optimal_beamforming}}\label{proof.prop.beamform}
{\color{\change}
Assume that $\mx{w}=\mx{w}^i$ is the beamforming vector corresponding to the symbol time $i$. Let $\mx{W}\triangleq\mx{w}\mx{w}^H$ and $\mx{Z}_1\triangleq \mathcal{A}_1(\mx{C}_{\widehat{\mx{h}}_1})$. Consider the function $f_1(\mx{Z}_1)\triangleq \log_2(1+\langle \mx{Z}_1, \mx{T} \rangle)-\frac{\langle \mx{Z}_1, \mx{T}\mx{Z}_1 \mx{T}\rangle}{2 \ln(2) (1+\langle \mx{Z}_1, \mx{T} \rangle)^2}$. The derivative of the latter function with respect to the positive definite matrix $\mx{W}$ is obtained by chain rule lemma $\frac{\partial f_1(\mx{Z}_1)}{\partial \mx{W}}=\frac{\partial f_1(\mx{Z}_1)}{\partial \mx{Z}_1}\frac{\partial \mx{Z}_1}{\partial \mx{W}}$ and the following relation:
\begin{align}
&\scalebox{.9}{$\frac{\partial f_1(\mx{Z}_1)}{\partial \mx{Z}_1}=
\frac{\mx{T}}{\ln(2)(1+\langle \mx{Z}_1, \mx{T} \rangle)}  -\frac{2 \mx{T} \mx{Z}_1 \mx{T}}{2\ln(2)(1+\langle \mx{Z}_1, \mx{T} \rangle)^2}
+\frac{2 \mx{T}(\langle \mx{Z}_1, \mx{T} \mx{Z}_1 \mx{T}  \rangle)}{2 \ln(2)(1+\langle \mx{Z}_1, \mx{T} \rangle)^3}$}\nonumber\\
&\scalebox{.9}{$=
\frac{(1+\langle \mx{Z}_1, \mx{T} \rangle)^2 \mx{T}-\mx{T} \mx{Z}_1 \mx{T} (1+\langle \mx{Z}_1, \mx{T} \rangle)+2 \mx{T} (\langle \mx{Z}_1, \mx{T} \mx{Z}_1 \mx{T}  \rangle)   }{\ln(2)(1+\langle \mx{Z}_1, \mx{T} \rangle)^3}$}.
\end{align}
Since $\mx{T}$ and $\mx{Z}_1$ are non-negative definite, ${\rm tr}(\mx{T} \mx{Z}_1\mx{T} \mx{Z}_1)$ is nonnegative. It remains to prove the positive definiteness of $(1+\langle \mx{Z}_1, \mx{T} \rangle)\mx{I}-\mx{T} \mx{Z}_1 $. For arbitrary vector $\mx{x}\in\mathbb{C}^{N_r\times 1}$, we have:
\begin{align}
  &  \mx{x}^H ((1+\langle \mx{Z}_1, \mx{T} \rangle) \mx{I}-\mx{T} \mx{Z}_1 )\mx{x}
    =\nonumber\\
    &\|\mx{x}\|_2^2+\mx{x}^H {\rm tr}(\mx{T} \mx{Z}_1) \mx{x}-\mx{x}^H\mx{T} \mx{Z}_1 \mx{x}\ge\nonumber\\
   & \|\mx{x}\|_2^2+\|\mx{x}\|_2^2tr(\mx{T} \mx{Z}_1)-\|\mx{T} \mx{Z}_1\|\|\mx{x}\|_2^2>0
\end{align}
where in the last relation, we used the fact that the sum of singular values of a matrix is greater than the maximum singular value.
This demonstrates that $f_1(\mx{Z}_1)$ increases with respect to $\mx{Z}_1$. Since $\frac{\partial \mx{Z}_1}{\partial \mx{W}}=\mx{C}_{\widehat{\mx{h}}_1}$ and nonnegative definite, this leads to the conclusion that the function $f_1$ is non-decreasing in $\mx{W}$. This proves that maximizing ${\rm SE}^{\circ}$ with respect to $\mx{W}\triangleq\mx{w}\mx{w}^H$ is equivalent to maximizing $\langle \mx{C}_{\mx{h}_1},\mx{W}\otimes \mx{T}\rangle$. Thus, the optimum $\mx{w}\mx{w}^H$ is the one that maximizes the term inside inner product.}
\begin{align}\label{eq:rell1}
    \left\langle \mx{C}_{{\widehat{\mx{h}}}_1}, \mx{W}\otimes \mx{T}\right\rangle=\sum_{i,l=1}^{N_r}T(i,l)\langle\mx{C}_{{\widehat{\mx{h}}}}^{i,l},\mx{W}\rangle,
\end{align}
where the second term is obtained by partitioning $\mx{C}_{z}$ as in \eqref{eq:R_z_partition} and using Kronecker product properties.
Moreover, the adjoint operator of $\mathcal{G}(\cdot)$ is given by
\begin{align}\label{eq:rell2}
   \langle \mathcal{G}(\mx{C}_{{\widehat{\mx{h}}_1}}), \mx{T}\rangle=\langle \mx{C}_{{\widehat{\mx{h}}}},\mathcal{G}^{\rm Adj}(\mx{T})\rangle. 
\end{align}
The optimal matrix $\mathbf{W}$ is acquired through performing \ac{SVD} on $\mathcal{G}(\mathbf{R}_{\mathbf{z}})$. Singular vectors associated with the maximum singular value are selectively retained. Consequently, the optimal vector $\mathbf{w}$ corresponds to the singular vector responsible for achieving this maximum value.

\section{Proof of Proposition \ref{prop.mmse_receiver} }\label{proof.prop_data_estimate}
 To find the optimal \ac{MMSE} estimate of the data symbol $s(t)$ of the tagged user, we need to solve the following optimization problem:
\begin{align}
     \mx{g}_1^{\star}(t)\triangleq \mathop{\arg\min}_{\mx{g}_1\in\mathbb{C}^{ N_r\times 1}} \mathds{E}_{s_1(t)|\widehat{\mx{h}}_1(t)}|\mx{g}^H\mx{y}_{\rm d}-s_1(t)|^2.
\end{align}
Since the objective function in the unconstrained optimization problem is convex, we can determine the optimal solution by calculating the derivative of the squared objective function with respect to $\mx{g}$ and setting it to zero. This process yields the following relationship:

\begin{align}
\tfrac{\partial \mathds{E}_{{s_1}| \widehat{\mx{h}}(t)}|\mx{g}^H \mx{y}_{\rm d}(t)-s_1|^2}{\partial \mx{g}}=2\mathds{E}\left\{ (\mx{g}^H\mx{y}_{\rm d}(t)-s_1)\mx{y}_{\rm d}(t)^{\rm H}\right\}=\mx{0}
\end{align}
which results in the following solution for the optimum receiver:
\begin{align}
 \mx{g}^{\star}(t)=(\mathds{E}\{\mx{y}_{\rm d}\mx{y}_{\rm d}^{\rm H}| \widehat{\mx{h}}_1(t)\})^{-1}\mathds{E}\{{s_1}\mx{y}_{\rm d}^{\rm H}| \widehat{\mx{h}}_1(t)\}.   
\end{align}
The latter solution requires us to explicitly calculate the expressions $\mathds{E}\{s_1\mx{y}_{\rm d}^{\rm H}| \widehat{\mx{h}}(t)\}$ and $\mathds{E}\{\mx{y}_{\rm d}\mx{y}_{\rm d}^{\rm H}| \widehat{\mx{h}}_1(t)\}$. For $\mathds{E}\{{s_1}\mx{y}_{\rm d}^{\rm H}|\widehat{\mx{h}}_1(t)\}$, it follows that
\begin{align}\label{eq:Exy^H|zeta}
   \mathds{E}\{{s_1}\mx{y}_{\rm d}^{\rm H}| \widehat{\mx{h}}_1(t)\} =\alpha_1 \sqrt{P}_{\rm d} \mx{w}_1^{\rm H} \mathds{E}\{\mx{h}_1(t)^{\rm H} | \widehat{\mx{h}}_1(t)\}
\end{align}
where we used Equation \eqref{eq:data_measurements}, the data symbol of the tagged user is zero mean and the data symbols of users are independent from each other.
To calculate $\mathds{E}\mx{y}_{\rm d}\mx{y}_{\rm d}^H$, it is convenient to first rewrite $\mx{y}_{\rm d}(t)$ in \eqref{eq:data_measurements} as follows:
\begin{align}\label{eq:data_measure_kronicker_form}
\mx{y}_{{\rm d}}(t)=\sum_{k=1}^K \alpha_k (\mx{x}_k^H\otimes \mx{I}) \mx{h}_k(t)  +\mx{n}_d(t) .
\end{align}
Then, by assuming that $\left\{\mx{x}_k\right\}_{k=1}^K$ are zero mean and independent from each other, each element of $\mathds{E}\mx{y}_{\rm d}\mx{y}_{\rm d}^H$ can be calculated as follows:
\begin{align}\label{eq:Eyjyl}
   &\mathds{E}\{y_j{y}_l^\ast| \widehat{\mx{h}}_k(t)\}=\nonumber\\
   &\sum_{k=1}^K \alpha_k^2 {P_d}_k \left\langle \mathds{E}\left\{\mx{h}_k(t)\mx{h}_k(t)^H| \widehat{\mx{h}}_k(t)\right\}, \mx{w}_k\mx{w}_k^H \otimes \mx{e}_l \mx{e}_j^H \right\rangle+\sigma^2_d
\end{align}
Define $\mx{D}_k\triangleq\mathds{E}\left\{\mx{h}_k(t)\mx{h}_k(t)^H| \widehat{\mx{h}}_k(t)\right\} \in\mathbb{C}^{N\times N}$ and also $\mx{D}_k^{\prime}\triangleq \mx{P}_{\rm c}^H \mx{D} \mx{P}_{\rm c}$ where $\mx{P}_{\rm c}$ is the commutation matrix defined in Proposition \ref{prop.mmse_receiver} . Write $\mx{D}_k$ in the form of a matrix with blocks that is divided into sub-matrices $\mx{D}_k^{i,j}\in\mathbb{C}^{N_t\times N_t}$ (see Equation \eqref{eq:D_k}). Then, \eqref{eq:Eyjyl} can be rewritten as below:

\begin{align}
 &\mathds{E}\{y_j{y}_l^\ast| \widehat{\mx{h}}_k(t)\}=\sum_{k=1}^K \alpha_k^2 P_{{\rm d},k} \langle {\mx{D}^{\prime}}_k^{(j,l)}, \mx{w}_k\mx{w}_k^H\rangle+\sigma_d^2
\end{align}
which then can be collected to form:
\begin{align}
    &\mathds{E}\mx{y}_{\rm d}\mx{y}_{\rm d}^H=\begin{bmatrix}
        \mathds{E}|y_1|^2&\hdots&\mathds{E} y_1y_{N_r}^\ast\\
        \vdots&\ddots&\hdots\\
         \mathds{E} y_{N_r} y_1^\ast&\hdots&\mathds{E} |y_{N_r}|^2\\
    \end{bmatrix}
    =\nonumber\\
    &\sum_{k=1}^K \alpha_k^2 \begin{bmatrix}
        \langle {\mx{D}^{\prime}}_k^{(1,1)}, \mx{C}_{\mx{x}_k} \rangle&\hdots&\langle {\mx{D}^{\prime}}_k^{(1,N_r)}, \mx{C}_{\mx{x}_k} \rangle\\
        \vdots&\ddots&\hdots\\
        \langle {\mx{D}^{\prime}}_k^{(N_r,1)}, \mx{C}_{\mx{x}_k} \rangle &\hdots&\langle {\mx{D}^{\prime}}_k^{(N_r,N_r)}, \mx{C}_{\mx{x}_k} \rangle
    \end{bmatrix}  +\sigma_d^2\mx{I}_{N_r},
\end{align}
 which according to the definition of the linear operator $\mathcal{A}_k(\mx{D}_k): \mathbb{C}^{N\times N}\rightarrow\mathbb{C}^{N_r\times N_r}$ results in
 \begin{align}\label{eq:Eyy^H|zeta}
     \mathds{E}\{\mx{y}_{\rm d}\mx{y}_{\rm d}^H|\widehat{\mx{h}}_k(t)\}=\sum_{k=1}^K \alpha_k^2 \mathcal{A}_k(\mx{D}_k)+\sigma_d^2\mx{I}_{N_r}.
 \end{align}
 Now, it remains to calculate $\mx{D}_k$ which is obtained as follows:
 \begin{align}\label{eq:D_k_expression}
 &\mx{D}_k\triangleq\mathds{E}\left\{\mx{h}_k(t)\mx{h}_k(t)^H| \widehat{\mx{h}}_k(t)\right\}  =\mx{C}_{\mx{h}_k(t)|\widehat{\mx{h}}_k(t)}+\nonumber\\
 &\mathds{E}\{\mx{h}_k(t)|\widehat{\mx{h}}_k(t)\}
 \mathds{E}\{\mx{h}_k(t)|\widehat{\mx{h}}_k(t)\}^H.
 \end{align}
 In what follows, we aim to obtain two terms in the above expression, i.e, $\mathds{E}\{\mx{h}_k(t)|\widehat{\mx{h}}_k(t)\}$ and $\mx{C}_{\mx{h}_k(t)|\widehat{\mx{h}}_k(t)}$.
 Rather than assuming any particular distribution on $\mx{h}_k(t)$, we only use its mean and covariance with the estimator. By the orthogonality principle of the LMMSE estimate, the best linear predictor of $\mx{h}_k(t)$ from $\widehat{\mx{h}}_k(t)$ is
 \begin{align}
  &   \mathds{E}\{\mx{h}_k(t)|\widehat{\mx{h}}_k(t)\}\approx\mx{C}_{\mx{h}_k(t),\widehat{\mx{h}}_k(t)}\mx{C}_{\widehat{\mx{h}}_k(t)}^{-1}\widehat{\mx{h}}_k(t)
 \end{align}
  and the corresponding conditional covariance is
  \begin{align}
    \mx{C}_{\mx{h}_k(t)|\widehat{\mx{h}}_k(t)}= \mx{C}_{{\mx{h}}_k(t)}-\mx{C}_{\mx{h}_k(t),\widehat{\mx{h}}_k(t)}\mx{C}_{\widehat{\mx{h}}_k(t)}^{-1}\mx{C}_{\widehat{\mx{h}}_k(t),\mx{h}_k(t)}
 \end{align}
 These expressions hold for any zero-mean process $\mx{h}_k(t)$, regardless of its full distribution profile.
 Due to the fact that the MMSE estimation error is orthogonal to any function of the measurements and the zero-mean assumption of the channel, we have that
\begin{align}\label{eq:orthogonality_principal}
    \mathds{E}\{ (\mx{h}_k(t) - \widehat{\mx{h}}_k(t)) \widehat{\mx{h}}_k(t)^H \} = \mx{0}.
\end{align}
This leads to
\begin{align}\label{eq:Eh|hhat}
    \mathds{E}\{\mx{h}_k(t)|\widehat{\mx{h}}_k(t)\}\approx\widehat{\mx{h}}_k(t)
\end{align}
and
\begin{align}
  \mx{C}_{\mx{h}_k(t)|\widehat{\mx{h}}_k(t)}=   \mx{C}_{{\mx{h}}_k(t)}- \mx{C}_{\widehat{\mx{h}}_k(t)} \triangleq \mx{Q}_k
\end{align}
which hold under the additional assumption that the conditional mean is well approximated as a linear function of $\widehat{\mx{h}}(t)$. One of the cases in which this happens is when the channel distribution behaves as a Gaussian distribution.
Replacing the above equations into \eqref{eq:D_k_expression}, we reach:
\begin{align}\label{eq:D_k_final}
    \mx{D}_k=\mx{Q}_k+\widehat{\mx{h}}_k(t)\widehat{\mx{h}}_k^{\rm H}(t).
\end{align}
Thus, by having $\mx{D}_k$ in \eqref{eq:D_k_final}, the expression $\mathds{E}\{\mx{y}_{\rm d}\mx{y}_{\rm d}^{\rm H}|\widehat{\mx{h}}_k(t)\}$ in \eqref{eq:Eyy^H|zeta} can be calculated.
Also for $\mathds{E}\{s\mx{y}_{\rm d}^{\rm H}|\widehat{\mx{h}}_k(t)\}$, we can use \eqref{eq:Eh|hhat} to have:
\begin{align}
\mathds{E}\{{s} \mx{y}_{\rm d}^{\rm H}|\widehat{\mx{h}}_k(t)\}=\alpha\sqrt{P}_{\rm d} \mx{w}_1^{\rm H}\textbf{vec}^{-H}(\widehat{\mx{h}}_k(t)) . 
\end{align}
The latter equation together with \eqref{eq:Eyy^H|zeta} gives the final result in Proposition \ref{prop.mmse_receiver}

\section{Proof of Lemma \ref{lem.stiel} }\label{proof.lem.stiel1}
Let $\mx{B}=\mx{U}^H\mx{\Lambda}\mx{U}$ and $\mx{C}=\mx{V}^H \mx{\Sigma} \mx{V}$ be the eigenvalue decomposition of $\mx{B}$ and $\mx{C}$ where $\mx{U}$ and $\mx{V}$ are some orthonormal matrices. It then follows that
\begin{align}
&\langle \mx{M} \mx{C} \mx{M}^H , (\mx{B}+\rho_{\rm d}  \mx{I})^{-1}\rangle =\langle \mathcal{A}(\mx{zz}^H), (\mx{B}+\rho_{\rm d}  \mx{I})^{-1}\rangle=
\nonumber\\
&\langle \mx{zz}^H, \mx{C}\otimes (\mx{B}+\rho_{\rm d}  \mx{I})^{-1} \rangle.
\end{align}
We can proceed by applying the eigenvalue decomposition of $\mx{B}$ and $\mx{C}$ to have :

\begin{align}\label{eq:stiel1}
&f(\rho_{\rm d} )=\langle \mx{M} \mx{C} \mx{M}^H , (\mx{B}+\rho_{\rm d}  \mx{I})^{-1}\rangle =\nonumber\\
&
{\widehat{\mx{h}}}^H \left(\mx{V}^H \mx{\Sigma} \mx{V}\otimes (\mx{U}^H \mx{\Lambda} \mx{U}+\rho_{\rm d}  \mx{I})^{-1} \right) {\widehat{\mx{h}}}=\nonumber\\
&{\widehat{\mx{h}}}^H (\mx{U}^H \otimes \mx{V}^H) \left(\mx{\Sigma}\otimes(\mx{\Lambda} +\rho_{\rm d}  \mx{I})^{-1}\right) (\mx{U}\otimes \mx{V}) {\widehat{\mx{h}}}\nonumber\\
&=
\sum_{i,j} \tfrac{|\mx{u}_i^H \mx{M} \mx{v}_j|^2}{\lambda+\rho_{\rm d} },
\end{align}
where $\mx{u}_i^T$ and $\mx{v}_j^T$ are the $i$-th and $j$-th rows of $\mx{U}$ and $\mx{V}$, respectively. As it turns out from \eqref{eq:stiel1}, $f(\rho_{\rm d} )$ is the Stieltjes transform of a finite measure in $\mathbb{C}^{-}$. Based on \cite[Corollary 3.1]{couillet2011random}, it holds that, for any $t>0$,
\begin{align}
|\tfrac{1}{1+t f(\rho_{\rm d} )}|\le 1.
\end{align}
\section{Proof of Lemma \ref{lem.SINR_equivalency}}\label{proof.lem.SINR_equivalency}
We can start the proof by writing the expression \eqref{eq:sinr_rel} as below:
\begin{align}
&\tfrac{{\widehat{\mx{h}}}^{\rm H} \mx{A} {\widehat{\mx{h}}}}{\beta-{\widehat{\mx{h}}}^{\rm H} \mx{A} {\widehat{\mx{h}}}}=\tfrac{{\widehat{\mx{h}}}^{\rm H} \mx{A} {\widehat{\mx{h}}}}{\beta}\left( 1+\tfrac{{\widehat{\mx{h}}}^{\rm H} \mx{A} {\widehat{\mx{h}}}}{\beta-{\widehat{\mx{h}}}^{\rm H} \mx{A} {\widehat{\mx{h}}}}\right).
\end{align}
We can write the latter equation as
\begin{align}\label{eq:sinr_rel2}
    &\tfrac{{\widehat{\mx{h}}}^{\rm H} \mx{A} {\widehat{\mx{h}}}}{\beta-{\widehat{\mx{h}}}^{\rm H} \mx{A} {\widehat{\mx{h}}}}=\tfrac{{\widehat{\mx{h}}}^{\rm H} \mx{A} {\widehat{\mx{h}}}}{\beta}+\tfrac{{\widehat{\mx{h}}}^{\rm H} \mx{A} {\widehat{\mx{h}}}}{\beta}\left(1-\tfrac{{\widehat{\mx{h}}}^{\rm H} \mx{A} {\widehat{\mx{h}}}}{\beta}\right)^{-1} \tfrac{{\widehat{\mx{h}}}^{\rm H} \mx{A} {\widehat{\mx{h}}}}{\beta}=\nonumber\\
    &{\widehat{\mx{h}}}^{\rm H}\left[ \tfrac{\mx{A}}{\beta}+\tfrac{\mx{A}}{\beta} {\widehat{\mx{h}}} \left(1-\tfrac{{\widehat{\mx{h}}}^{\rm H} \mx{A} {\widehat{\mx{h}}}}{\beta}\right)^{-1}{\widehat{\mx{h}}}^{\rm H} \tfrac{\mx{A}}{\beta}\right] {\widehat{\mx{h}}}
\end{align}
Now, by using Woodbury matrix lemma \cite{horn2013matrix}, we can rewrite \eqref{eq:sinr_rel2} as 
\begin{align}
 &{\widehat{\mx{h}}}^{\rm H}\left[ \tfrac{A}{\beta}+\tfrac{\mx{A}}{\beta} {\widehat{\mx{h}}} \left(1-\tfrac{{\widehat{\mx{h}}}^{\rm H} \mx{A} {\widehat{\mx{h}}}}{\beta}\right)^{-1}{\widehat{\mx{h}}}^{\rm H} \tfrac{\mx{A}}{\beta}\right] {\widehat{\mx{h}}}=\nonumber\\
 &{\widehat{\mx{h}}}^{\rm H}  \left(\mx{A}^{-1}-\tfrac{{\widehat{\mx{h}}} {\widehat{\mx{h}}}^{\rm H}}{\beta} \right)^{-1}{\widehat{\mx{h}}} =\beta {\widehat{\mx{h}}}^{\rm H} \left(\beta \mx{A}^{-1}-{\widehat{\mx{h}}} {\widehat{\mx{h}}}^{\rm H}\right)^{-1} {\widehat{\mx{h}}}, 
\end{align}
which proves the result.
\section{Proof of Proposition \ref{prop.A(zz^T)}}\label{proof.prop.A(zz^T)}
Consider a matrix $\widehat{\mx{H}}_k=[{\bs{\iota}^1},...,{\bs{\iota}^{N_r}}]^T
\in\mathbb{C}^{N_r\times N_t}$ and its vectorized version $\mx{z}\triangleq \textbf{vec}(\widehat{\mx{H}}_k)$. Here, $\bs{\iota}^i,i=1,..., N_r$ are column vectors of size $N_t\times 1$ and ${{\mx{z}}}\in\mathbb{C}^{N_r N_t\times 1}$ can be partitioned as $\mx{z}=[{\mx{z}^1}^T,..., {\mx{z}^{N_r}}^T]^T$,
where $\mx{z}_i$s are of size $N_t\times 1$. Due to the definition of operator $\mathcal{A}$ in Equation \eqref{eq:A_operator} of the paper, the left-hand side of Equation \eqref{eq:A(zz^T)} can be stated as follows:
\begin{align}
  \mathcal{A}_k(\mx{z}\mx{z}^T)=\begin{bmatrix}
        \langle \mx{z}^1{\mx{z}^1}^{\rm H},\mx{C}_{\mx{x}_k} \rangle& \hdots& \langle \mx{z}^1{\mx{z}^{N_r}}^{\rm H},\mx{C}_{\mx{x}_k} \rangle\\  
        \vdots &\ddots&\vdots\\
         \langle \mx{z}^{N_r}{\mx{z}^1}^{\rm H},\mx{C}_{\mx{x}_k} \rangle& \hdots& \langle \mx{z}^{N_r}{\mx{z}^{N_r}}^{\rm H},\mx{C}_{\mx{x}_k} \rangle
    \end{bmatrix}.
\end{align}
By the definition of $\textbf{vec}(\cdot)$ operator, we have $\mx{z}^i=\bs{\iota}^i$. Thus, this leads to
\begin{align}
  &\mathcal{A}_k(\mx{z}\mx{z}^T)=\begin{bmatrix}
        \langle {\bs{\iota}^1}^T\mx{C}_{\mx{x}_k}\bs{\iota}^1 \rangle& \hdots& \langle {\bs{\iota}^1}^T\mx{C}_{\mx{x}_k}\bs{\iota}^{N_r} \rangle\\  
        \vdots &\ddots&\vdots\\
         \langle {\bs{\iota}^{N_r}}^T\mx{C}_{\mx{x}_k}\bs{\iota}^1 \rangle& \hdots& \langle {\bs{\iota}^{N_r}}^T\mx{C}_{\mx{x}_k}\bs{\iota}^{N_r} \rangle
    \end{bmatrix}=\nonumber\\
    &\widehat{\mx{H}}\mx{C}_{\mx{x}_k} \widehat{\mx{H}}^T,
\end{align}
which is exactly the right-hand side of Equation \ref{eq:A(zz^T)}.

\section{Proof of Lemma \ref{lem.adjoint}}\label{proof.lem.adjoint}
For any matrices $\mx{D}\in\mathbb{C}^{N_tN_r\times N_tN_r}$ and $\mx{Y}\in\mathbb{C}^{N_r\times N_r}$, it holds that

\begin{align}
 \langle \mathcal{A}_k(\mx{D}),\mx{Y} \rangle=\langle \mx{D},\mathcal{A}_k^{\rm Adj} \mx{Y} \rangle.     
\end{align}
This leads to
{
\begin{align}
   &\sum_{i,j=1}^{N_r} Y^{\ast}_{i,j} \langle \mx{D}^{i,j}, \mx{C}_{\mx{x}_k}\rangle=\sum_{i,j=1}^{N_r}\langle \mx{D}^{i,j}, \mx{C}_{\mx{x}_k} Y_{i,j}  \rangle   =\nonumber\\
   &
  \scalebox{.8}{$ \left\langle \begin{bmatrix}
         \mx{D}_k^{{(1,1)}}&\hdots&\mx{D}_k^{(1,N_r)}\\
         \vdots&\ddots&\hdots\\
         \mx{D}_k^{{(N_r,1)}}&\hdots&\mx{D}_k^{(N_r,N_r)}
     \end{bmatrix}, \begin{bmatrix}
         \mx{C}_{\mx{x}_k}Y_{1,1}&\hdots&\mx{C}_{\mx{x}_k}Y_{1,N_r}\\
         \vdots&\ddots&\hdots\\
         \mx{C}_{\mx{x}_k}Y_{N_r,1}&\hdots&\mx{C}_{\mx{x}_k}Y_{N_r,N_r}
     \end{bmatrix}  \right\rangle $}\nonumber\\
     &=\left\langle \mx{D}_k, \mx{C}_{\mx{x}_k}\otimes\mx{Y}  \right\rangle,
\end{align}}
\noindent where $Y_{i,j}$ is the $(i,j)$ element of $\mx{Y}$ and $Y^{\ast}_{i,j}$ is the complex conjugate of $Y_{i,j}$. This shows that
\begin{align}
    \mathcal{A}^{\rm Adj}_k(\mx{Y})=\mx{C}_{\mx{x}_k}\otimes \mx{Y}.
\end{align}

\bibliography{references}

\begin{thebibliography}{10}
\providecommand{\url}[1]{#1}
\csname url@samestyle\endcsname
\providecommand{\newblock}{\relax}
\providecommand{\bibinfo}[2]{#2}
\providecommand{\BIBentrySTDinterwordspacing}{\spaceskip=0pt\relax}
\providecommand{\BIBentryALTinterwordstretchfactor}{4}
\providecommand{\BIBentryALTinterwordspacing}{\spaceskip=\fontdimen2\font plus
\BIBentryALTinterwordstretchfactor\fontdimen3\font minus
  \fontdimen4\font\relax}
\providecommand{\BIBforeignlanguage}[2]{{%
\expandafter\ifx\csname l@#1\endcsname\relax
\typeout{** WARNING: IEEEtran.bst: No hyphenation pattern has been}%
\typeout{** loaded for the language `#1'. Using the pattern for}%
\typeout{** the default language instead.}%
\else
\language=\csname l@#1\endcsname
\fi
#2}}
\providecommand{\BIBdecl}{\relax}
\BIBdecl

\bibitem{rhee2004optimality}
W.~Rhee, W.~Yu, and J.~M. Cioffi, ``The optimality of beamforming in uplink
  multiuser wireless systems,'' \emph{IEEE Transactions on Wireless
  Communications}, vol.~3, no.~1, pp. 86--96, 2004.

\bibitem{rusek2012scaling}
F.~Rusek, D.~Persson, B.~K. Lau, E.~G. Larsson, T.~L. Marzetta, O.~Edfors, and
  F.~Tufvesson, ``Scaling up {MIMO}: Opportunities and challenges with very
  large arrays,'' \emph{IEEE signal processing magazine}, vol.~30, no.~1, pp.
  40--60, 2012.

\bibitem{jafar2001channel}
S.~A. Jafar, S.~Vishwanath, and A.~Goldsmith, ``Channel capacity and
  beamforming for multiple transmit and receive antennas with covariance
  feedback,'' in \emph{ICC 2001. IEEE International Conference on
  Communications. Conference Record (Cat. No. 01CH37240)}, vol.~7.\hskip 1em
  plus 0.5em minus 0.4em\relax IEEE, 2001, pp. 2266--2270.

\bibitem{jafar2004transmitter}
S.~A. Jafar and A.~Goldsmith, ``Transmitter optimization and optimality of
  beamforming for multiple antenna systems,'' \emph{IEEE Transactions on
  Wireless Communications}, vol.~3, no.~4, pp. 1165--1175, 2004.

\bibitem{lu2014overview}
L.~Lu, G.~Y. Li, A.~L. Swindlehurst, A.~Ashikhmin, and R.~Zhang, ``An overview
  of massive {MIMO}: Benefits and challenges,'' \emph{IEEE journal of selected
  topics in signal processing}, vol.~8, no.~5, pp. 742--758, 2014.

\bibitem{soysal2009optimality}
A.~Soysal and S.~Ulukus, ``Optimality of beamforming in fading {MIMO} multiple
  access channels,'' \emph{IEEE Transactions on Communications}, vol.~57,
  no.~4, pp. 1171--1183, 2009.

\bibitem{Hassibi:03}
B.~Hassibi and B.~M. Hochwald, ``How much training is needed in
  multiple-antenna wireless links ?'' \emph{IEEE Transactions on Information
  Theory}, vol.~49, no.~4, pp. 951--963, April 2003.

\bibitem{Hoydis:13}
J.~Hoydis, S.~T. Brink, and M.~Debbah, ``Massive {MIMO} in the {UL/DL} of
  cellular networks: How many antennas do we need ?'' \emph{IEEE Journal on
  Selected Areas in Communications}, vol.~31, no.~2, pp. 160--171, Feb. 2013.

\bibitem{Fodor:16}
G.~Fodor, P.~D. Marco, and M.~Telek, ``On the impact of antenna correlation and
  {CSI} errors on the pilot-to-data power ratio,'' \emph{IEEE Transactions on
  Communications}, vol.~64, no.~6, pp. 2622--2633, 2016.

\bibitem{jorswieck2004channel}
E.~A. Jorswieck and H.~Boche, ``Channel capacity and capacity-range of
  beamforming in {MIMO} wireless systems under correlated fading with
  covariance feedback,'' \emph{IEEE Transactions on Wireless Communications},
  vol.~3, no.~5, pp. 1543--1553, 2004.

\bibitem{marzetta2010noncooperative}
T.~L. Marzetta, ``Noncooperative cellular wireless with unlimited numbers of
  base station antennas,'' \emph{IEEE transactions on wireless communications},
  vol.~9, no.~11, pp. 3590--3600, 2010.

\bibitem{papazafeiropoulos2015deterministic}
A.~K. Papazafeiropoulos and T.~Ratnarajah, ``Deterministic equivalent
  performance analysis of time-varying massive {MIMO} systems,'' \emph{IEEE
  Transactions on Wireless Communications}, vol.~14, no.~10, pp. 5795--5809,
  2015.

\bibitem{Kobyashi:11}
M.~Kobayashi, N.~Jindal, and G.~Caire, ``Training and feedback optimization for
  multiuser {MIMO} downlink,'' \emph{IEEE Transactions on Communications},
  vol.~59, no.~8, pp. 2228--2240, 2011.

\bibitem{Khalilsarai:23}
M.~B. Khalilsarai, Y.~Song, T.~Yang, and G.~Caire, ``Fdd massive mimo channel
  training: Optimal rate-distortion bounds and the spectral efficiency of
  one-shot schemes,'' \emph{IEEE Transactions on Wireless Communications},
  vol.~22, no.~9, pp. 6018--6032, 2023.

\bibitem{daei2024improved}
S.~Daei, M.~Skoglund, and G.~Fodor, ``Improved downlink channel estimation in
  time-varying fdd massive mimo systems,'' in \emph{2024 IEEE 25th
  International Workshop on Signal Processing Advances in Wireless
  Communications (SPAWC)}.\hskip 1em plus 0.5em minus 0.4em\relax IEEE, 2024,
  pp. 571--575.

\bibitem{non_stationary_model}
Z.~Lian, L.~Jiang, C.~He, and D.~He, ``A non-stationary {3-D} wideband {GBSM}
  for {HAP-MIMO} communication systems,'' \emph{IEEE Transactions on Vehicular
  Technology}, vol.~68, no.~2, pp. 1128--1139, 2018.

\bibitem{Banerjee:20}
S.~Banerjee, R.~Bhattacharjee, and A.~Sinha, ``Fundamental limits of
  age-of-information in stationary and non-stationary environments,'' in
  \emph{2020 IEEE International Symposium on Information Theory (ISIT)}, 2020,
  pp. 1741--1746.

\bibitem{Iimori:21}
H.~Iimori, G.~T.~F. deAbreu, D.~Gonz\'alez~G., and O.~Gonsa, ``Mitigating
  channel aging and phase noise in millimeter wave {MIMO} systems,'' \emph{IEEE
  Transactions on Vehicular Technology}, vol.~70, no.~7, pp. 7237--7242, 2021.

\bibitem{Loschenbrand:22}
D.~L\"oschenbrand, M.~Hofer, L.~Bernado, S.~Zelenbaba, and T.~Zemen, ``Towards
  cell-free massive {MIMO}: A measurement-based analysis,'' \emph{IEEE Access},
  vol.~10, pp. 89\,232--89\,247, 2022.

\bibitem{Song:22}
Z.~Song, T.~Yang, X.~Wu, H.~Feng, and B.~Hu, ``Regret of age-of-information
  bandits in nonstationary wireless networks,'' \emph{IEEE Wireless
  Communications Letters}, vol.~11, no.~11, pp. 2415--2419, 2022.

\bibitem{cheng2022channel}
X.~Cheng, Z.~Huang, and L.~Bai, ``Channel nonstationarity and consistency for
  beyond 5g and 6g: A survey,'' \emph{IEEE Communications Surveys \&
  Tutorials}, vol.~24, no.~3, pp. 1634--1669, 2022.

\bibitem{zou2023joint}
Z.~Zou, M.~Careem, A.~Dutta, and N.~Thawdar, ``Joint spatio-temporal precoding
  for practical non-stationary wireless channels,'' \emph{IEEE Transactions on
  Communications}, vol.~71, no.~4, pp. 2396--2409, 2023.

\bibitem{bian2021general}
J.~Bian, C.-X. Wang, X.~Gao, X.~You, and M.~Zhang, ``A general {3D}
  non-stationary wireless channel model for {5G} and beyond,'' \emph{IEEE
  Transactions on Wireless Communications}, vol.~20, no.~5, pp. 3211--3224,
  2021.

\bibitem{daei2024towards}
S.~Daei, G.~Fodor, M.~Skoglund, and M.~Telek, ``Toward optimal pilot spacing
  and power control in multi-antenna systems operating over non-stationary
  rician aging channels,'' \emph{IEEE Transactions on Communications}, vol.~73,
  no.~6, pp. 3761--3777, 2025.

\bibitem{Fodor:2021}
G.~Fodor, S.~Fodor, and M.~Telek, ``Performance analysis of a linear {MMSE}
  receiver in time-variant rayleigh fading channels,'' \emph{IEEE Transactions
  on Communications}, vol.~69, no.~6, pp. 4098--4112, 2021.

\bibitem{Fodor:22}
G.~{Fodor}, S.{Fodor}, and M.{Telek}, ``{MU-MIMO} receiver design and
  performance analysis in time-varying {R}ayleigh fading,'' \emph{IEEE
  Transactions on Communications}, vol.~70, no.~2, pp. 1214--1228, 2022.

\bibitem{Abeida:10}
H.~Abeida, ``Data-aided {SNR} estimation in time-variant {Rayleigh} fading
  channels,'' \emph{IEEE Transactions on Signal Processing}, vol.~58, no.~11,
  pp. 5496--5507, Nov. 2010.

\bibitem{Hijazi:10}
H.~Hijazi and L.~Ros, ``Joint data {QR}-detection and {Kalman} estimation for
  {OFDM} time-varying {Rayleigh} channel complex gains,'' \emph{IEEE
  Transactions on Communications}, vol.~58, no.~1, pp. 170--177, Jan. 2010.

\bibitem{Truong:13}
K.~T. {Truong} and R.~W. {Heath}, ``Effects of channel aging in massive {MIMO}
  systems,'' \emph{Journal of Communications and Networks}, vol.~15, no.~4, pp.
  338--351, 2013.

\bibitem{Kong:2015}
C.~{Kong}, C.~{Zhong}, A.~K. {Papazafeiropoulos}, M.~{Matthaiou}, and
  Z.~{Zhang}, ``Sum-rate and power scaling of massive {MIMO} systems with
  channel aging,'' \emph{IEEE Transactions on Communications}, vol.~63, no.~12,
  pp. 4879--4893, 2015.

\bibitem{Yuan:20}
J.~{Yuan}, H.~Q. {Ngo}, and M.~{Matthaiou}, ``Machine learning-based channel
  prediction in massive {MIMO} with channel aging,'' \emph{IEEE Transactions on
  Wireless Communications}, vol.~19, no.~5, pp. 2960--2973, 2020.

\bibitem{Kim:20}
H.~{Kim}, S.~{Kim}, H.~{Lee}, C.~{Jang}, Y.~{Choi}, and J.~{Choi}, ``Massive
  {MIMO} channel prediction: {Kalman} filtering vs. machine learning,''
  \emph{IEEE Transactions on Communications}, pp. 1--1, 2020, early access.

\bibitem{Fodor:23}
S.~Fodor, G.~Fodor, D.~G{\"u}rg{\"u}no{\u{g}}lu, and M.~Telek, ``Optimizing
  pilot spacing in {MU-MIMO} systems operating over aging channels,''
  \emph{IEEE Transactions on Communications}, vol.~71, pp. 3708--3720, 2023.

\bibitem{daei2024optimaltransmitter}
S.~Daei, G.~Fodor, and M.~Skoglund, ``Optimal transmitter design and pilot
  spacing in mimo non-stationary aging channels,'' in \emph{2024 32nd European
  Signal Processing Conference (EUSIPCO)}.\hskip 1em plus 0.5em minus
  0.4em\relax IEEE, 2024, pp. 2127--2131.

\bibitem{Savazzi:09}
S.~Savazzi and U.~Spagnolini, ``On the pilot spacing constraints for continuous
  time-varying fading channels,'' \emph{IEEE Transactions on Communications},
  vol.~57, no.~11, pp. 3209--3213, 2009.

\bibitem{Savazzi:09B}
S.~{Savazzi} and U.~{Spagnolini}, ``Optimizing training lengths and training
  intervals in time-varying fading channels,'' \emph{IEEE Transactions on
  Signal Processing}, vol.~57, no.~3, pp. 1098--1112, 2009.

\bibitem{bai2008clt}
J.~W. Silverstein, ``Clt for linear spectral statistics of large-dimensional
  sample covariance matrices,'' in \emph{Advances In Statistics}.\hskip 1em
  plus 0.5em minus 0.4em\relax World Scientific, 2008, pp. 281--333.

\bibitem{wagner2012large}
S.~Wagner, R.~Couillet, M.~Debbah, and D.~T. Slock, ``Large system analysis of
  linear precoding in correlated miso broadcast channels under limited
  feedback,'' \emph{IEEE transactions on information theory}, vol.~58, no.~7,
  pp. 4509--4537, 2012.

\bibitem{tulino2004random}
A.~M. Tulino, S.~Verd{\'u} \emph{et~al.}, ``Random matrix theory and wireless
  communications,'' \emph{Foundations and Trends{\textregistered} in
  Communications and Information Theory}, vol.~1, no.~1, pp. 1--182, 2004.

\bibitem{adhikary2013joint}
A.~Adhikary, J.~Nam, J.~Y. Ahn, and G.~Caire, ``Joint spatial division and
  multiplexing:the large-scale array regime,'' \emph{IEEE Transactions on
  Information Theory}, vol.~59, no.~10, pp. 6441--6463, 2013.

\bibitem{couillet2011deterministic}
R.~Couillet, M.~Debbah, and J.~W. Silverstein, ``A deterministic equivalent for
  the analysis of correlated {MIMO} multiple access channels,'' \emph{IEEE
  Transactions on Information Theory}, vol.~57, no.~6, pp. 3493--3514, 2011.

\bibitem{asgharimoghaddam2018decentralizing}
H.~Asgharimoghaddam, A.~T{\"o}lli, L.~Sanguinetti, and M.~Debbah,
  ``Decentralizing multicell beamforming via deterministic equivalents,''
  \emph{IEEE Transactions on Communications}, vol.~67, no.~3, pp. 1894--1909,
  2018.

\bibitem{marzetta2016fundamentals}
T.~L. Marzetta, E.~G. Larsson, H.~Yang, and H.~Q. Ngo, \emph{Fundamentals of
  massive MIMO}.\hskip 1em plus 0.5em minus 0.4em\relax Cambridge University
  Press, 2016.

\bibitem{ngo2013energy}
H.~Q. Ngo, E.~G. Larsson, and T.~L. Marzetta, ``Energy and spectral efficiency
  of very large multiuser mimo systems,'' \emph{IEEE Transactions on
  Communications}, vol.~61, no.~4, pp. 1436--1449, 2013.

\bibitem{marzetta2006much}
T.~L. Marzetta, ``How much training is required for multiuser mimo?'' in
  \emph{2006 fortieth asilomar conference on signals, systems and
  computers}.\hskip 1em plus 0.5em minus 0.4em\relax IEEE, 2006, pp. 359--363.

\bibitem{couillet2012random}
R.~Couillet, J.~Hoydis, and M.~Debbah, ``Random beamforming over quasi-static
  and fading channels: A deterministic equivalent approach,'' \emph{IEEE
  Transactions on Information Theory}, vol.~58, no.~10, pp. 6392--6425, 2012.

\bibitem{baddour2004accurate}
K.~E. Baddour and N.~C. Beaulieu, ``Accurate simulation of multiple
  cross-correlated rician fading channels,'' \emph{IEEE transactions on
  communications}, vol.~52, no.~11, pp. 1980--1987, 2004.

\bibitem{jakes1974mobile}
W.~C. Jakes, ``Mobile microwave communication,'' 1974.

\bibitem{bai2010spectral}
Z.~Bai and J.~W. Silverstein, \emph{Spectral analysis of large dimensional
  random matrices}.\hskip 1em plus 0.5em minus 0.4em\relax Springer, 2010,
  vol.~20.

\bibitem{jose2011pilot}
J.~Jose, A.~Ashikhmin, T.~L. Marzetta, and S.~Vishwanath, ``Pilot contamination
  and precoding in multi-cell tdd systems,'' \emph{IEEE Transactions on
  Wireless Communications}, vol.~10, no.~8, pp. 2640--2651, 2011.

\bibitem{bjornson2017massive}
E.~Bj{\"o}rnson, J.~Hoydis, and L.~Sanguinetti, ``Massive mimo has unlimited
  capacity,'' \emph{IEEE Transactions on Wireless Communications}, vol.~17,
  no.~1, pp. 574--590, 2017.

\bibitem{bjornson2016massive}
E.~Bj{\"o}rnson, L.~Sanguinetti, and M.~Debbah, ``Massive mimo with imperfect
  channel covariance information,'' in \emph{2016 50th Asilomar Conference on
  Signals, Systems and Computers}.\hskip 1em plus 0.5em minus 0.4em\relax IEEE,
  2016, pp. 974--978.

\bibitem{onlineCCS}
C.~Park and B.~Lee, ``Online compressive covariance sensing,'' \emph{Signal
  Processing}, vol. 162, pp. 1--9, 2019.

\bibitem{koopa_theory_ML}
Y.~Liu, C.~Li, J.~Wang, and M.~Long, ``Koopa: Learning non-stationary time
  series dynamics with koopman predictors,'' \emph{Advances in Neural
  Information Processing Systems}, vol.~36, 2024.

\bibitem{Wagner:2012}
S.~{Wagner}, R.~{Couillet}, M.~{Debbah}, and D.~T.~M. {Slock}, ``Large system
  analysis of linear precoding in correlated {MISO} broadcast channels under
  limited feedback,'' \emph{IEEE Transactions on Information Theory}, vol.~58,
  no.~7, pp. 4509--4537, 2012.

\bibitem{pap2002convergence}
P.~D. L.-E. PAP, ``Convergence theorems for set functions,'' \emph{Handbook of
  Measure Theory}, vol.~1, pp. 125--178, 2002.

\bibitem{harmonic2017}
E.~W. Weisstein, ``Harmonic addition theorem,'' \emph{Mathworld-A Wolfram Web
  Resource}, 2017.

\bibitem{continous_mapp}
H.~B. Mann and A.~Wald, ``On stochastic limit and order relationships,''
  \emph{The Annals of Mathematical Statistics}, vol.~14, no.~3, pp. 217--226,
  1943.

\bibitem{second_order_delta}
A.~vd~Vaart, ``Asymptotic statistics. cambridge series in statistical and
  probabilistic mathematics,'' 1998.

\bibitem{silverstein1995empirical}
J.~W. Silverstein and Z.~Bai, ``On the empirical distribution of eigenvalues of
  a class of large dimensional random matrices,'' \emph{Journal of Multivariate
  analysis}, vol.~54, no.~2, pp. 175--192, 1995.

\bibitem{couillet2011random}
R.~Couillet and M.~Debbah, \emph{Random matrix methods for wireless
  communications}.\hskip 1em plus 0.5em minus 0.4em\relax Cambridge University
  Press, 2011.

\bibitem{bai2007signal}
Z.~Bai and J.~W. Silverstein, ``On the signal-to-interference ratio of cdma
  systems in wireless communications,'' \emph{Ann. Appl. Probab.}, vol.~17,
  no.~1, pp. 81--101, 2007.

\bibitem{matrix_codebook}
K.~B. Petersen, M.~S. Pedersen \emph{et~al.}, ``The matrix cookbook,''
  \emph{Technical University of Denmark}, vol.~7, no.~15, p. 510, 2008.

\bibitem{horn2013matrix}
R.~A. Horn and C.~R. Johnson, \emph{Matrix Analysis}, 2nd~ed.\hskip 1em plus
  0.5em minus 0.4em\relax Cambridge University Press, 2013.

\end{thebibliography}
\end{document}